\documentclass[structabstract]{aa}  
\usepackage{graphicx}
\usepackage{txfonts}
\usepackage{amssymb}
\usepackage{amstext}
\usepackage{rotate}
\usepackage{rotating}
\usepackage{supertabular}
\usepackage{epsfig}
\usepackage{lscape}
\usepackage{enumerate}
\usepackage{natbib}

\newcommand{\gsim}{\hbox{\rlap{\lower.55ex\hbox{$\sim$}} \kern-.3em
\raise.4ex \hbox{$>$}}}
\newcommand{\lsim}{\hbox{\rlap{\lower.55ex\hbox{$\sim$}} \kern-.3em
\raise.4ex \hbox{$<$}}}

\newcommand{\sha}{\textsc{[S\,ii]}$\lambda\lambda$6717,6731/H$\alpha$}

\newcommand{\ohb}{\textsc{[O\,iii]}$\lambda$5007/H$\beta$}
\newcommand{\hb}{H$\beta$}

\newcommand{\ha}{H$\alpha$}

\newcommand{\hi}{\textsc{H\,i}}
\newcommand{\hei}{He\,\textsc{i}}

\newcommand{\sii}{\textsc{[S\,ii]}}

\newcommand{\ariv}{[Ar\,\textsc{iv}]}

\newcommand{\oiii}{\textsc{[O\,iii]}}
\newcommand{\heii}{He\,\textsc{ii}}

\newcommand{\ghiir}{GH\,\textsc{ii}R}
\newcommand{\hii}{H\,\textsc{ii}}

\begin{document}
   \title{He\,\textsc{i} in the central Giant \hii\ Region of NGC~5253}

   \subtitle{A 2D observational approach to collisional and radiative transfer  effects\thanks{Based on observations  
       collected at the European Organisation for Astronomical
       Research in the Southern Hemisphere, Chile (ESO Programme
       078.B-0043 and 383.B-0043) and at the Gemini South Telescope (Programme
GS-2008A-Q-25).}}

   \author{A. Monreal-Ibero\inst{1}
          \and
         J. R. Walsh\inst{2} 
         \and
         M. S. Westmoquette\inst{2}
         \and
         J. M. V\'{\i}lchez\inst{1}
         }

   \offprints{A. Monreal-Ibero}

   \institute{Instituto de Astrof\'{\i}sica de Andaluc\'{\i}a (CSIC), C/
              Camino Bajo de Hu\'etor, 50, 18008 Granada, Spain. 
              \email{ami@iaa.es}
   \and
   European Southern Observatory, Karl-Schwarzschild Strasse 2, D-85748 Garching bei M\"unchen, Germany.
} 
   \date{Received: 1 March 2013 ; Accepted: 19 March 2013}

 
  \abstract
   {NGC~5253 is a nearby peculiar Blue Compact Dwarf (BCD) galaxy that, on account of its proximity,
   provides an ideal laboratory for detailed spatial study of starburst galaxies. An open issue not addressed so far is how the collisional and self-absorption effects on \hei\ emission influence the determination of the He$^+$ abundance in 2D and what is the relation to
   the physical and chemical properties of the ionized gas.}
   {A 2D, imaging spectroscopy, study of the  spatial behavior of collisional and radiative transfer effects in He$^+$ and their impact on the determination of He$^+$ abundance is presented for the first time in a starburst galaxy.}
   {The \hei\ lines are analysed based on previously presented optical Integral Field Spectroscopy (IFS) data, obtained with FLAMES at the VLT and lower resolution gratings of the Giraffe spectrograph as well as with GMOS at Gemini and the R381~grating.
  }
   {Collisional effects reproduce the electron density structure. They are  negligible (i.e. $\sim$0.1-0.6\%) for transitions in the singlet cascade while relatively important  for those  in the triplet cascade. In particular, they can contribute up to 20\% of the flux in the \hei$\lambda$7065 line. Radiative transfer effects  are important over an extended and circular area of $\sim$30~pc in diameter centered at the Super Star Clusters. 
The singly ionized helium abundance, $y^+$, has been mapped using extinction corrected fluxes of six \hei\ lines, realistic assumptions for $T_e$, $n_e$, and the stellar absorption equivalent width as well as the most recent emissivities. We found a mean($\pm$ standard deviation) of $10^3 y^+ \sim80.3(\pm2.7)$ over the mapped area.
The relation between the excitation and the total helium abundance, $y_{tot}$, is consistent with no abundance gradient.
Uncertainties in the derivation of helium abundances are dominated by the adopted assumptions.
We illustrated the difficulty of detecting a putative helium enrichment due to the presence of Wolf-Rayet stars in the main \ghiir. Data are marginally consistent with an excess in the $N/He$ ratio in the nitrogen enriched area of the order of both, the atmospheric $N/He$ ratios in WR stars and the uncertainties estimated for the $N/He$ ratios.
We explored the influence of the kinematics in the evaluation of the \hei\ radiative transfer effects. Our data empirically support  the use of the traditional assumption that motions in an extragalactic \hii\ region have a negligible effect in the estimation of the global optical depths. Individually, the broad kinematic component (associated with an outflow) is affected by radiative transfer effects in a much more significant way than the narrow one.
We found a relation between the amount of extra nitrogen and the upper limit of the contribution from radiative transfer effects that requires further investigation. We suggest the electron temperature as perhaps a common agent causing this relation.
    }
   {}

   \keywords{Galaxies: starburst  ---  Galaxies: dwarf --- Galaxies:
   individual, NGC~5253 --- Galaxies: ISM --- Galaxies: abundances --- Galaxies: kinematics and dynamics} 

   \titlerunning{He\,\textsc{i} in the central \hii\ region of NGC~5253}

   \maketitle
%

\section{Introduction}

At a distance of 3.8~Mpc \citep{sak04}, \object{NGC~5253}, in the Centaurus A / M~83 group \citep{kar07}, is one of the closest Blue Compact Dwarf (BCD) galaxies. This galaxy is well known for presenting several peculiarities, whose detailed study are closely connected to its proximity and high surface brightness. For example, it contains a deeply embedded very dense compact \hii\ region at its nucleus (hereafter, "the supernebula"), detected in the radio at 1.3~cm and 2~cm \citep{tur00} that host two very massive Super Star Clusters \citep[SSCs,][]{alo04} and is embedded in a larger (i.e. $\sim$100~pc$\times$80~pc) Giant \hii\ Region (hereafter, the central \ghiir). Recently, mid-infrared observations showed how its kinematics is compatible with a model for the supernebula in which gas is outflowing from the molecular cloud \citep{bec12}. Indeed, the whole central region of the galaxy is dominated by an intense burst of star formation in the form of a large collection of compact young ($\sim1-12$~Myr) star clusters \citep[e.g.][]{har04}. In contrast to this, the main body of \object{NGC~5253} resembles that of a dwarf elliptical galaxy and recently, three potentially massive ($\gsim10^5$~M$_\odot$) and old ($1-2$ Gyr) star clusters have been found in the outskirts of the galaxy \citep{har12}.
Finally, \object{NGC~5253} is best-known  for being one of the few examples (and the closest) of a galaxy presenting  a confirmed local excess in nitrogen \citep[see e.g.][]{wal89}.

\defcitealias{mon10}{Paper~I}
\defcitealias{mon12}{Paper~II}
\defcitealias{wes13}{Paper~III}

We are carrying out  a detailed study of this galaxy using Integral Field Spectroscopy (IFS). The results obtained so far have further highlighted its peculiar nature.
In \citet[][hereafter Paper I]{mon10}, we found that the emission line profiles were complex and consistent with an scenario where the two SSCs produce an outflow \citep[see also][]{bec12}. Also, we delimited very precisely the area polluted with extra nitrogen. Moreover, we detected nebular \heii$\lambda$4686 in several locations, some associated with WN-type Wolf-Rayet (WR) stars (as traced by the blue bump at around 4680\AA) and some not, but not necessarily coincident with the area exhibiting extra nitrogen. In \citet[][hereafter Paper II]{mon12}, we studied the 2D distribution of  the main physical (electron temperature and density, degree of ionization) and chemical properties (metallicity and relative abundances of several light elements) of the ionized gas. A new area of enhanced nitrogen abundance at $\sim$130~pc from the main area of enhancement and not reported so far was found. In \citet[][hereafter Paper III]{wes13} several locations showing emission characteristic of WC-type WR stars (via the red bump at around 5810\AA) were identified. The fact that WR stars are spread over $\sim$350~pc gives an idea of the area over which the recent starburst has occurred.
The chemical analysis was extended with the finding that, with the exception of the aforementioned localised N excess, the $O/H$ and $N/H$ distributions are flat within the whole central 250~pc.

An issue not addressed in detail so far in NGC~5253 is the 2D determination and distribution of the He$^+$ abundance. In a cosmological context, this is particularly relevant since the joint determination of metallicity (as traced by the $O/H$ abundance) and $^4He$ abundance ($Y$\footnote{Here, we use $Y$ for  the helium mass fraction and $y$ for the number density of helium relative to hydrogen. Assuming $Z = 20(O/H)$, they are related as $Y = \frac{4y(1-20(O/H))}{1+4y}$. }) for extragalactic \hii\ regions and star-forming galaxies at low-metallicity was proposed as a means to estimate the primordial helium abundance, $Y_P$ \citep{pei76,pag86}
 serving as a test-bench for the standard hot big band model of nucleosynthesis. However, the  density of baryonic matter depends weakly on $Y_P$. Therefore, to put useful constrains on $Y_P$, $^4He$ abundance of  individual objects  has to be determined with accuracies $\lsim$1\%. Nowadays, emission
line flux data of this quality can be achieved and, indeed, the astronomical community is actively working on getting and improving the estimation of $Y_P$ \citep[see e.g. ][for recent estimations by the different groups]{ave10,izo10,fuk06,pei07,izo07,oli01,pei02}.
However, He abundance determinations are influenced by several effects and systematic errors which are not, in principle, straightforward to quantify and correct  \citep[see for example][]{oli01}. Specifically, the intensity of \hei\ emission lines may intrinsically deviate from the recombination values due to collisional and radiative transfer effects. Moreover, the emitted spectrum also depends on the physical conditions of the ionized gas (e.g. temperature, density and ionization structure). Also, on top of these specific properties of the \hii\ region, extinction by dust and a possible underlying stellar absorption component can also affect the observed spectrum. 
All these effects contribute to the uncertainties associated with the estimation of \emph{ionized} helium abundance ($y^+=He^+/H^+$).
A final extra source of uncertainty is associated with the estimation of the amount of existing neutral helium (i.e. the estimation of the ionization correction factor, icf(He)) and the calculation of the total helium abundance, $y=y_{tot}=\rm{icf(He)}\times y^+$.

Tentative values for  $y^+$ were presented in \citetalias{mon10} based on the \hei$\lambda$6678 line, which is almost insensitive to collisional and self-absorption effects. However, 2D distributions of all the relevant physical properties for the ionized gas were not available at that time. Moreover, other \hei\ lines, in particular  \hei$\lambda$7065, can indeed be affected by collisional and self-absorption effects, specially in conditions of relatively high electron temperature ($T_e$) and density ($n_e$), as it is the case in the main \ghiir\ of NGC~5253. Supporting this expectation, long-slit measurements predict too high He$^+$ abundances from the  \hei$\lambda$7065 when only recombination effects are taken into account \citep{sid10} and a non-negligible optical depth $\tau$(3889) when the He$^+$ abundance is derived using several emission lines in a consistent manner \citep{lop07}.

At present, a complete 2D characterization of the physical properties of the ionized gas in the main \ghiir\ of \object{NGC~5253} is available. Therefore, we are in an optimal situation for both mapping the collisional and radiative transfer effects in 2D, and re-visiting the derivation of the He$^+$ abundance map taking into account many \hei\ lines. This will be the main purpose of this work. Given 
that the metallicity of the object  \citepalias[$12+\log(O/H)=8.26$,][]{mon12} is only moderately low, our focus will not to be in achieving the $\lsim1\%$ accuracy required in the determination of the primordial He abundance, but to explore the effects of a parameter not taken into account so far: namely spatial resolution.
In addition, we will explore the 2D relationship between the properties of the ionized gas derived so far and the \hei\ collisional and self-absorption effects. To our knowledge this work constitutes the first attempt to study in 2D the collisional and self-absorption effects  in He\,\textsc{i} in any extra-galactic object.
Moreover, irrespective of the spatial resolution,  due to the characteristics inherent in IFS data, there is the guarantee that the set of $\sim100$ spectra utilized in this work have been processed in a homogeneous manner all the way from the observations (i.e. a given observable was taken with the same observing conditions for the whole set of data) to the final $y^+$ and $y_{tot}$ derivations. 

The paper is structured as follows:  Sec. 2 describes the characteristics of the data utilized in our analysis; Sec. 3 contains an evaluation of the \hei\ collisional and self-absorption effects in 2D as well as the derivation of the $y^+$ and $y$ maps. Sec. 4 discusses the relation between radiative transfer effects and other quantities (e.g. kinematics of the gas, relative $N/O$ abundance). Our main conclusions are summarized in Sec. 5.

\section{The data}

\begin{figure}[th!]
   \centering
\includegraphics[angle=0,width=0.48\textwidth,clip=]{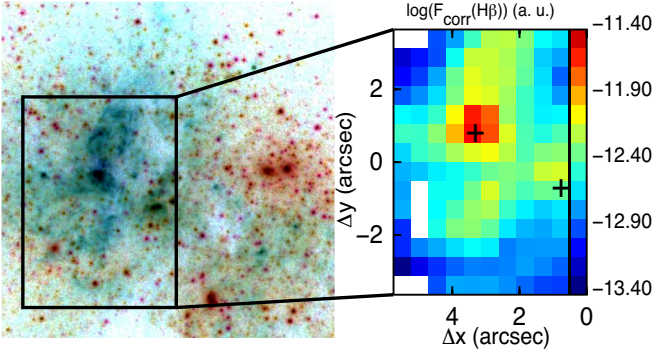}
   \caption[Area under study]{
\emph{Left:} False colour image in filters $F658N$ (\ha, cyan channel),  $F550M$ ($V$, yellow channel), and $F814W$ ($I$, magenta channel) for the central part of \object{NGC~5253} using the HST-ACS images (programme 10608, P.I.: Vacca). The area  studied here is marked with a black rectangle.
\emph{Right:}   
   Ionized gas  distribution as traced by the extinction corrected \hb\ map derived from a portion of the original FLAMES data.   The position of the two main peaks of continuum emission are marked with crosses.
The map is presented in logarithmic scale in order to emphasize  the relevant morphological features and cover  a range of 2.0~dex.
Flux units are arbitrary. Note the existence of three dead spaxels at $\sim[5\farcs0,-1\farcs0]$ as well as absence of signal in the spaxels at the two left corners of the field of view.  \label{apuntado}}
 \end{figure}

We will focus our study on the area associated with the main \ghiir\ in \object{NGC~5253} (see Fig. \ref{apuntado}).
This is a portion of the full area studied in \citetalias{mon10}
and the location where: i) the gas presents relatively high electron temperature and density, and therefore, important collisional and radiative transfer effects are expected; and ii) several  helium lines can be detected over a relatively large area with sufficient quality, and therefore a 2D analysis based on the information in individual spaxels is feasible. The utilized data were collected during several observing runs using the FLAMES-Argus and the GMOS Integral Field Units (IFUs) and together cover the whole optical spectral range. In the following, we briefly describe the basic instrumental characteristics of each set of data and compile the information that was extracted.

  \begin{figure}[th!]
   \centering
\includegraphics[angle=0,width=0.24\textwidth,clip=,bb = 45 30 405 390]{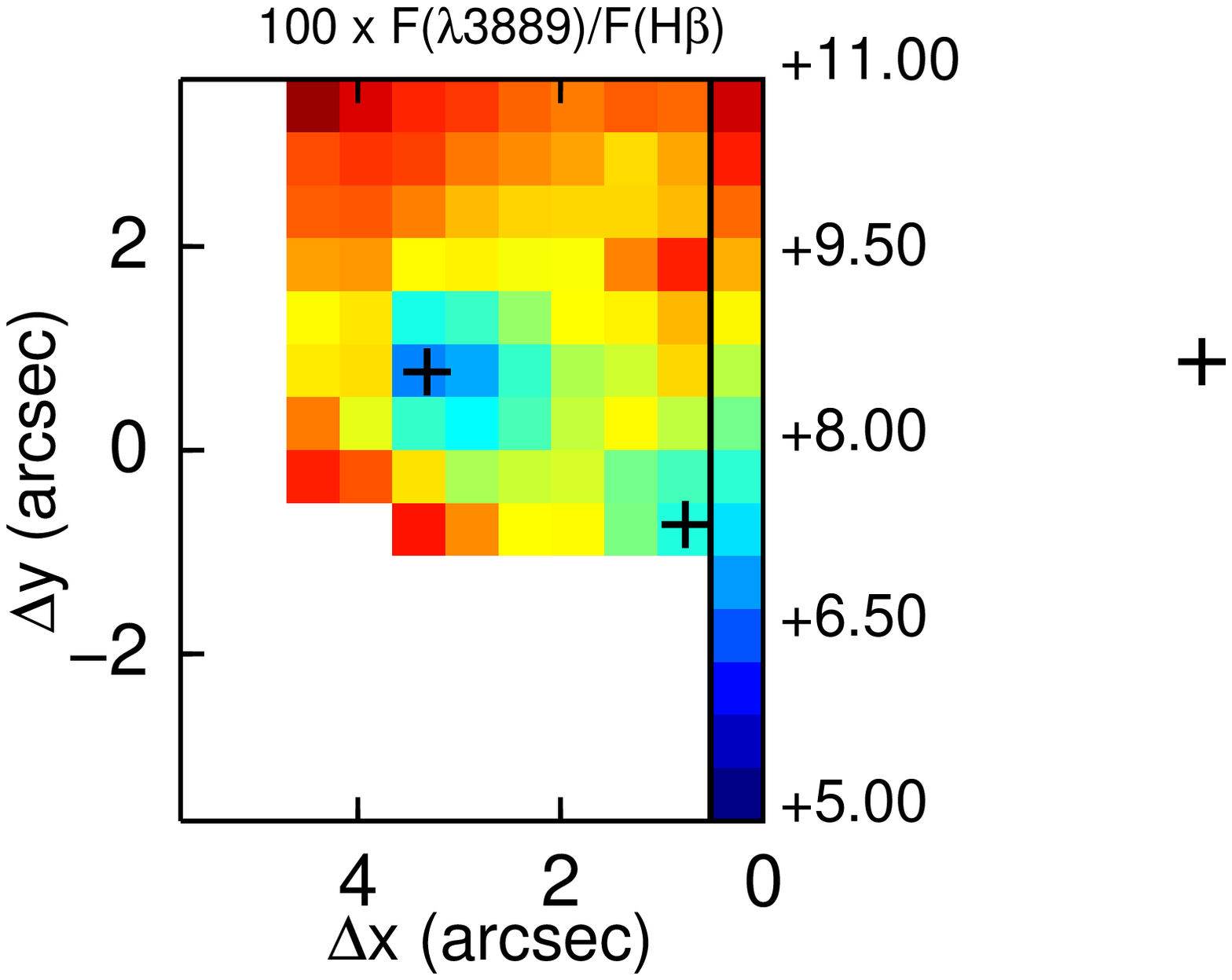}
\includegraphics[angle=0,width=0.24\textwidth,clip=,bb = 45 30 405 390]{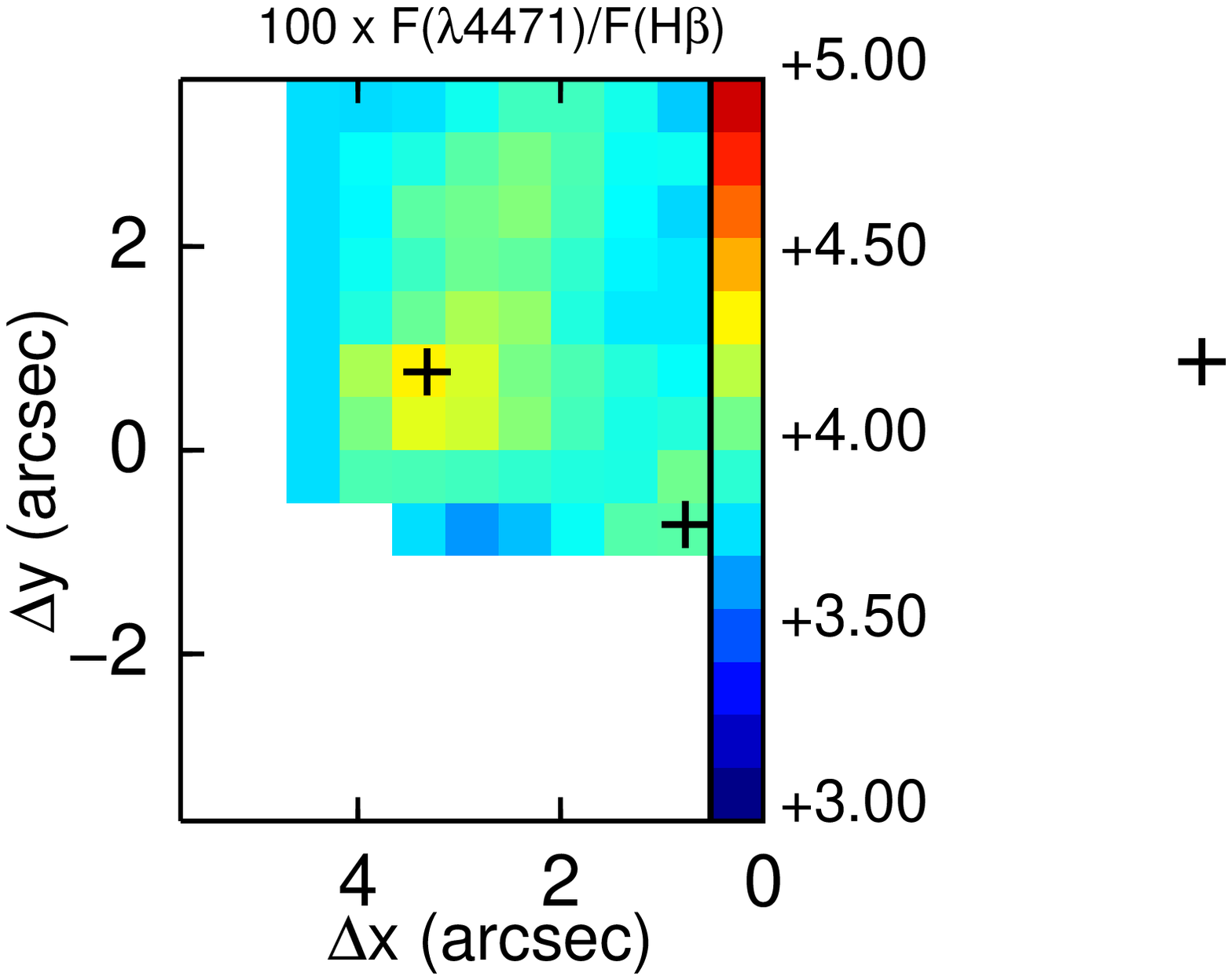}\\
\includegraphics[angle=0,width=0.24\textwidth,clip=,bb = 45 30 405 390]{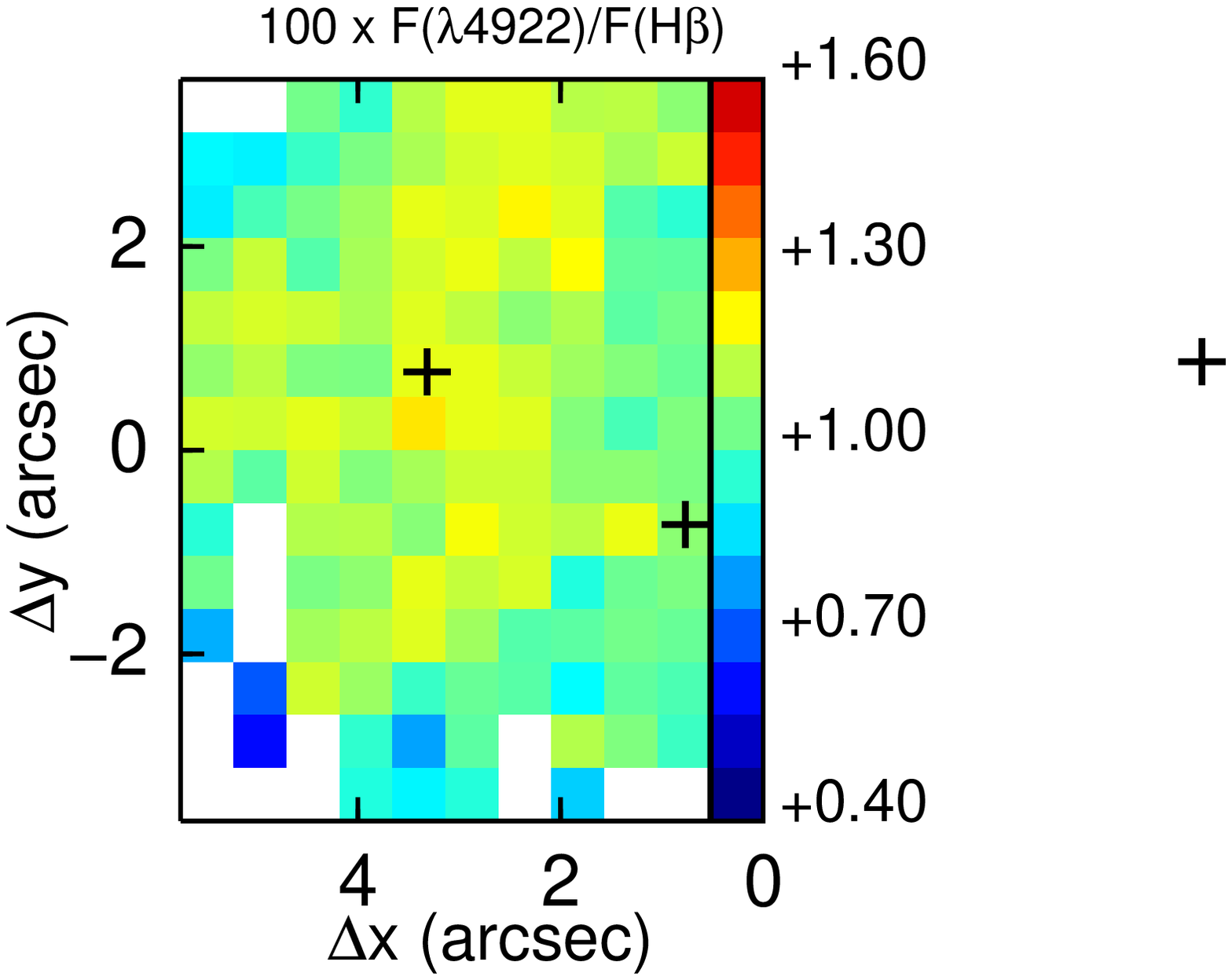}
\includegraphics[angle=0,width=0.24\textwidth,clip=,bb = 45 30 405 390]{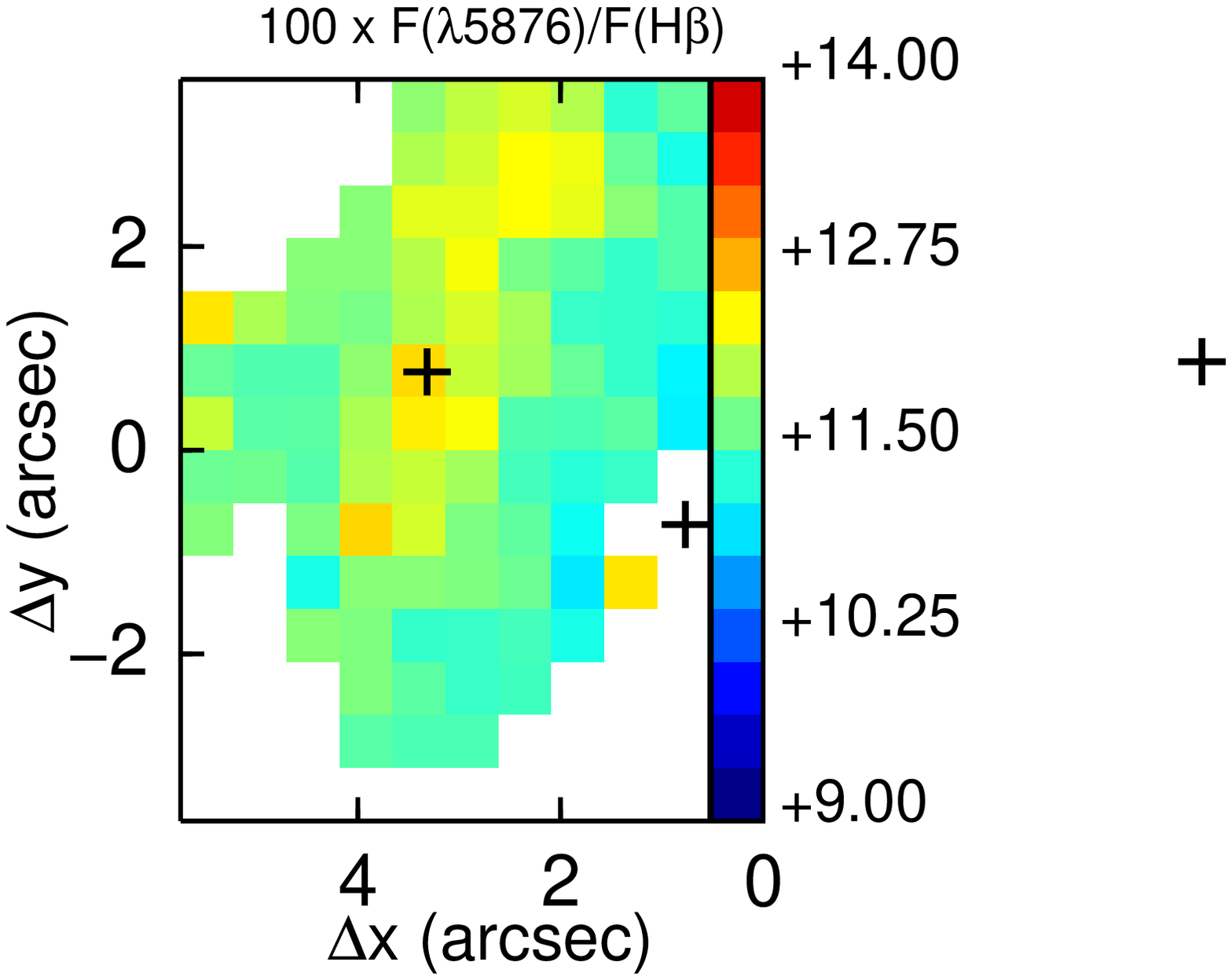}\\
\includegraphics[angle=0,width=0.24\textwidth,clip=,bb = 45 30 405 390]{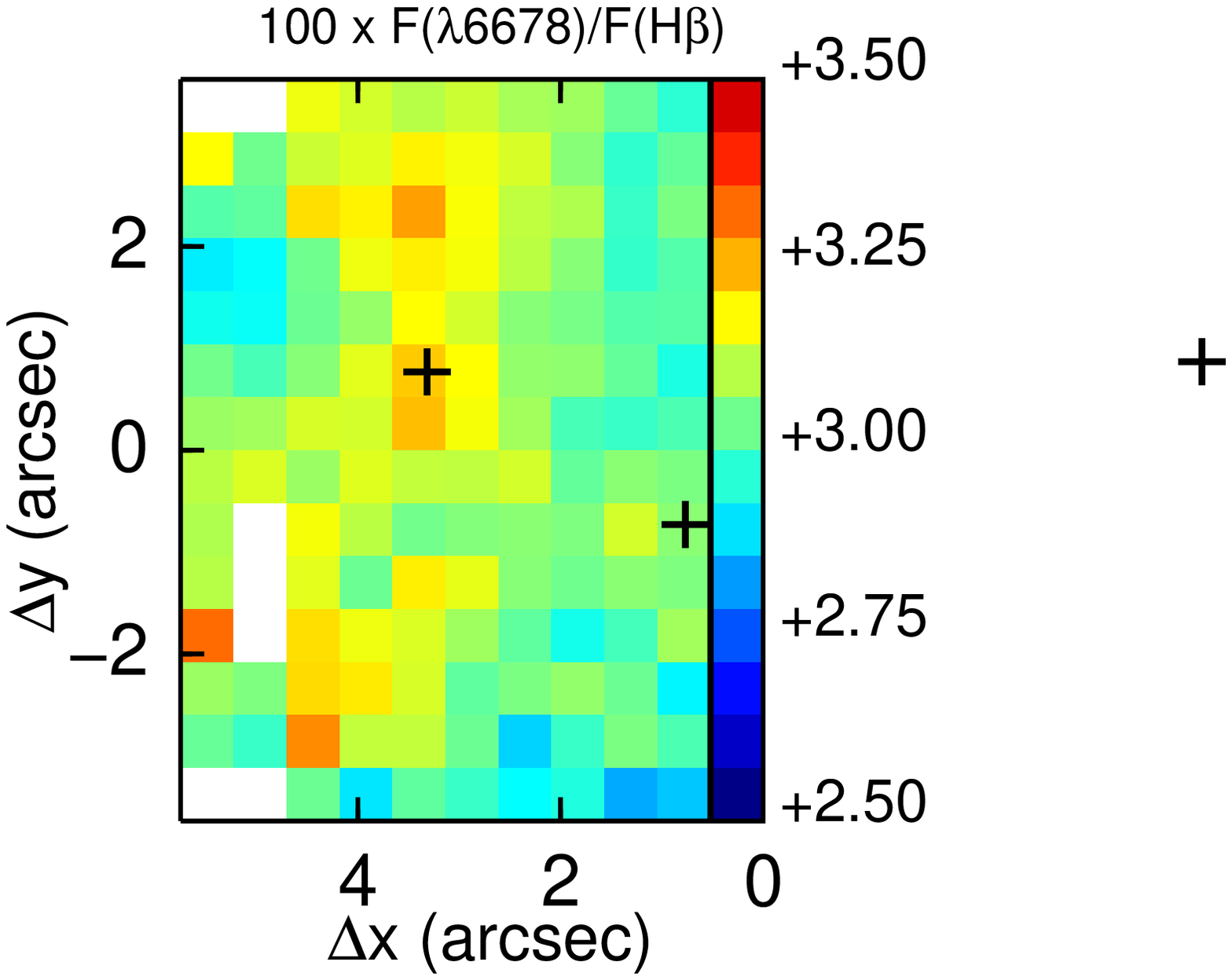}
\includegraphics[angle=0,width=0.24\textwidth,clip=,bb = 45 30 405 390]{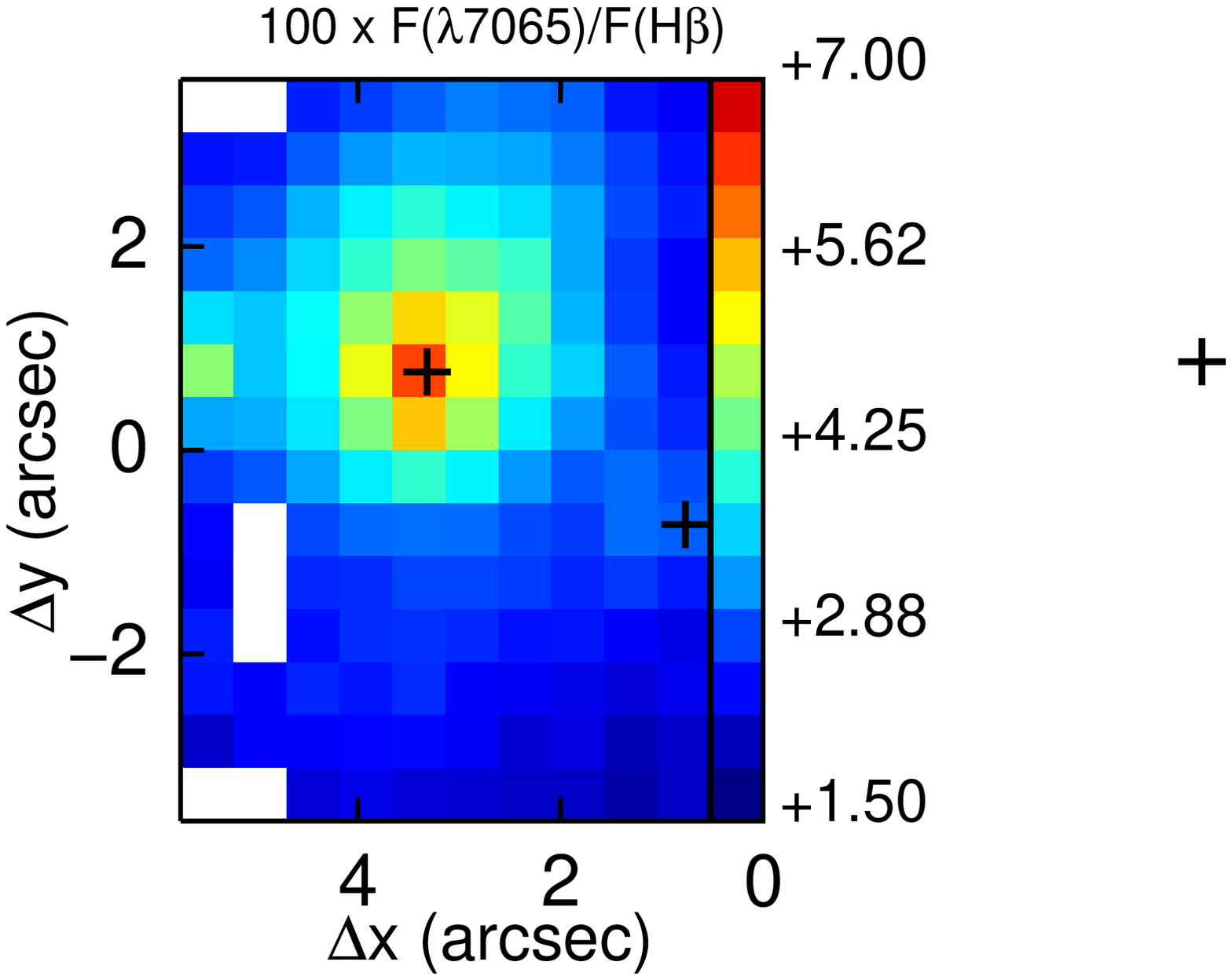}\\
   \caption[Maps of the extinction corrected fluxes of the utilized lines]{Extinction corrected flux maps normalized to \hb\ for the lines utilized in this work.  The position of the two main peaks of continuum emission are marked with crosses in this and all subsequent maps.
We marked in white the areas where a given line was not observed. Specifically, for the \hei$\lambda$6678,  \hei$\lambda$4922 and \hei$\lambda$7065, they correspond to dead fibers. For \hei$\lambda$5876, \hei$\lambda$4471 and \hei$\lambda$3889, these areas were not covered by the corresponding GMOS or FLAMES-Argus field of view.
 \label{hetohb}}
 \end{figure}

\subsection{FLAMES-Argus data}

Data were obtained with FLAMES \citep{pas02} at VLT UT2 in Paranal. We used the Argus IFU with the sampling of 0.52$^{\prime\prime}$/lens, covering a field of view (f.o.v.) of 11\farcs5$\times$7\farcs3, and four low resolution gratings (LR1, LR2, LR3, and LR6).
All together, they  offer a spectral coverage of $3\,610-5\,070$~\AA\ plus  $6\,440-7\,180$~\AA\ at a dispersion of 0.2~\AA~pix$^{-1}$.
 Details of the observations, data reduction, and cube processing,
as well as maps for most of the physical and chemical properties utilized in this work, have already been presented in \citetalias{mon10} and \citetalias{mon12}. 
Maps for the helium lines were derived by independently fitting in each spaxel the  \hei\ line profiles with a single Gaussian function with MPFITEXPR \citep{mar09}. 
For the particular case of \hei$\lambda$3889,  which is blended with H8 at our spectral resolution, we created an extinction corrected map by subtracting from the H8+\hei$\lambda$3889 map, a map of 0.659$\times$ the H7 map. This H8 line intensity is that predicted by \citet{sto95}  for Case B, $T_e = 10^4$~K and $n_e = 100$~cm$^{-3}$.
The final set of FLAMES-Argus maps utilized here are:
\begin{enumerate}[i)]
\item an extinction map derived from the \ha/\hb\ line ratio;
\item a map for the \hb\ equivalent width  ($EW$(\hb));
\item a map of electron temperature $T_e$ as derived from the \oiii$\lambda\lambda$4959,5007/ \oiii$\lambda$4363 line ratio: $T_e$(\oiii). In those spaxels where no determination of $T_e(\oiii)$ was available, we assumed $T_e(\oiii)=10\,500$~K \citepalias[see][for typical $T_e$(\oiii) values outside the main \ghiir]{mon12};
\item a map for the electron density ($n_e$) as derived from the \sii$\lambda$6717/\sii$\lambda$6731 line ratio; 
\item maps for the $O^+/H^+$, $O/H$ and $S^+/H^+$ abundances, as derived from collisional lines using the direct method, to estimate the icf(He);
\item maps for different tracers of the excitation degree (i.e. \sha\ and \ohb\ line ratio) to be used in the estimation of the icf(He) at those locations where no measure of the $O^+/H^+$, $O/H$ and $S^+/H^+$ abundances is available and to explore the dependence of $y^+$ and $y_{tot}$ on the excitation;
\item a map of the relative abundance of nitrogen, $N/O$, as derived from collisional lines using the direct method;
\item maps for the $\lambda$3889, $\lambda$4471, $\lambda$4922, $\lambda$6678, $\lambda$7065 \hei\ equivalent widths and extinction corrected  line fluxes using our extinction map and  the extinction curve of \citet{flu94}. The extinction corrected line flux maps normalized to \hb\ are presented in Fig. \ref{hetohb}. 
Note that in order to minimize uncertainties associated to aperture matching, absolute flux calibration and extinction, lines were measured relative to a bright Balmer line observed simultaneously with a given helium line. Then, we assumed the theoretical Balmer line intensities obtained from \citet{sto95} for Case B, $T_e = 10^4$~K and $n_e = 100$~cm$^{-3}$. 
Additional \hei\ lines were covered by the FLAMES set-up but not used here. Specifically, \hei$\lambda$5016 and \hei$\lambda$4713 emission lines were detected over most of the FLAMES f.o.v. \hei$\lambda$5016 is relatively close to the much brighter (i.e. $\sim$150-400 times)  \oiii$\lambda$5007 line and, in the main \ghiir, the wings of the \oiii\ line profile prevented us from measuring a reliable line flux. Regarding the relatively weak $\lambda$4713 line, at the spectral resolution of these data, this line is strongly blended with \ariv$\lambda$4711 and  the uncertainties associated with the deblending of these lines in the \ghiir\ are relatively large due to the existence of several distinct kinematic components \citepalias[see][]{mon10,wes13};
\item maps with the extinction corrected line fluxes in $\lambda$6678, $\lambda$7065 \hei\  and \ha, for the different kinematic components presented in \citetalias{mon10}.
\end{enumerate}

\subsection{GMOS data}

The Gemini-South Multi-Object Spectrograph (GMOS) data were taken using the one-slit mode of its IFU \citep{all02}. In this mode, the IFU covers a f.o.v. of $5\farcs0\times3\farcs5$ sampled by 500 contiguous hexagonal lenslets of 0\farcs2 diameter.
The utilized grating (R381) gives a spectral coverage of $4\,750-6\,850$~\AA\ at a dispersion of 0.34~\AA~pix$^{-1}$,  thus complementing the FLAMES-Argus data. We refer to \citetalias{wes13}  for further details on the observations and data reduction. The product of the reduction is a datacube per pointing with a uniformly sampled grid of 0\farcs1. In this work, we utilized the two pointings (out of four) that mapped the central \ghiir. As we did with the FLAMES's data, in each spaxel all the lines of interest were independently fit with a single Gaussian function with MPFITEXPR. 
The final set of GMOS maps utilized here are:

\begin{enumerate}[i)]
\item Maps for \ha\ and \hb\ fluxes. These images were utilized to check the consistency between the FLAMES and GMOS data, both in terms of observed structure and derived extinction map, to estimate the offset and rotation that was necessary to be applied to the GMOS data, and to correct for extinction the  \hei$\lambda$5876 map;
\item A map for \hei$\lambda$5876 flux. This is the strongest \hei\ line and therefore, one of the key observables for the present study;
\item  Equivalent widths and extinction corrected line flux maps of \hei$\lambda$5876 flux normalized to \hb. These were derived from the maps previously mentioned and were rotated and reformatted to match the FLAMES data using the \texttt{drizzle} task of the Space Telescope Science Data Analysis System (STSDAS) package of IRAF\footnote{The Image Reduction and Analysis Facility \emph{IRAF} is distributed by the National Optical Astronomy Observatories which is operated by the association of Universities for Research in Astronomy, Inc. under cooperative agreement with the National Science Foundation.}. They were the only GMOS maps utilized jointly with the FLAMES's maps. The flux map is shown in Fig. \ref{hetohb} together with the other \hei\ line flux maps. 
\end{enumerate}

\section{Results}

\subsection{Collisional effects as traced by the theoretical C/R ratio \label{seccoli}}

Collisional excitation in hydrogen can be important in regions of very low metallicities ($Z\lsim1/6$~Z$_\odot$)
 due to their relatively high temperatures \citep{lur09}. However, for the typical temperatures and densities found in H\,\textsc{ii} regions in general, and in the main Giant H\,\textsc{ii} Region (\ghiir) of \object{NGC~5253} in particular, the collisional excitation in hydrogen is negligible in comparison with recombination \citep[see e.g. Fig. 9 of ][]{ave10}.  This is not the case for  helium.  The \hei\ $2^3S$ level \citep[see Fig. 1 in ][for a representation of the Grotian diagram for \hei\ singlet and triplet ladders]{ben99} is highly metastable, and collisional transition from it can be important \citep{ost06}. Specifically, at relatively high densities this level can be depopulated via collisional transitions to the $2^3P^0$, $2^1P^0$ and $2^1S$ and, to a lesser extent, to higher singlets and triplets (mainly $3^3P^0$). The effect on the observed \hei\ emission lines is always an increase in the observed flux. The relative importance of the collisional effects on a given emission line is characterized by the $C/R$ factor, i.e. the ratio of the collisional component to that arising from recombination which is given by:

\begin{equation}
\frac{C}{R} = \frac{n_{2^3S}k_{eff}}{n^+_{He}\alpha_{eff}}
\end{equation}

where $n_{2^3S}$, and $n^+_{He}$ are the densities of the $2^3S$ state and He$^+$ respectively, $\alpha_{eff}$ is the effective recombination coefficient for the line, and $k_{eff}$ is the effective collisional rate coefficient.

  \begin{figure}[th!]
   \centering
\includegraphics[angle=0,width=0.24\textwidth,clip=,bb = 45 30 405 390]{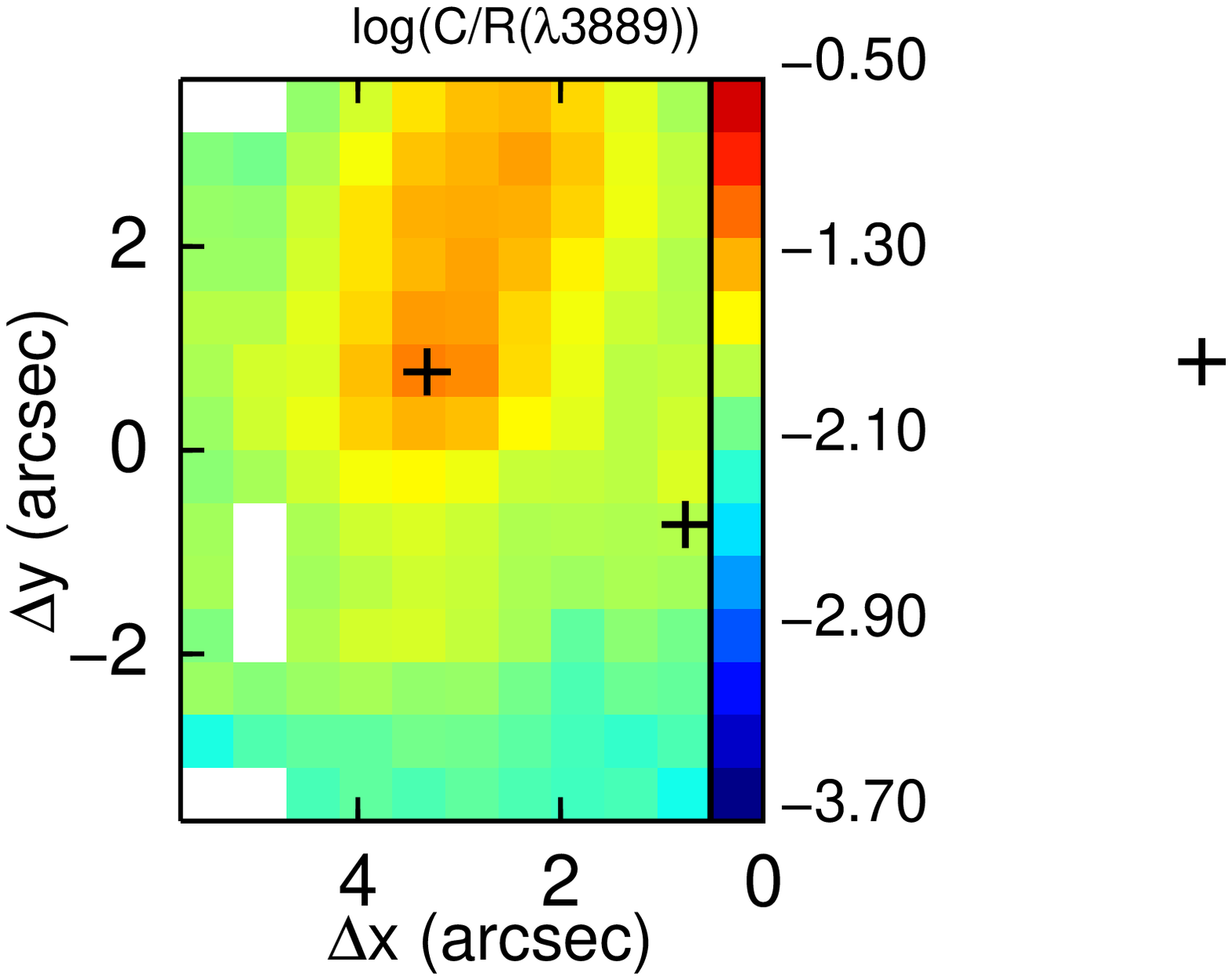}
\includegraphics[angle=0,width=0.24\textwidth,clip=,bb = 45 30 405 390]{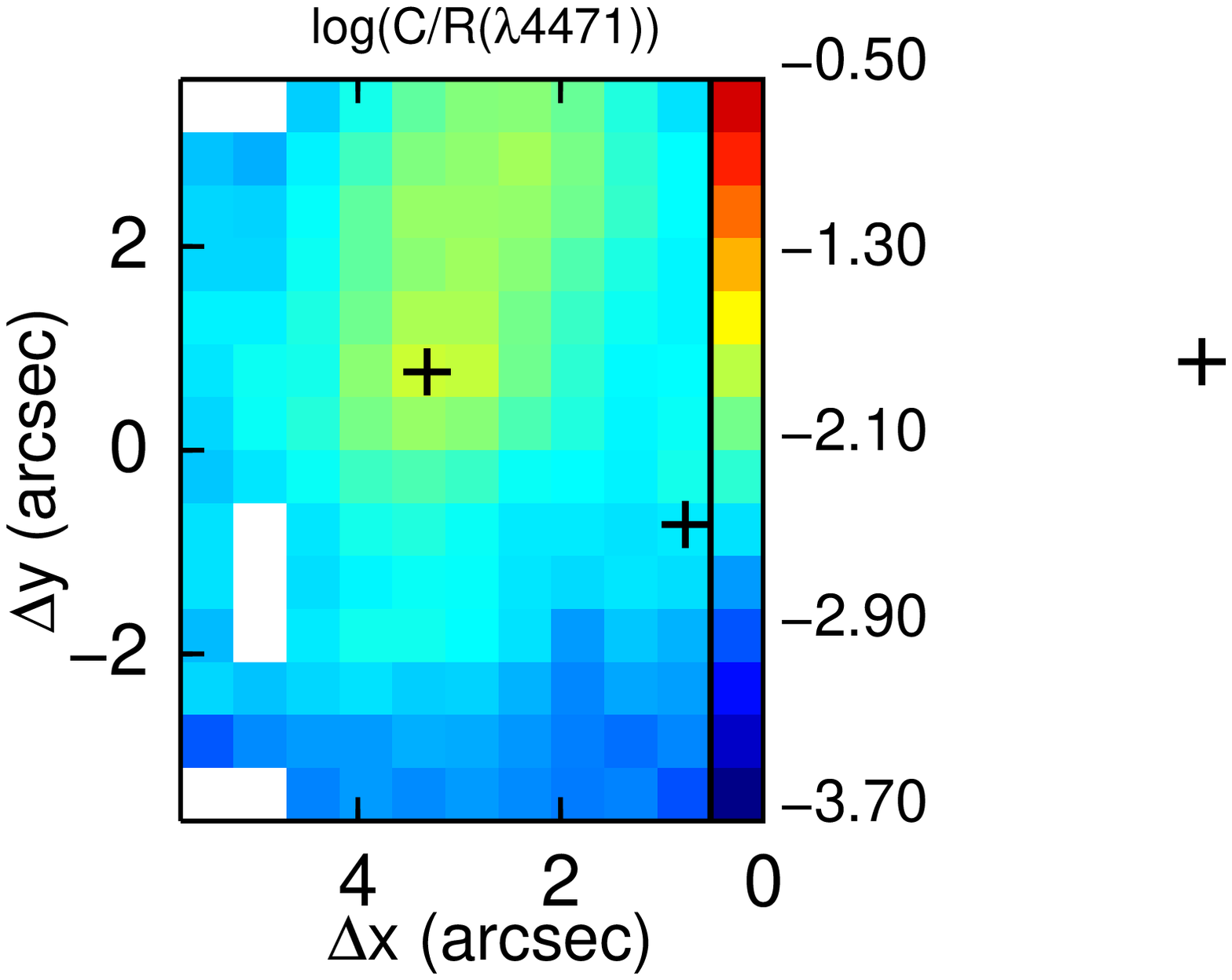}\\
\includegraphics[angle=0,width=0.24\textwidth,clip=,bb = 45 30 405 390]{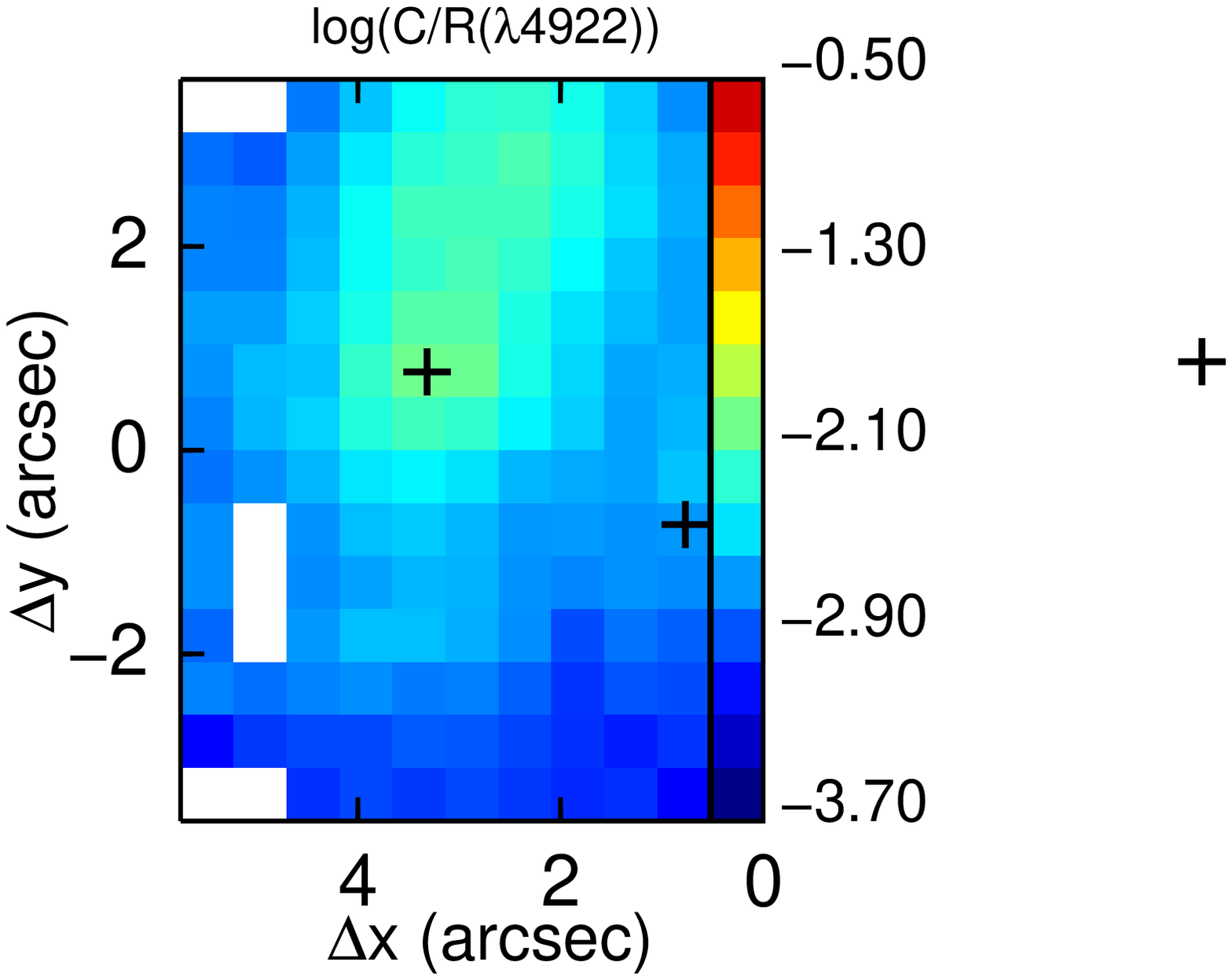}
\includegraphics[angle=0,width=0.24\textwidth,clip=,bb = 45 30 405 390]{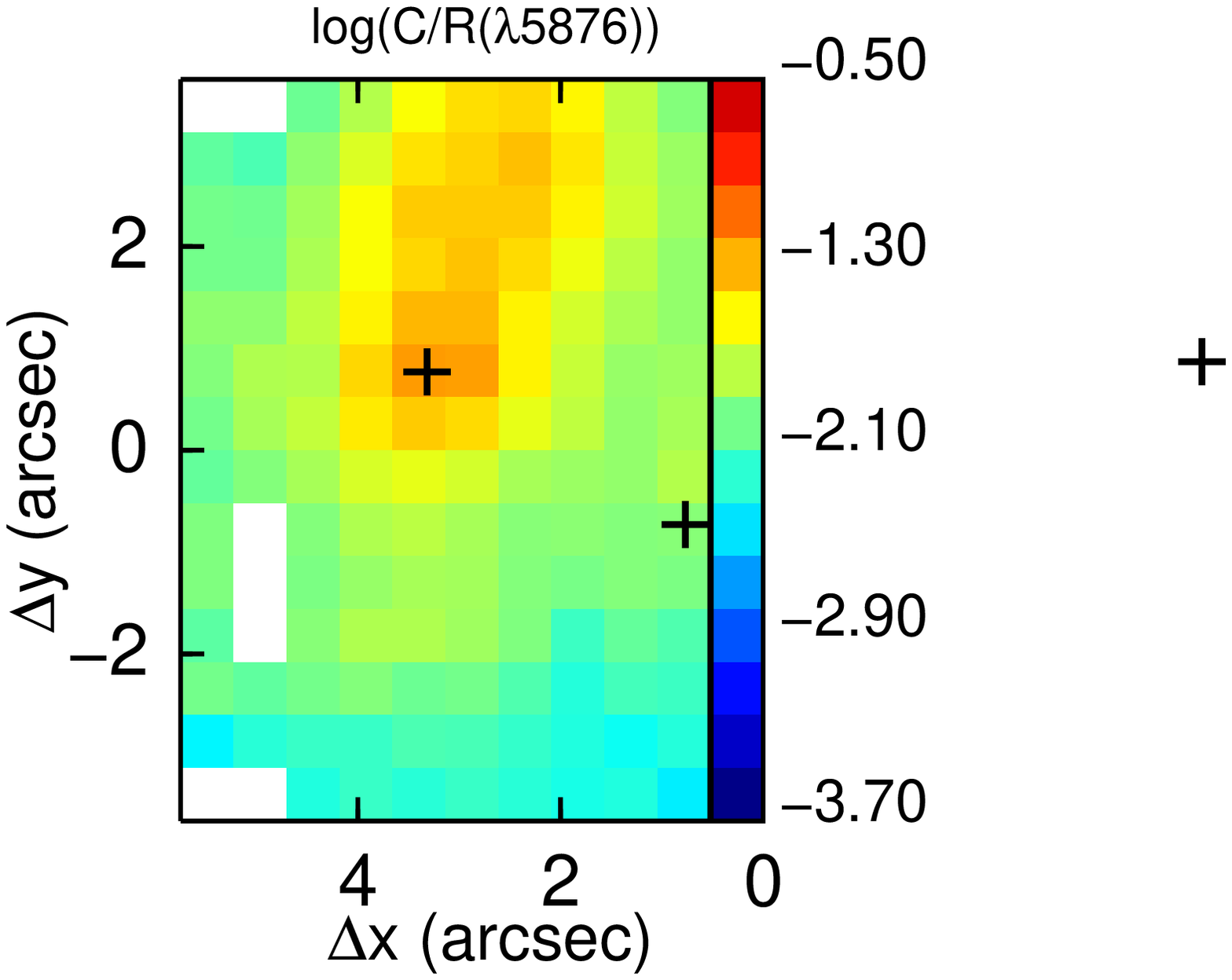}\\
\includegraphics[angle=0,width=0.24\textwidth,clip=,bb = 45 30 405 390]{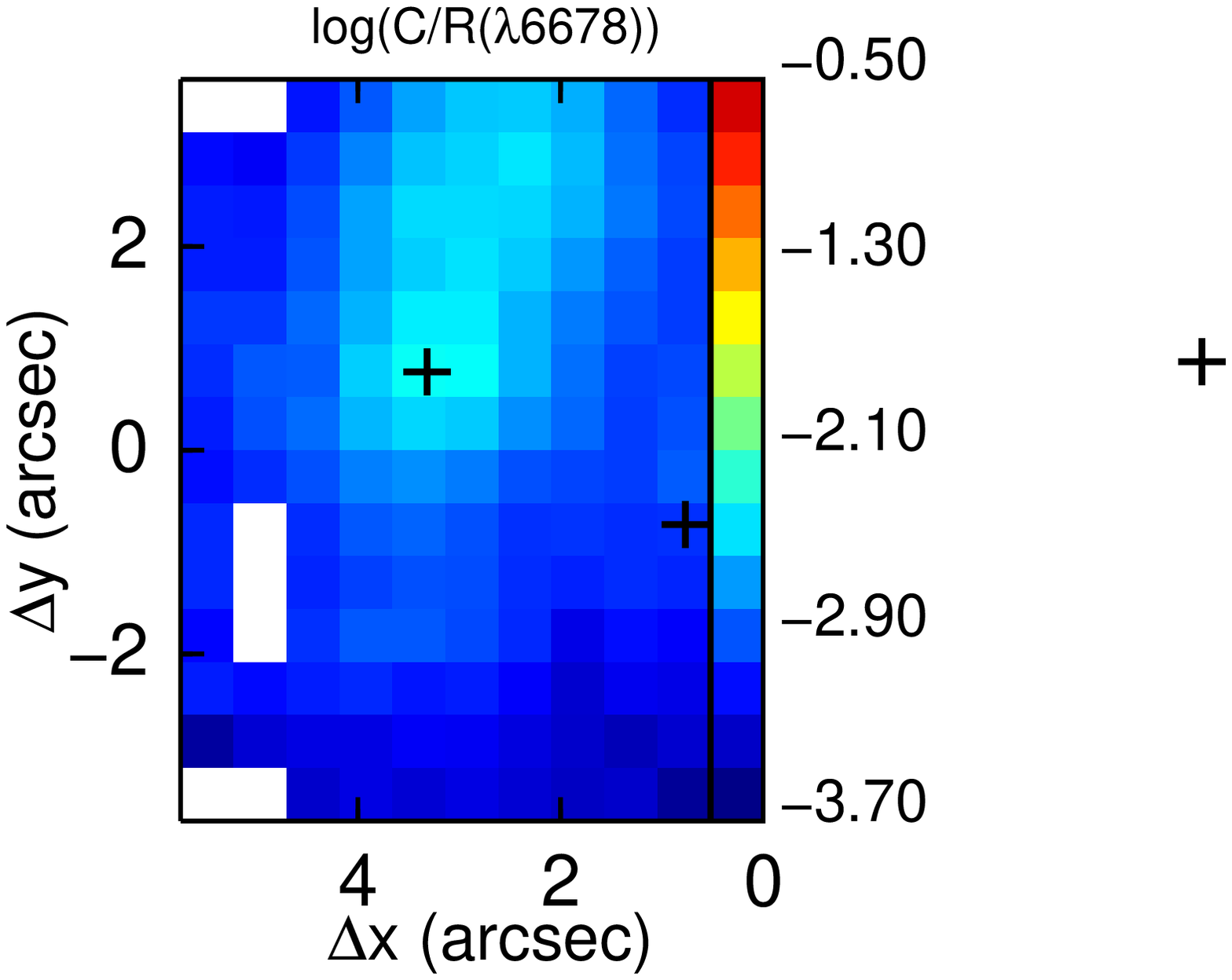}
\includegraphics[angle=0,width=0.24\textwidth,clip=,bb = 45 30 405 390]{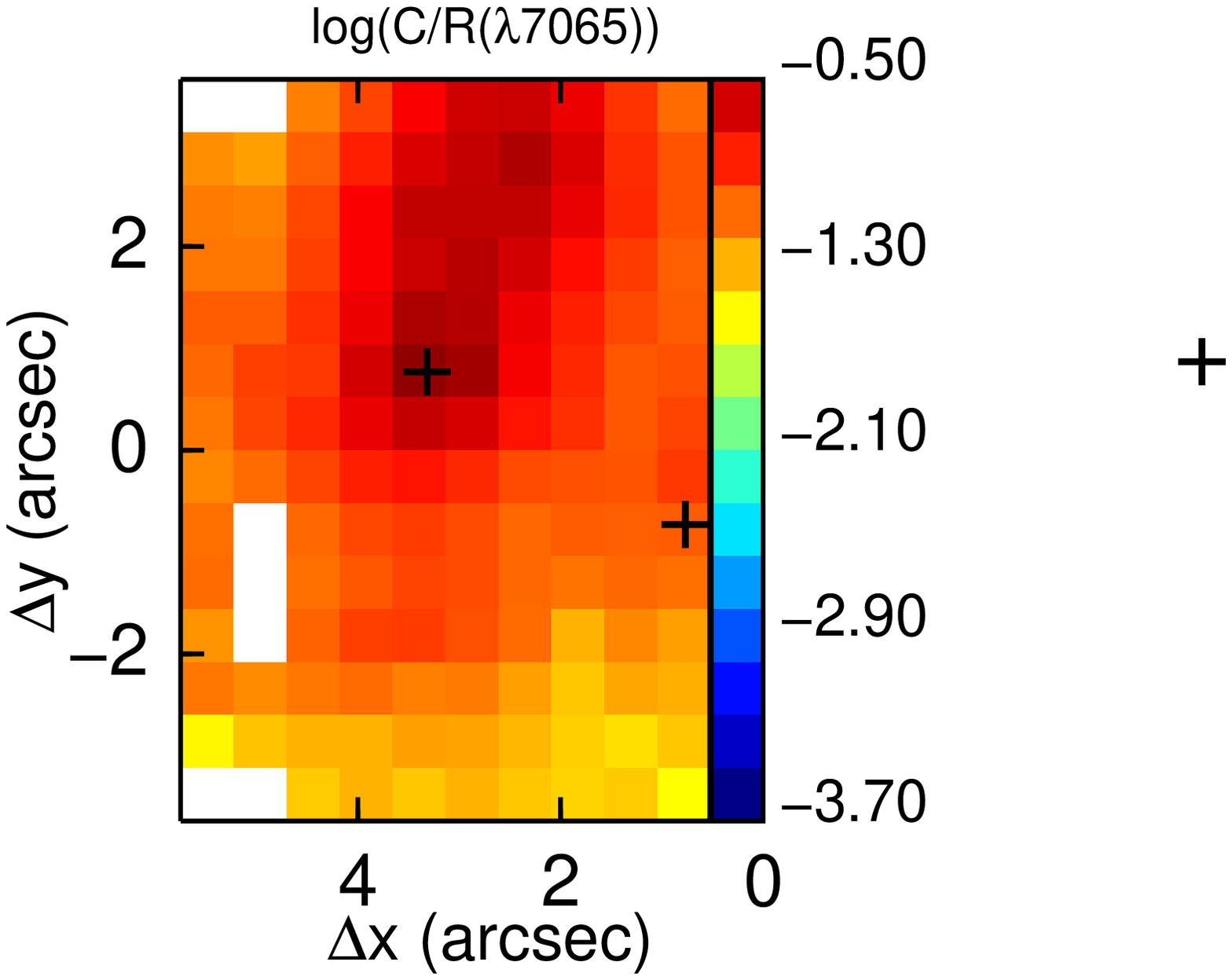}\\
   \caption[C/R ratios]{Collisional effects as traced by the $C/R$ ratio for the \hei\ lines utilized in this work. Note that a common scale was used for all the lines in order to facilitate comparison of the relative effects between the different lines. Also, a logarithmic color stretch is used to emphasize the variations \emph{within} the region for a given line.
 \label{c2rratios}}
 \end{figure}

  \begin{figure}[th!]
   \centering
\includegraphics[angle=0,width=0.24\textwidth,clip=,bb = 45 30 405 390]{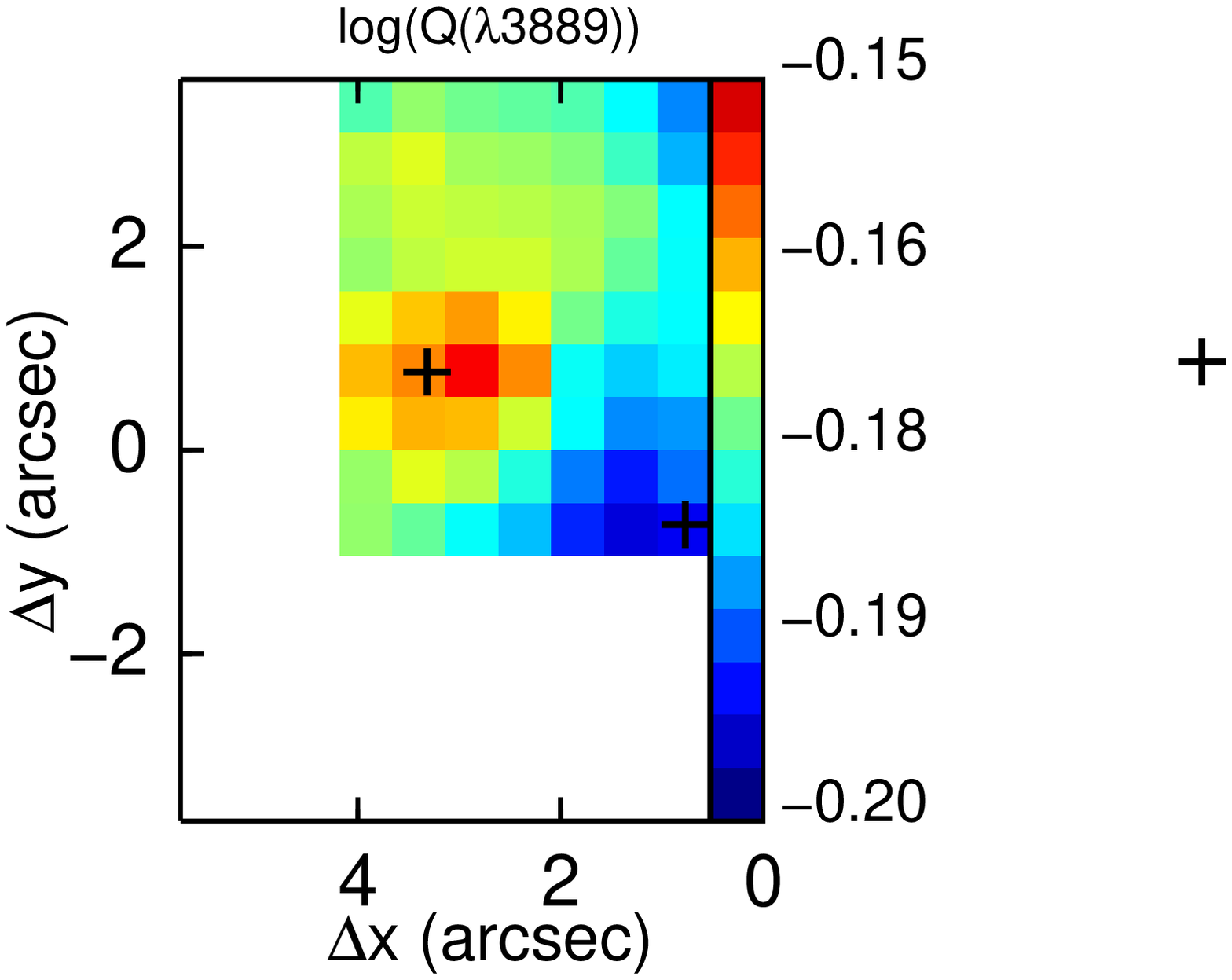}
\includegraphics[angle=0,width=0.24\textwidth,clip=,bb = 45 30 405 390]{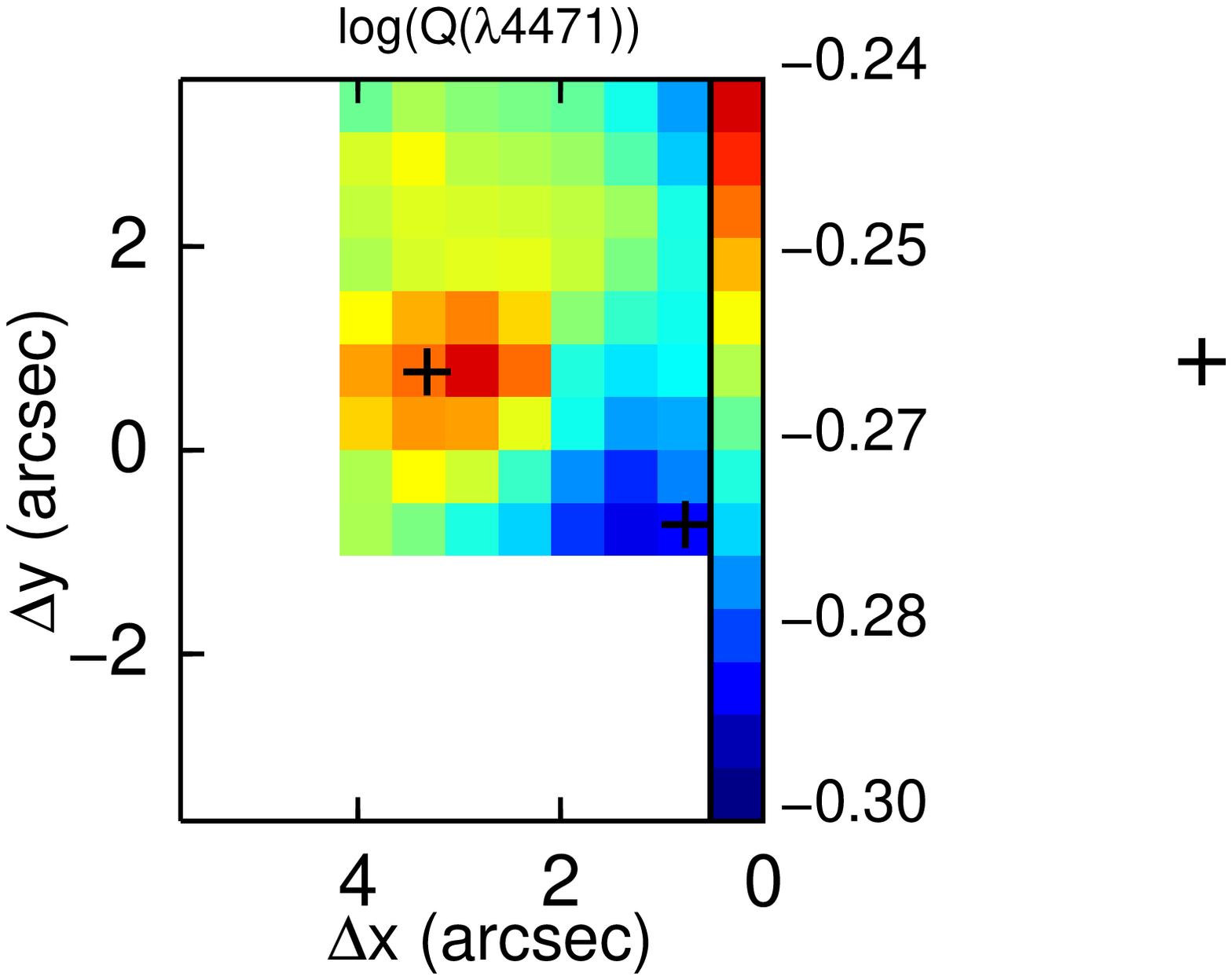}\\
\includegraphics[angle=0,width=0.24\textwidth,clip=,bb = 45 30 405 390]{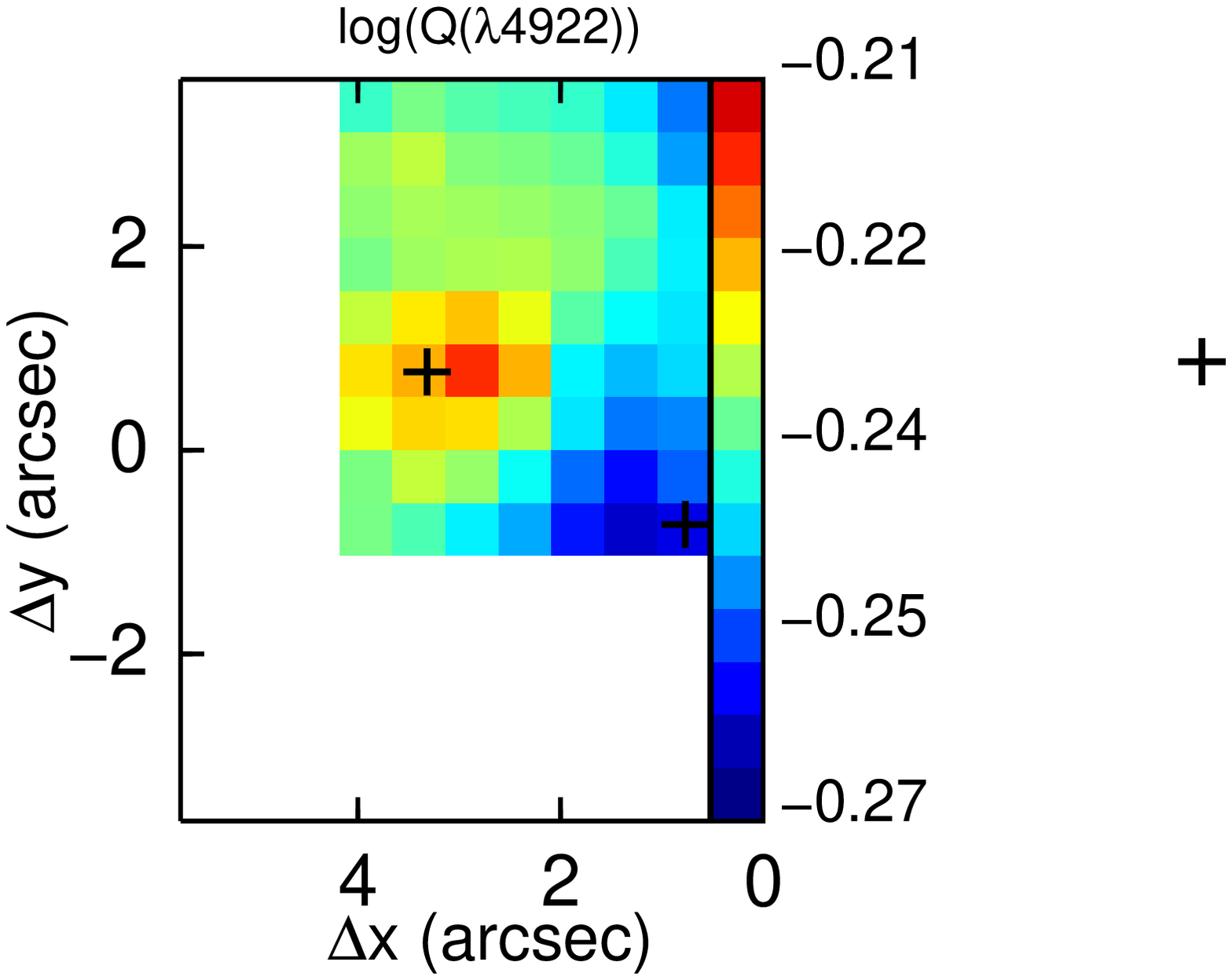}
\includegraphics[angle=0,width=0.24\textwidth,clip=,bb = 45 30 405 390]{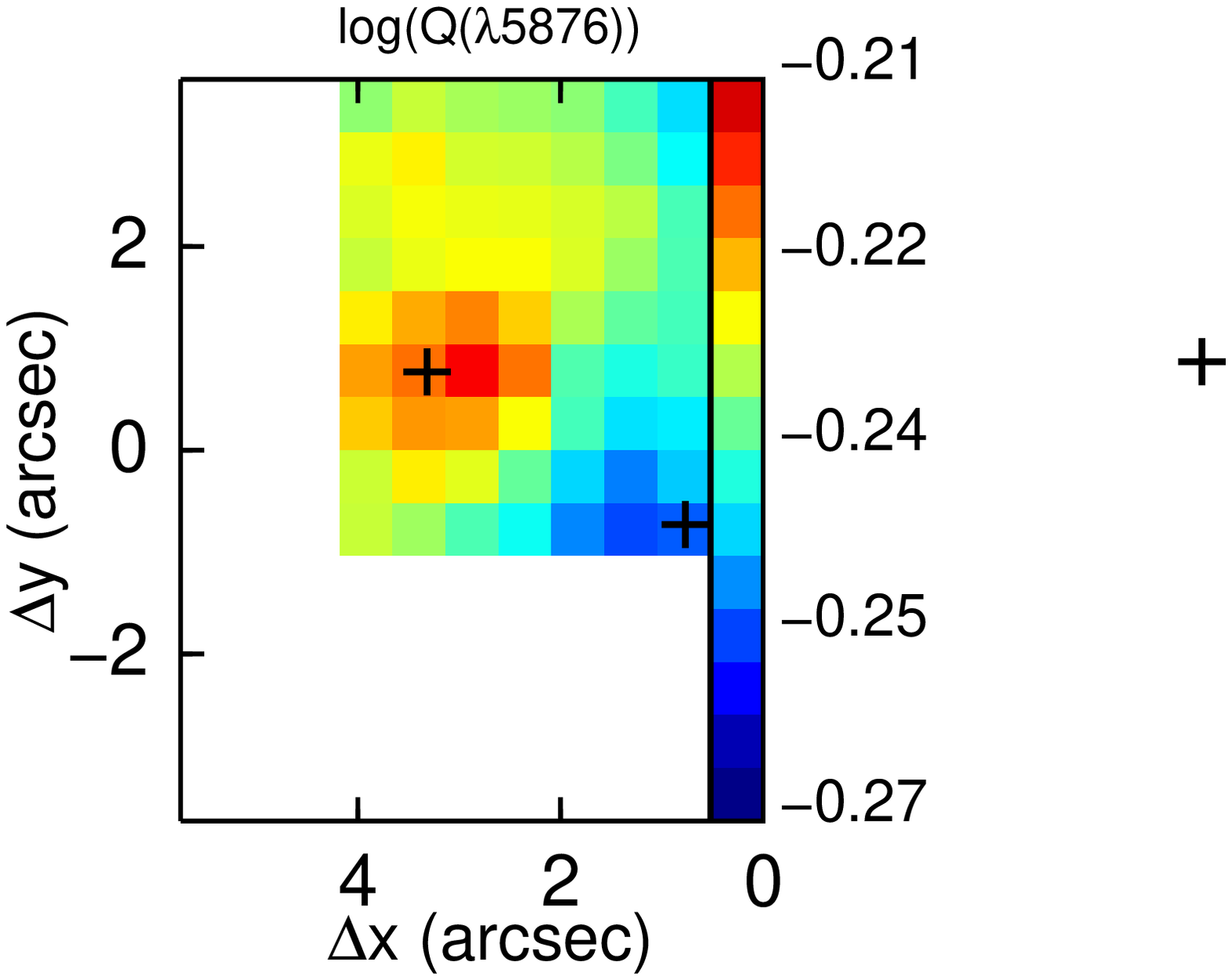}\\
\includegraphics[angle=0,width=0.24\textwidth,clip=,bb = 45 30 405 390]{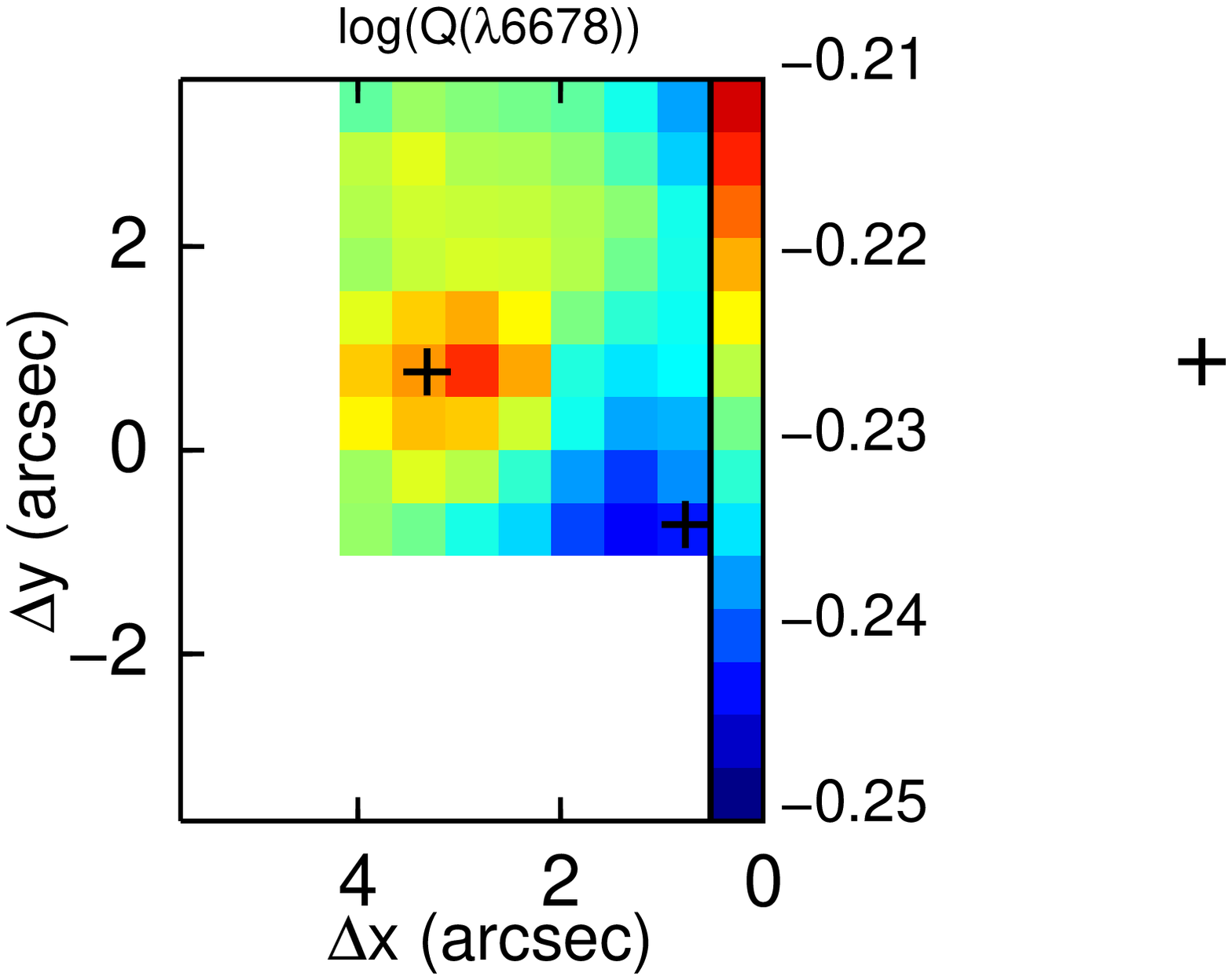}
\includegraphics[angle=0,width=0.24\textwidth,clip=,bb = 45 30 405 390]{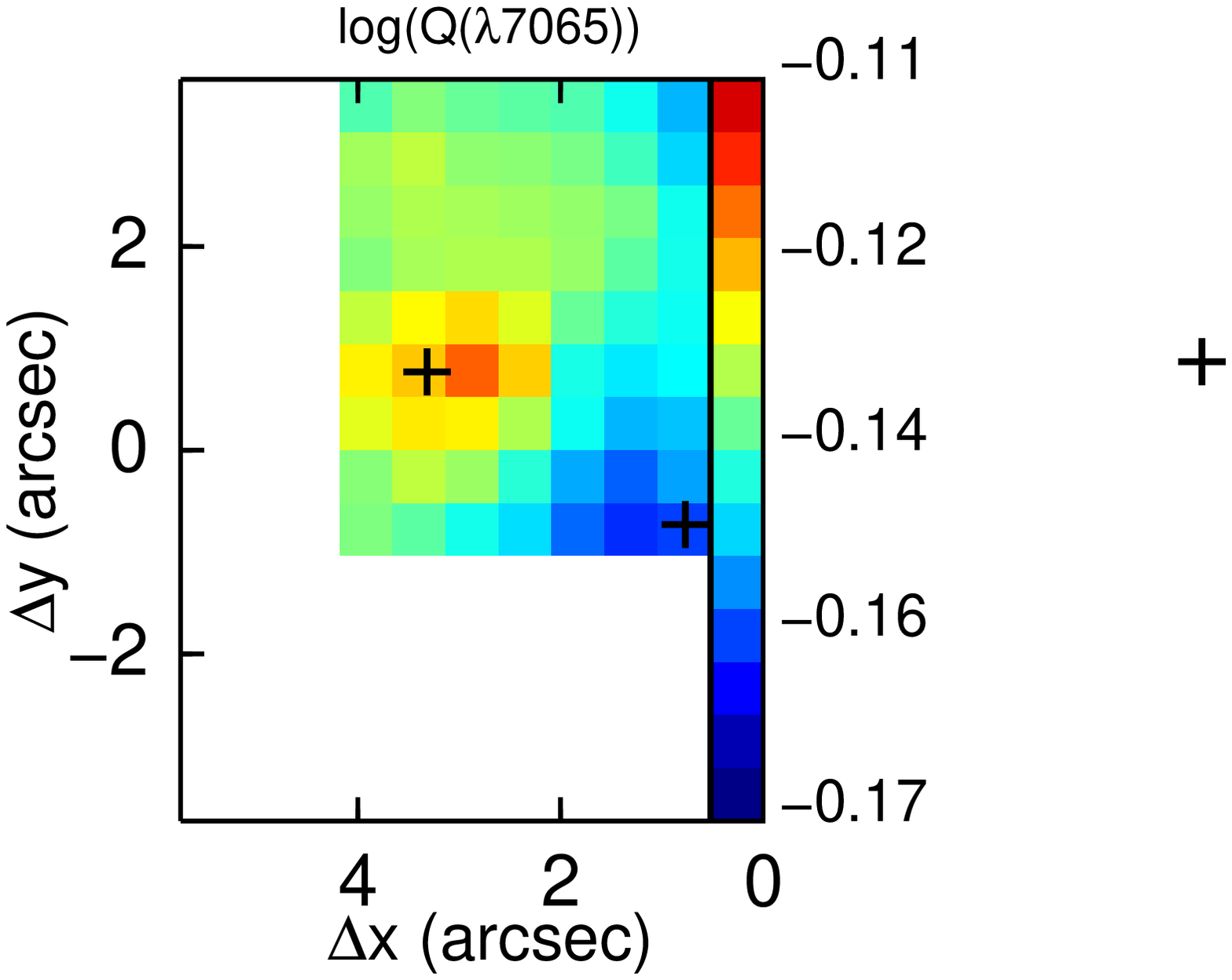}\\
   \caption[Differences on the $C/R$ factors  for different electron temperature]{
 Maps for the ratio between the $C/R$ factors derived for two assumptions of the $T_e(\hei)$: $Q(\lambda)=\frac{C/R(\lambda)_{Case 1}}{C/R(\lambda)_{Case 2}}$. Note that areas with constant values for which a $T_e$(\oiii)=10\,500~K was assumed have not been included in the comparison.
 A logarithmic color stretch is employed to emphasize the variations \emph{within} the region for a given line.
 \label{c2rratiosdif}}
 \end{figure}

Here, we will estimate the 2D contribution to the collisional component using the relations derived by  \citet{por07}, where  theoretical $C/R$ factors are calculated as functions of $n_e$ and $T_e$, and assuming that the density from \sii\ traces well that from helium.
Plasma temperatures as traced by helium lines can typically be $\sim$50\% of those traced by oxygen in planetary nebulae \citep{zha05}. For \hii\ the situation is not so clear. For the specific case of \object{NGC~5253}, values between 82\% and 96\% have been reported in specific apertures \citep{lop07}. We made two assumptions for the electron temperature:
\begin{itemize}
\item Case 1: The temperature from oxygen traces well that from helium: $T_e$(\hei) = $T_e$(\oiii);
\item Case 2: The temperature from helium is proportional to the temperature from oxygen, with the constant of proportionality estimated as the mean of the ratios between both temperatures provided by \citet{lop07}:  $T_e$(\hei) = 0.87\,$T_e$(\oiii).
\end{itemize}

Maps for the $C/R$ factors using the first assumption are presented in Fig. \ref{c2rratios} while the ratio between the two estimations is presented in Fig. \ref{c2rratiosdif}.
This is the first time that collisional effects for a set of helium lines have been mapped in an
extra-galactic source. Several results can be extracted from these maps.

Firstly, all the $C/R$ maps display the same structure. That is: higher ratios at the peak of emission for the ionized gas and towards the northwest half of the \ghiir\ and a decrease of the collisional contribution outwards. This reproduces the observed density structure \citepalias[see e. g. Fig. 6 in ][]{mon10}.

Secondly, 
lines corresponding to transitions in the singlet cascade have a negligible contribution from collisional effects (e.g. $C/R$ factor for $\lambda$4922 varies between  $\sim0.001-0.006$) while for those lines in the triplet cascade ($\lambda$7065, $\lambda$5876, $\lambda$4471,  and $\lambda$3889), the contributions from  collisional effects can be important.  In particular,  the  $C/R$ factor  for $\lambda$7065 ranges between $\sim0.02$ and $\sim0.22$, meaning it reaches  $\sim$20\% in the nucleus of the galaxy. 

Thirdly, the assumed temperature has some influence in the estimation of the collisional effects. In our particular case the assumed temperature in Case 2 was only $\sim15\%$ smaller than that in Case 1. However, this implies a smaller contribution of the collisional effects by $\sim$25-30~\% for $\lambda$7065  and $\sim$30-35~\% for $\lambda$3889 (the two lines most affected by collisional effects) and up to $\sim50$\% for the other lines under study.  Interestingly, areas of lower temperature are more sensitive to the assumption on $T_e$. 
It is important to note that the uncertainties associated to the errors due to the measurement of the line fluxes involved in the determination of $T_e$(\oiii) were typically $\lsim$1\,000~K \citepalias{mon12}. This is smaller than the difference  in the assumed temperature between the two reasonable assumptions, Case 1 and 2, which range between $\sim$1\,500~K at the peak of emission to $\sim$1\,300~K in the areas of lowest surface brightness. 
This implies that, at this level of data quality, more than errors associated to line fluxes, it is systematic errors associated with assumptions in density and temperature that dominate the uncertainties in the evaluation of the collisional effects.

\subsection{Radiative transfer effects as traced by $\tau(3889)$ and derivation of He$^+$ abundance \label{secymas}}

Single ionized helium abundance, $y^+$,   can be calculated as follows:

\begin{equation}
y^+(\lambda) = \frac{F(\lambda)}{F(H\beta)}
\frac{E(H\beta)}{E(\lambda)}
\frac{\frac{EW(\lambda) + a_{He\,I}(\lambda)}{EW(\lambda)}}{\frac{EW(H\beta) + a_{H}(H\beta)}{EW(\lambda)}} \frac{1}{f_\tau(\lambda)}
\label{eqymas}
\end{equation}

where $F(\lambda)/F$(\hb) is the extinction corrected flux for the \hei\ line scaled to \hb, $E$(\hb) and $E$(\hei) are the theoretical emissivities; $f_{\tau}(\lambda)$ is a factor that takes into account  radiative transfer effects. The remaining term in Eq. \ref{eqymas}, with $a(\lambda)$ is the equivalent width in absorption, which takes into account the effect of the underlying stellar population.
In the following, we will describe how each of these terms were evaluated.

\subsubsection{Calculation of the emissivities}

For \hb\  emissivities, we utilized the function from \cite{sto95} which is, in units of 10$^{25}$ erg cm$^{-3}$ s$^{-1}$:

\begin{equation}
E(H\,\beta) =  4\pi j_{H\,\beta}/n_e n_{H^+} = 1.37 t_e^{-0.982} \exp(-0.104/t_e)
\end{equation}

with $t_e = T_e/10^4 $.
For the \hei\ lines, we utilized those emissivities originally provided by \citet{por12} and recently corrected by \citet{por13}. These are the most recent \hei\ emissivities and have the collisional effects included. Therefore, there was no need to include an extra term in Eq. \ref{eqymas} to take into account collisional effects. They are tabulated for discrete values of $n_e$ and $T_e$. However, the $n_e$ in the main \ghiir\  varies between $\lsim100$~cm$^{-3}$ and $\sim660$~cm$^{-3}$ and $T_e$ ranges between $\sim9\,000$ and $\sim12\,000$~K. Therefore, to evaluate the emissivities at each individual spaxel, we fitted the values provided for $n_e= 100$ and 1\,000~cm$^{-3}$  and $T_e$ ranging from 5\,000~K to 25\,000~K to functions with the same parametrization as $E(H\,\beta)$, $a  t_e^{b} \exp(c/t_e)$, and then interpolated on a logarithmic scale as follows:

\begin{equation}
E(\lambda,\log n_e) = E(\lambda,2) + (E(\lambda,3) - E(\lambda,2)) (\log n_e -2)
\end{equation}

\begin{figure}[th!]
   \centering
\includegraphics[angle=0,width=0.24\textwidth,clip=]{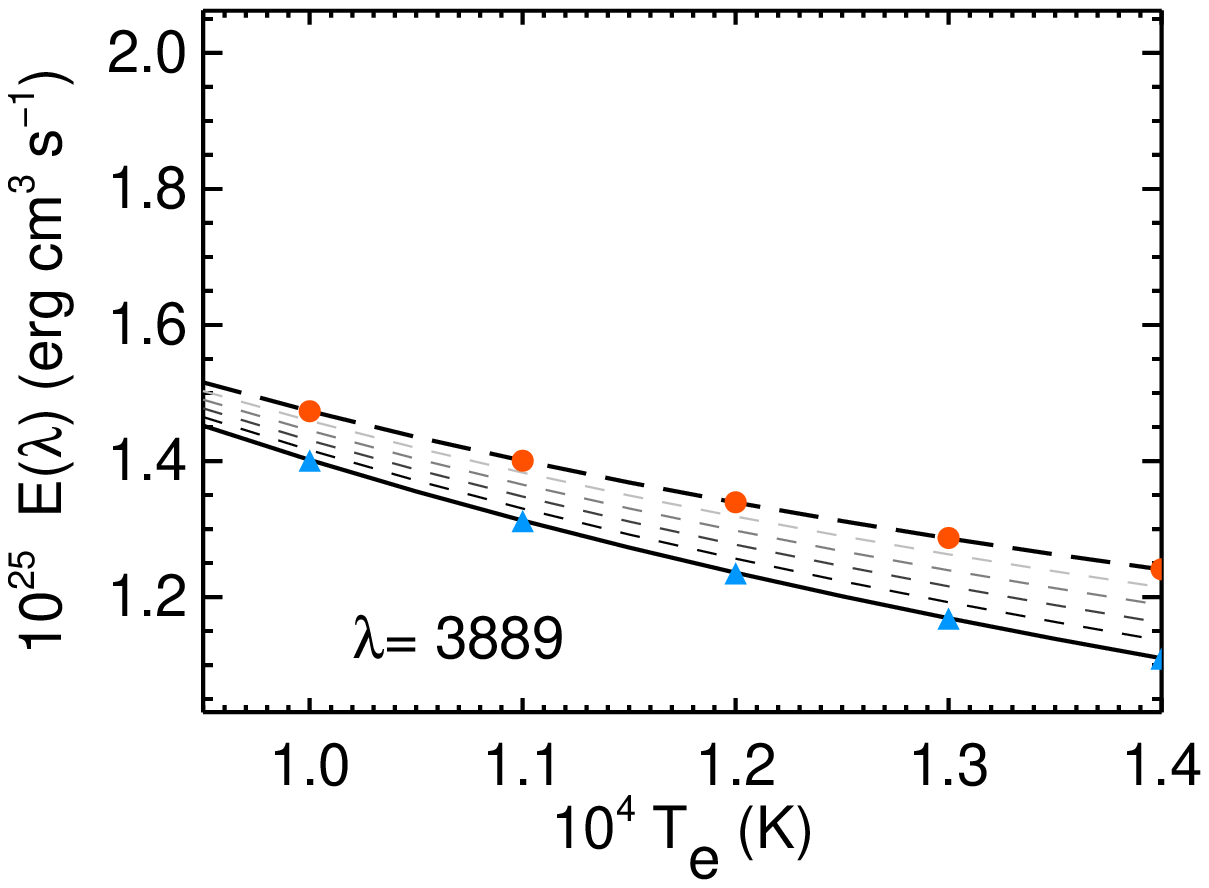}
\includegraphics[angle=0,width=0.24\textwidth,clip=]{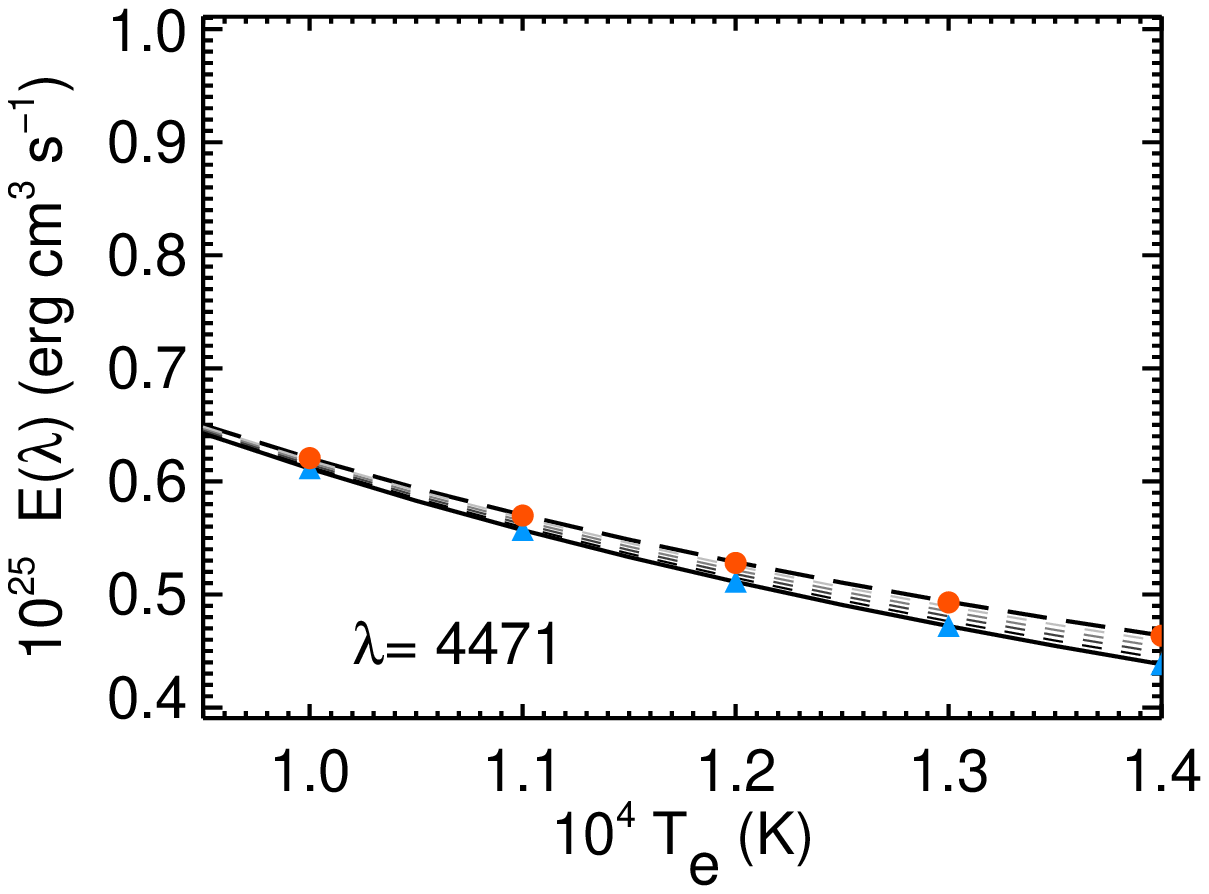}\\
\includegraphics[angle=0,width=0.24\textwidth,clip=]{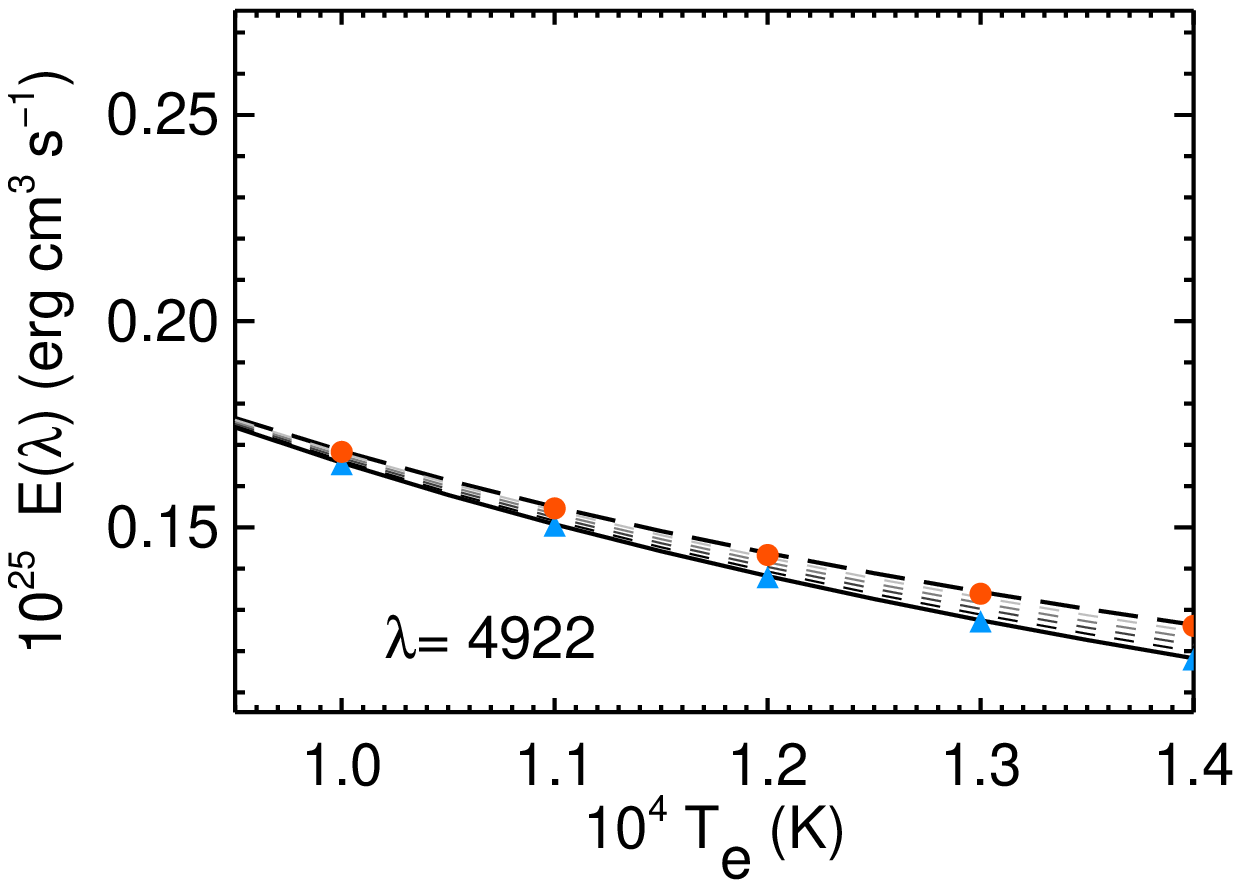}
\includegraphics[angle=0,width=0.24\textwidth,clip=]{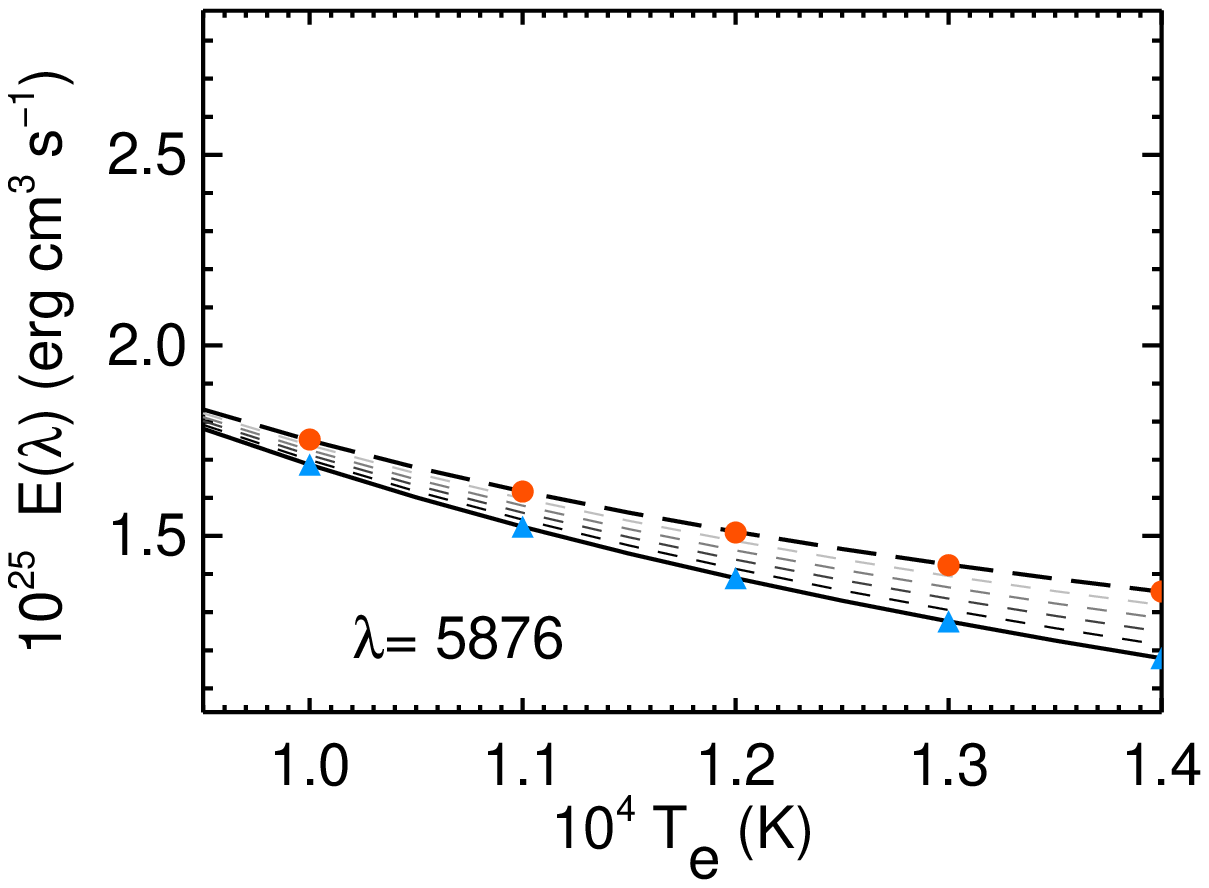}\\
\includegraphics[angle=0,width=0.24\textwidth,clip=]{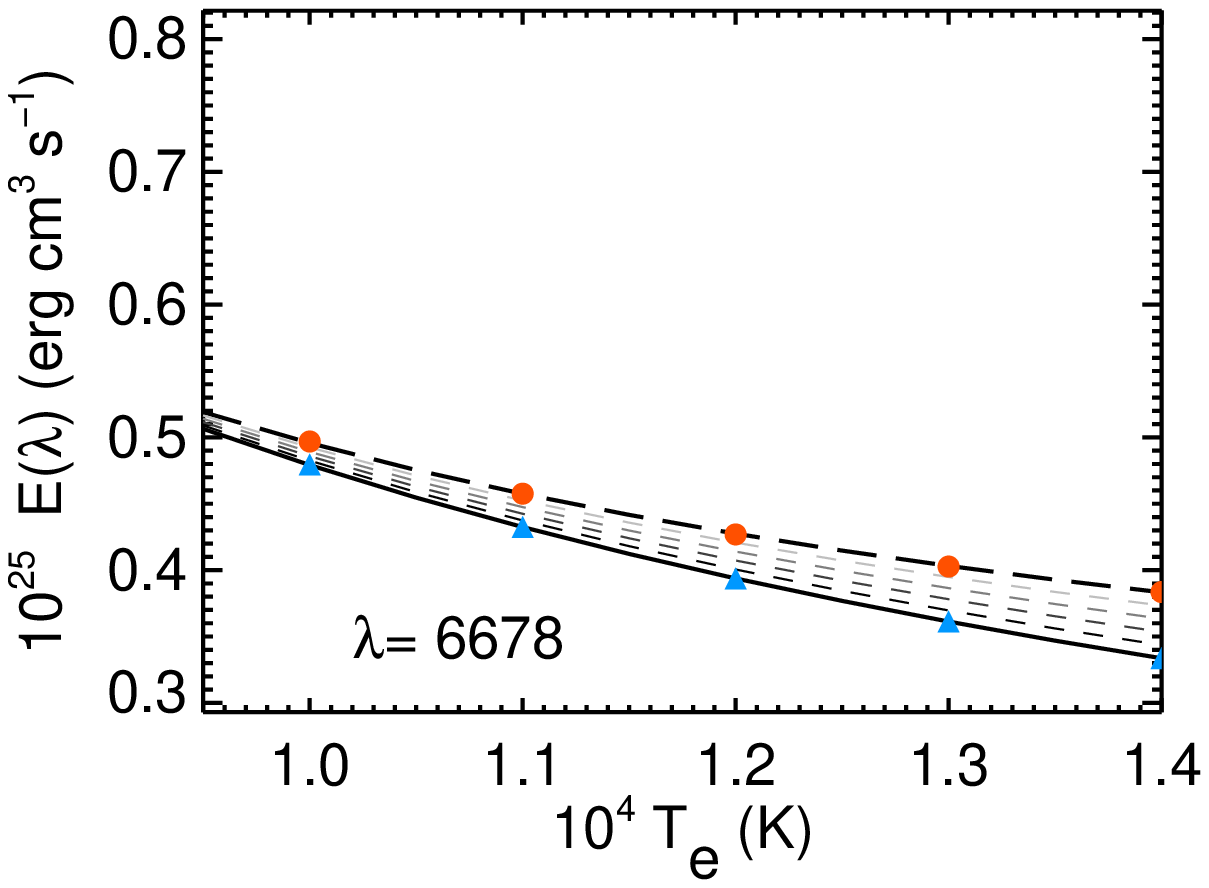}
\includegraphics[angle=0,width=0.24\textwidth,clip=]{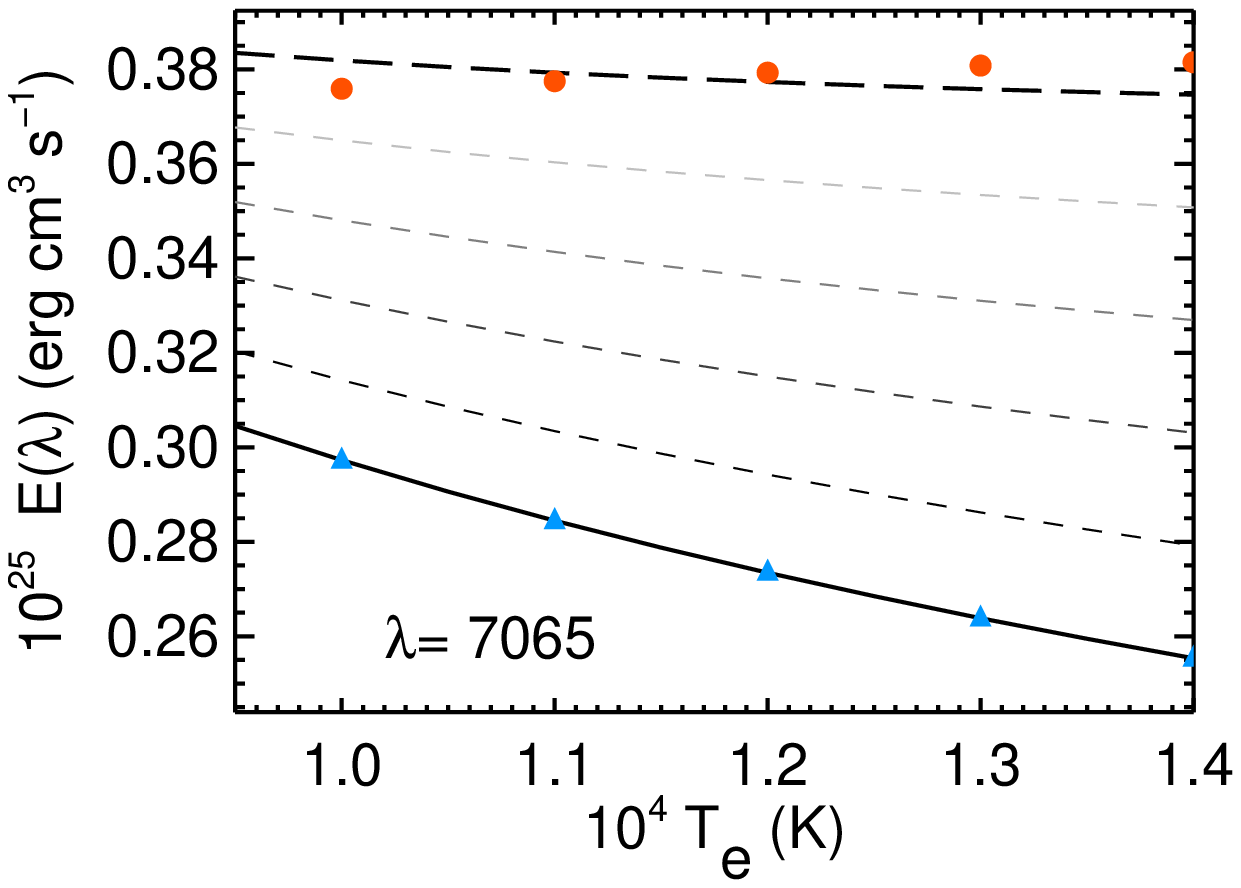}\\
   \caption[Fitted function to the Porter et al. 2012 emissivities]{Fitted functions to the \citet{por12} emissivities. The black lines are the functions for $n_e=100$~cm$^{-3}$ (continuous line) and for $n_e=1000$~cm$^{-3}$ (long dashed line). The intemediate short dashed lines represent the interpolated emissivities for $\log(n_e) = 2.2, 2.4, 2.6, 2.8$. 
 \label{compaemi}} 
 \end{figure}

The coefficients and standard deviations of the fits are compiled in Table \ref{coeficientes} while Fig. \ref{compaemi} shows a 
comparison between the fitted function and the discrete values provided by \citet{por12} for the range of densities and temperatures covered in the \ghiir. A comparison between this figure and Fig. \ref{c2rratios} shows the correspondence between the degree of dependence on the density and the contribution of the collisional effects.

\begin{table}
\small
     \centering
     \caption[]{Coefficients for the equations fitted to the \citet{por12} Table 2. \label{coeficientes}}
            \begin{tabular}{lcccccccccccc}
            \hline
            \noalign{\smallskip}
  Line                   & $a$  &  $b$  &  $c$  & Std Dev (\%)  \\
\hline
\multicolumn{5}{c}{$n_e=100$~cm$^{-3}$}\\
\hline
\noalign{\smallskip}  
3889 & 1.453 & $-0.724$ & $-0.036$ & 0.010 \\
4471 & 0.679 & $-1.078$ & $-0.105$ & 0.003\\
4922 & 0.184 & $-1.091$ & $-0.105$ & 0.001\\
5876 & 1.680 & $-1.061$ & $+0.004$ & 0.042\\
6678 & 0.473 & $-1.065$ & $+0.013$ & 0.015\\
7065 & 0.269 & $-0.367$ & $+0.100$ & 0.007\\ 
\hline
\multicolumn{5}{c}{$n_e=1000$~cm$^{-3}$}\\
\hline
3889 &  1.198 & $-0.336$ & $+0.207$ & 0.032\\
4471 &  0.507 & $-0.694$ & $+0.202$ & 0.048\\
4922 & 0.131 & $-0.642$ & $+0.252$ & 0.016\\
5876 & 0.895 & $-0.195$ & $+0.671$ & 0.255\\
6678 & 0.237 & $-0.138$ & $+0.738$ & 0.094\\
7065 & 0.320 &  $+0.171$ & $+0.163$ & 0.096\\ 

\noalign{\smallskip}  
\hline

\end{tabular}
\end{table}


\subsubsection{Correction for underlying stellar population}

To take into account the effect of the underlying stellar population it is necessary to estimate the equivalent width both in emission and absorption for \hb\ and the helium lines.
The equivalent width of \hb\ in emission (not shown) ranges typically from $\sim$240~\AA\ at the peak of emission for the ionized gas to $\sim$65~\AA\ in the most outer regions of the area sampled here.
We assumed a component in absorption of 2~\AA, which is adequate for very young starbursts, as the one at the nucleus of \object{NGC~5253}  \citep[e.g.][]{gon05,alo10}. 
This implies a correction for \hb\ in absorption from a negligible value (i.e. $\lsim1\%$) at the peak of emission to $\sim$4\% in the outer parts of the covered area. 

  \begin{figure}[th!]
   \centering
\includegraphics[angle=0,width=0.24\textwidth,clip=,bb = 45 30 405 390]{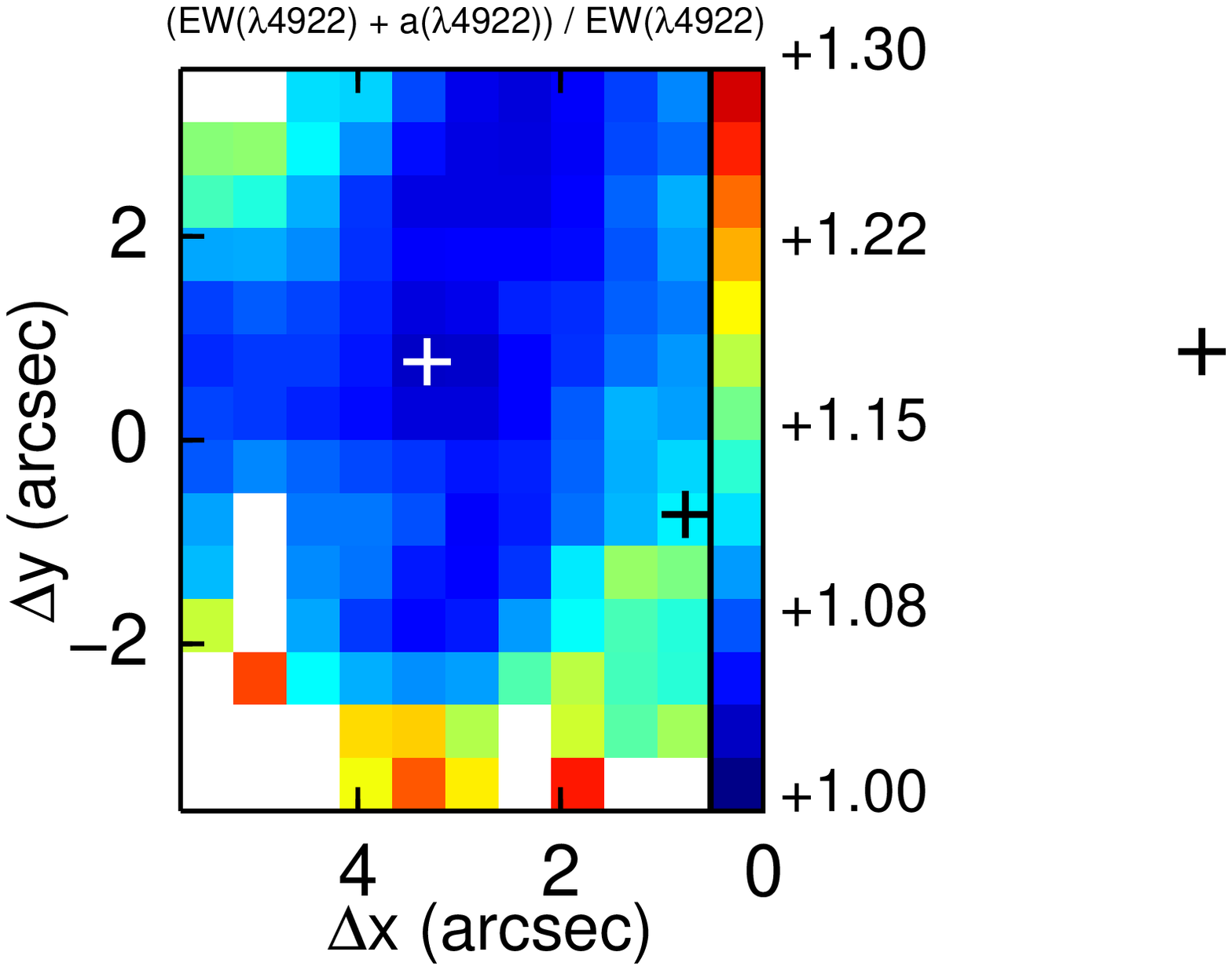}
\includegraphics[angle=0,width=0.24\textwidth,clip=,bb = 45 30 405 390]{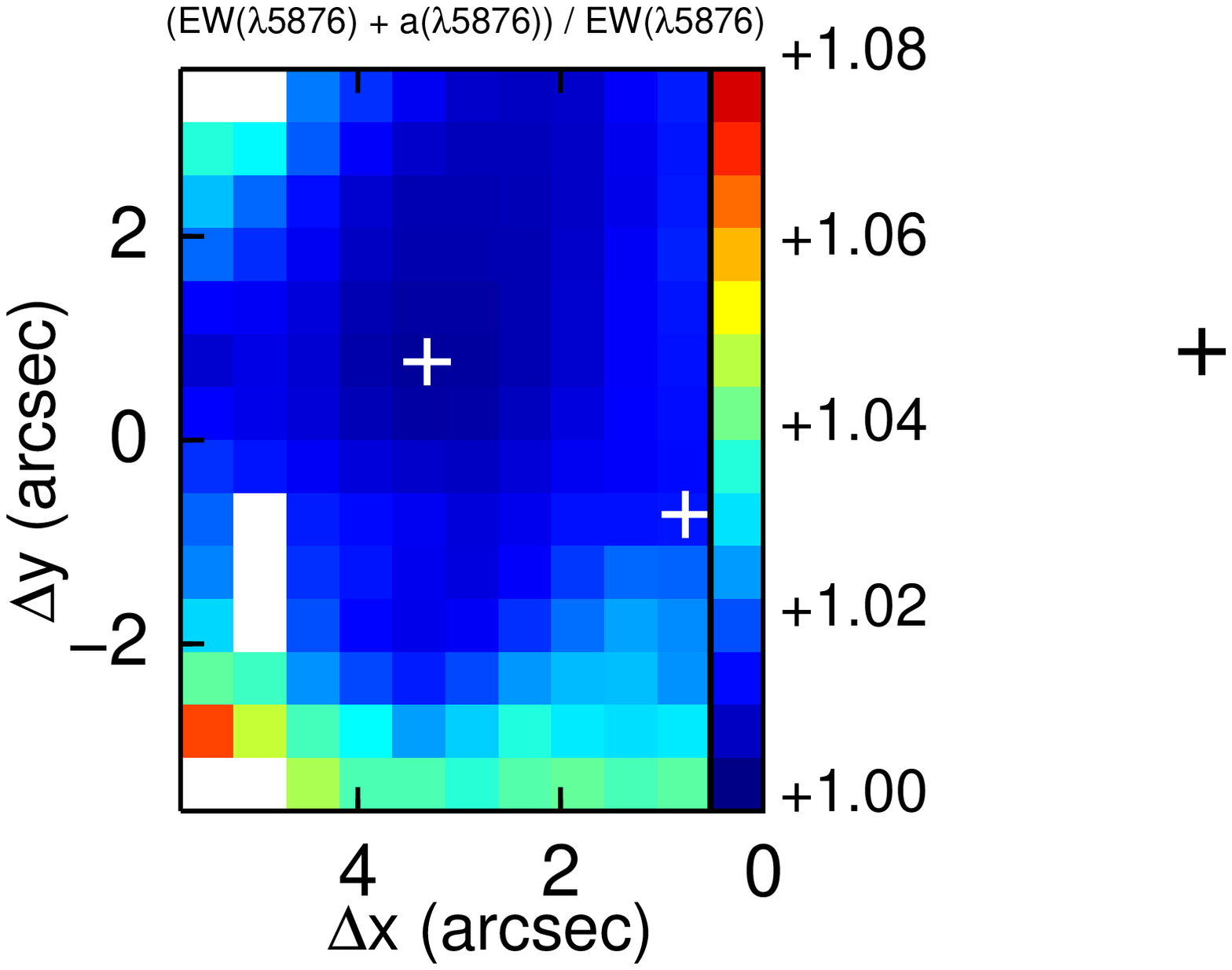}\\
\includegraphics[angle=0,width=0.24\textwidth,clip=,bb = 45 30 405 390]{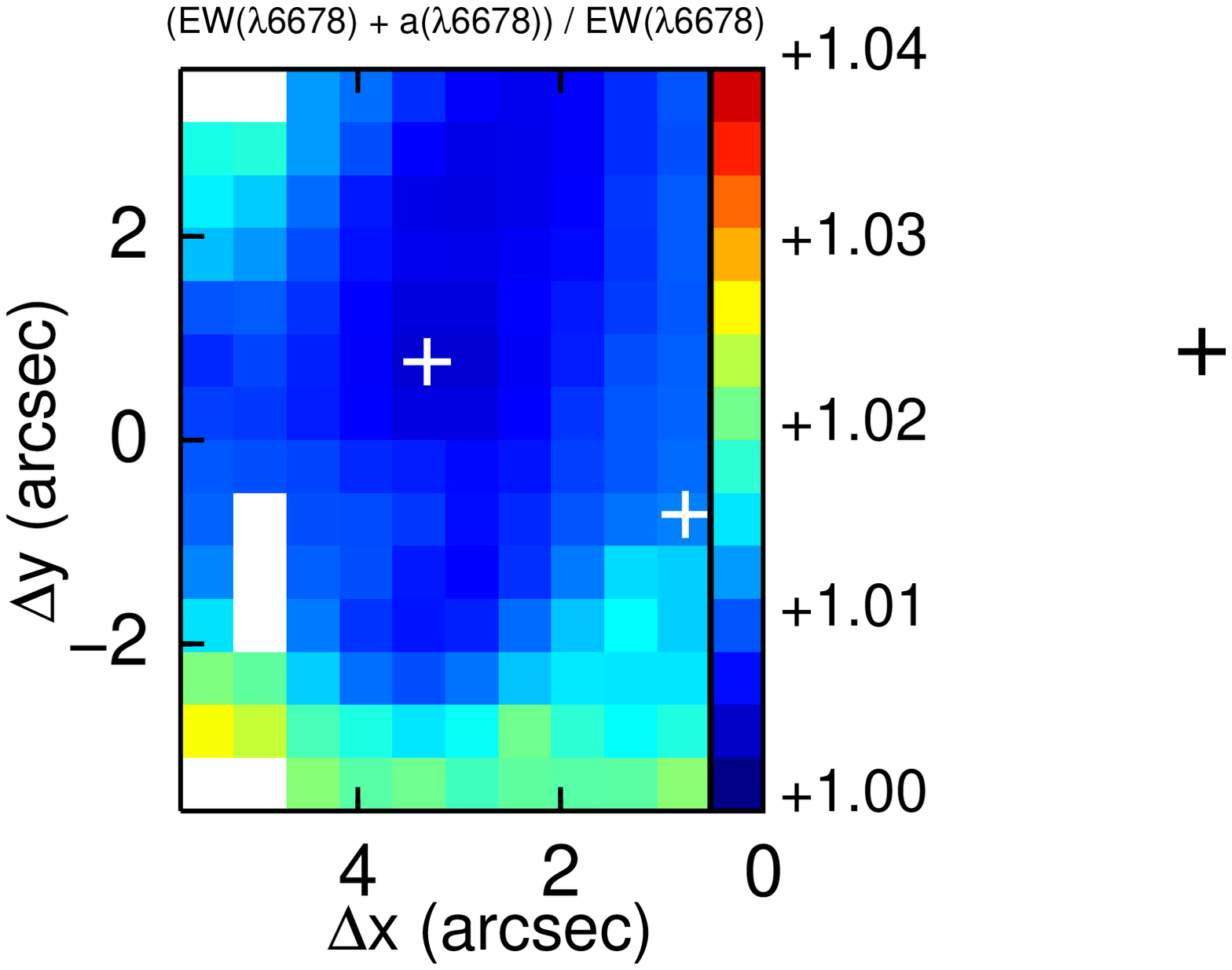}
\includegraphics[angle=0,width=0.24\textwidth,clip=,bb = 45 30 405 390]{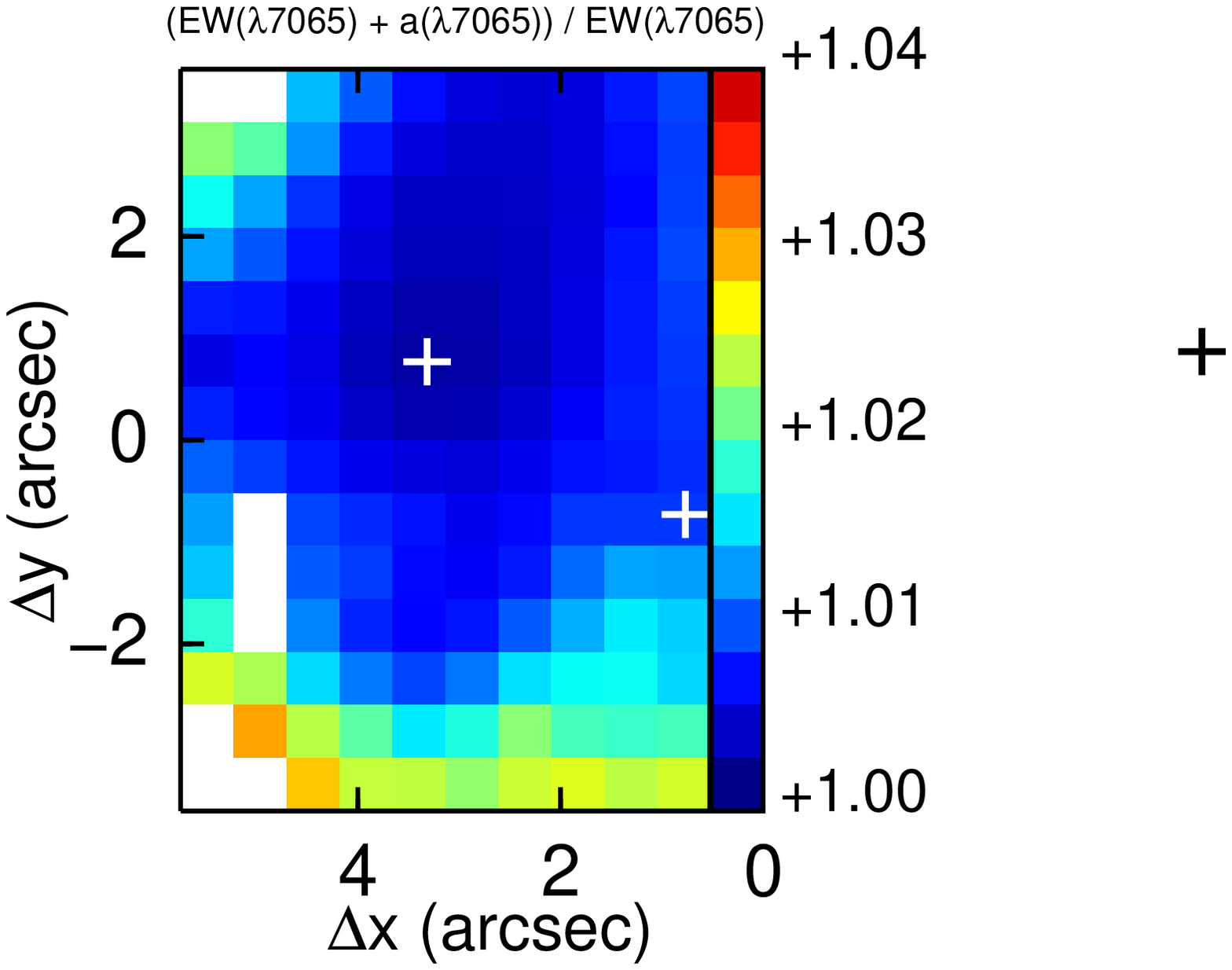}\\
   \caption[Maps for the estimated correction factors for an underlying stellar population]{
 Maps of the correction factor due to a component in absorption in the \hei\ lines.
Different scales were utilized in the different maps to emphasize the structure.
 The corresponding map for the \hb\ correction factor (not shown) is similar to those for $\lambda$6678 and $\lambda$7065.
 \label{corrabs}}
 \end{figure}

The correction due to the underlying stellar population is not straightforward to estimate for the helium lines. 
As explained in \citetalias{mon12}, for the 
data obtained with FLAMES Giraffe and the LR1 or LR2 gratings, the contribution of the stellar population was separated from that of the gas using the STARLIGHT code \citep{cid05,cid09} to match the stellar continuum. Therefore, 
we can assume $a_{HeI}(\lambda4471)=a_{HeI}(\lambda3889)=0$.
For the other lines,  a common approach assumes identical stellar equivalent widths for all the helium lines. However,  \hei\ lines are produced by the bluest stars. Therefore, in the context of a galaxy  suffering a burst of star formation on top of an older population, redder lines should have smaller equivalent widths in absorption, since older stars contribute more to the stellar continuum.
Specifically, typical estimations of the equivalent width in absorption for redder lines would be $\sim40-80$\% that of $a_{\hei}(\lambda4471)$ \citep{ave10}. This is one of the lines observed with the LR2+FLAMES configuration for which nebular and stellar information have been disentangled. Therefore $a_{\hei}(\lambda4471)$ could be measured from our emission line free cube. Typical values were extremely low (i.e. $\sim0.11\pm0.03$~\AA) and without any obvious variation following the structure of the \ghiir\ or the location of the Super Star Clusters.  Taking this as reference, and the relative values between the different $a_{\hei}(\lambda)$'s reported by \citet{ave10}, which were in turn derived using the models presented by \citet{gon05} and \citet{mar05b}, we assumed $a_{\hei}(\lambda)=0.09, 0.08, 0.06$ and 0.05~\AA\ for $\lambda$4922, $\lambda$5876, $\lambda$6678, $\lambda$7065, respectively. 
Equivalent widths for the  $\lambda$4922, $\lambda$5876, $\lambda$6678, $\lambda$7065 \hei\ emission lines were $\sim0.3-4$~\AA, $\sim20-60$~\AA, $\sim3-20$~\AA, and $\sim3-35$~\AA, respectively. With these values, the largest correction for absorption was for \hei$\lambda$4922, with typical values between 3\% and 6\%. However, corrections in the most external spaxels could reach up to  $\sim$25\%. For the other lines the correction was more moderate, with values between $\sim$1\% and $\lsim$8\% for \hei$\lambda$5876 and comparable to the correction in \hb\ for the  \hei$\lambda$6678 and  \hei$\lambda$7065 lines. This is illustrated in Fig. \ref{corrabs}.

\subsubsection{Estimation of radiative transfer effects}

Radiative transfer effects can be important in recombination radiation. Given the metastable character of the $2^3 S$ level, under certain conditions, the optical depths in lower $2^3S - n^3 P^0$ lines imply non-negligible effects on the emission line strengths \citep{ost06}. Specifically $\lambda$10\,830 photons are only scattered, but absorbed photons, corresponding to transitions to higher levels, can be converted to several photons of lower energy. The most prominent example is
$\lambda$3889 photons that can be converted to $\lambda4.3~\mu$m $3^3 S-3^3 P^0$, plus $\lambda 7065\, 2^3 S-3^3 P^0$, plus $\lambda10\,830\, 2^3 S-2^3 P^0$. The net effect  is that lines associated with transitions from the $2^3 P$ level upwards (e.g. $\lambda3889$) are weakened by self-absorption, while lines associated with several transitions from higher levels (e.g. $\lambda$7065) are strengthened by resonance fluorescence. Contrary to collisional effects, radiative transfer effects do not affect photons in the singlet cascade.
The relative importance of radiative transfer effects is quantified by a correction factor, $f_\tau(\lambda)$,  for each line which is a function of the optical depth at $\lambda$3889, $\tau(3889)$. Here, we  parametrized these factors by fitting the values provided by \citet{rob68} to a non-expanding nebula to the functional form $f_\tau(\lambda) _{\omega=0} = 1 + a \tau^b$. The corresponding functions for the lines utilized in this work are:

\begin{equation}
f_\tau(7065)_{\omega=0} = 1 + 0.741\tau^{0.341} \label{eqfac7065}
\end{equation}

\begin{equation}
f_\tau(6678)_{\omega=0} = 1    \label{eqfac6678}
\end{equation}

\begin{equation}
f_\tau(5876)_{\omega=0} = 1 + 0.0126\tau^{0.496}   \label{eqfac5876}
\end{equation}

\begin{equation}
f_\tau(4922)_{\omega=0} = 1    \label{eqfac4922}
\end{equation}

\begin{equation}
f_\tau(4471)_{\omega=0} = 1 + 0.0022\tau^{0.728}   \label{eqfac4471}
\end{equation}

\begin{equation}
f_\tau(3889)_{\omega=0} = 1 - 0.261\tau^{0.305}  \label{eqfac3889}
\end{equation}

Note that no values were provided for \hei$\lambda$4922 in the original work of \citet{rob68}. However, as it happens for \hei$\lambda$6678, this is a singlet line, and therefore $f_\tau(4922) = 1$ can be assumed. 

Typically, optical depth ($\tau(3889)$) and \hei\ abundance ($y^+$)  (and other parameters) are determined simultaneously by minimizing $\chi^2$, defined as the difference between each helium line's abundance (weighted according to a reasonable criterion like the line flux) and the average. In this methodology it is implicit that all the lines trace the same location in the nebula/galaxy. However, the area under study in this work suffers from heavy extinction \citepalias[see Fig. 3 in][]{mon10}.
Therefore,  \emph{a priori}, it is not possible to assume that all the lines equally penetrate the nebula interior and that bluer and redder lines trace zones with the same $y^+$. Because of that, we grouped the lines in two sets according to their wavelengths, called hereafter the \emph{blue} ($\lambda$3889, $\lambda$4471, $\lambda$4922) and \emph{red} ($\lambda$5876, $\lambda$6678, $\lambda$7065) \emph{sets}, which will be analyzed independently. Each set is made out of: i) a line from the singlet cascade, and therefore not affected by radiative transfer effects; ii) a line from the triplet cascade, thus highly sensitive to radiative transfer effects; iii) a line from the triplet cascade mildly sensitive to radiative transfer effects.

  \begin{figure}[th!]
   \centering
\includegraphics[angle=0,width=0.24\textwidth,clip=,bb = 45 30 405 390]{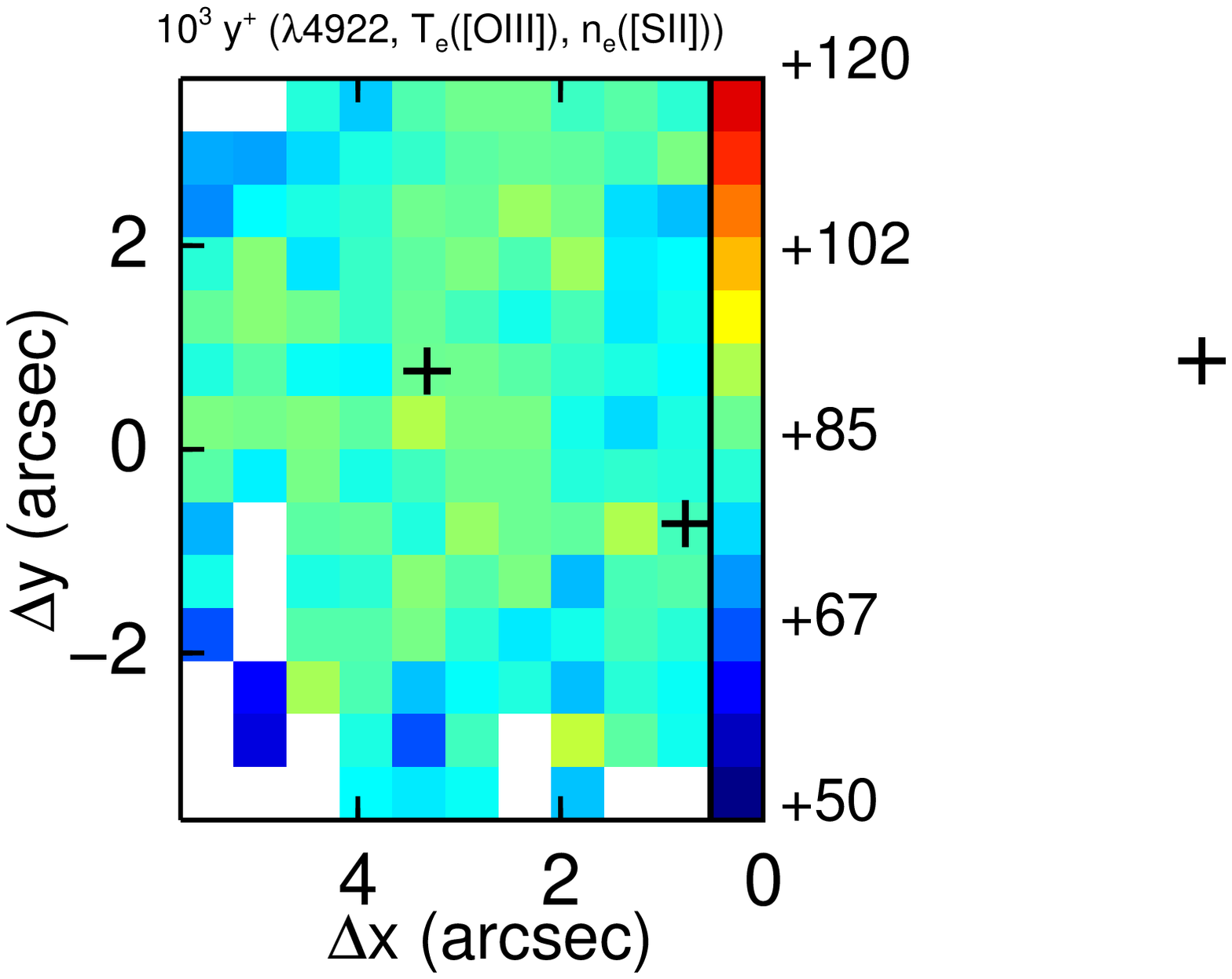}
\includegraphics[angle=0,width=0.24\textwidth,clip=,bb = 45 30 405 390]{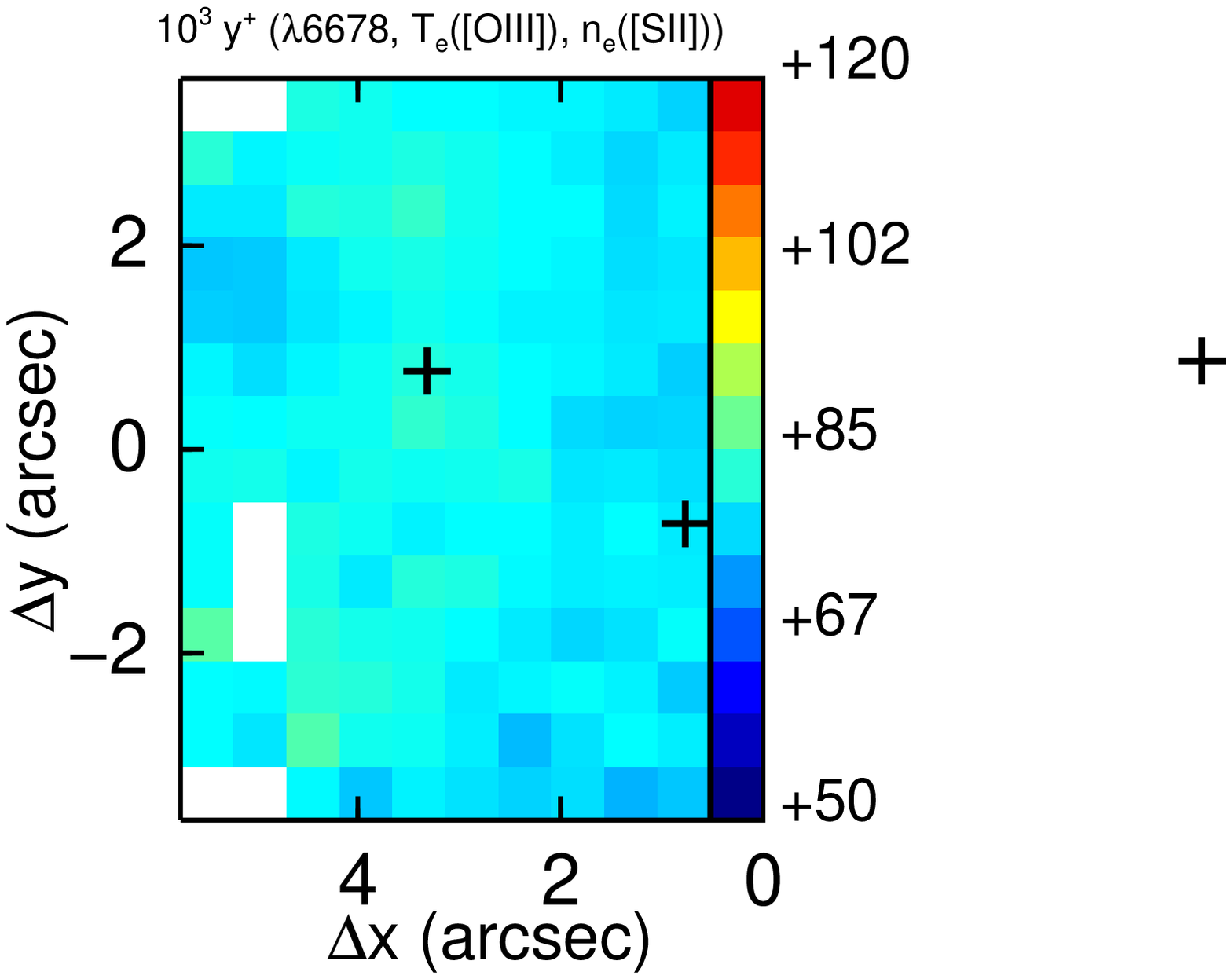}\\
   \caption[Abundance maps from singlets]{Maps for $y^+$ derived from lines of the singlet cascade for the blue (left) and red (right) sets.
A common scale was used in both maps to facilitate comparison between them.
  \label{abunmapsinglet}}
 \end{figure}

A first estimation of the $y^+$ abundance structure as traced by the red and blue lines can be obtained from the singlet lines, since they are not affected by radiative transfer effects. This is shown in Fig. \ref{abunmapsinglet}. The mean ($\pm$ standard deviation) are 
80.7($\pm$5.1) and 76.8($\pm$1.8)
for the $\lambda$4922 and $\lambda$6678 lines averaged over the mapped area. These values indicate that even if the red and blue lines are not tracing exactly the same gas columns, at least they sample areas with a the same  $y^+$ within $\sim$5\%. 

\begin{figure}[th!]
   \centering
\includegraphics[angle=0,width=0.24\textwidth,clip=,bb = 45 30 405 390]{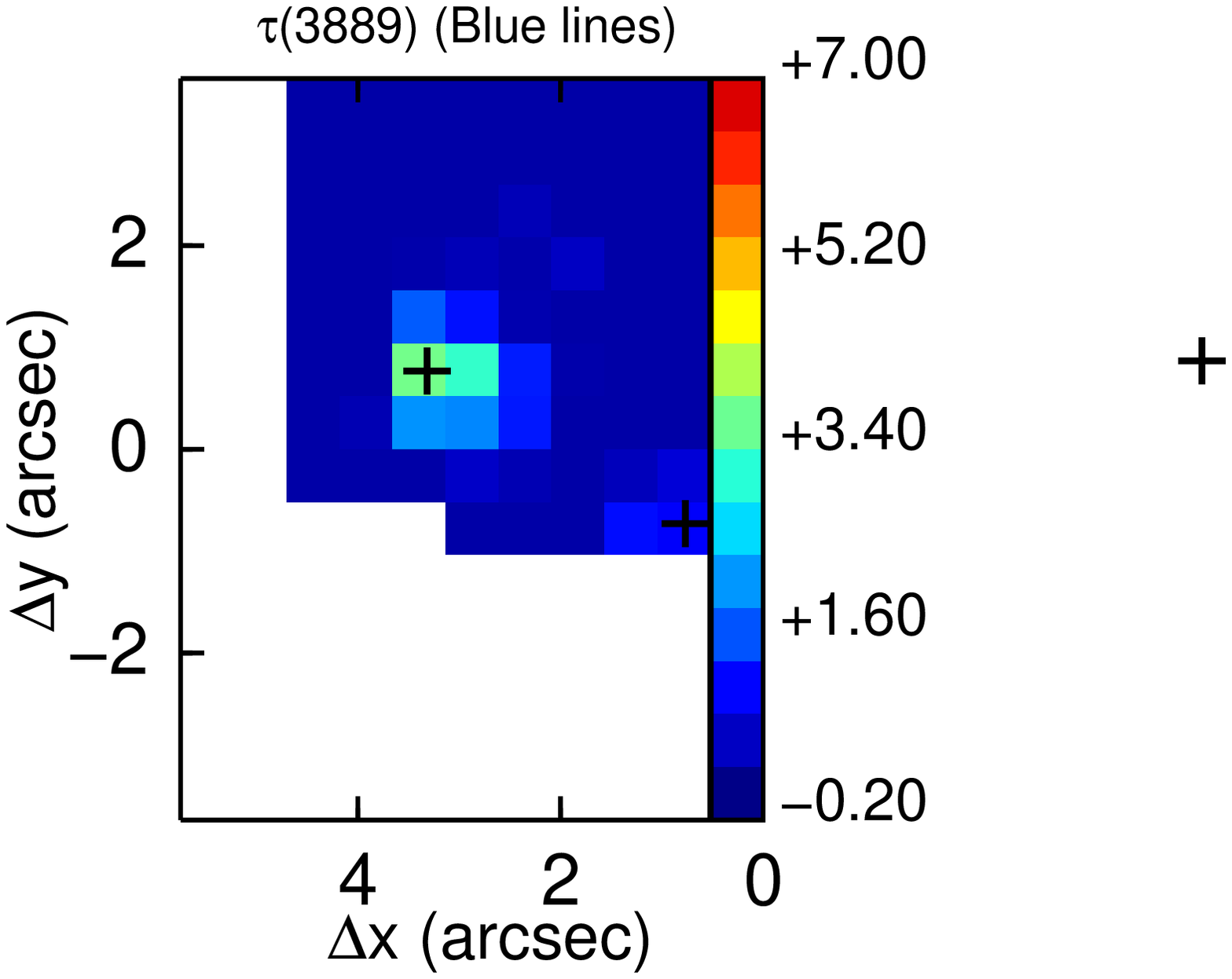}
\includegraphics[angle=0,width=0.24\textwidth,clip=,bb = 45 30 405 390]{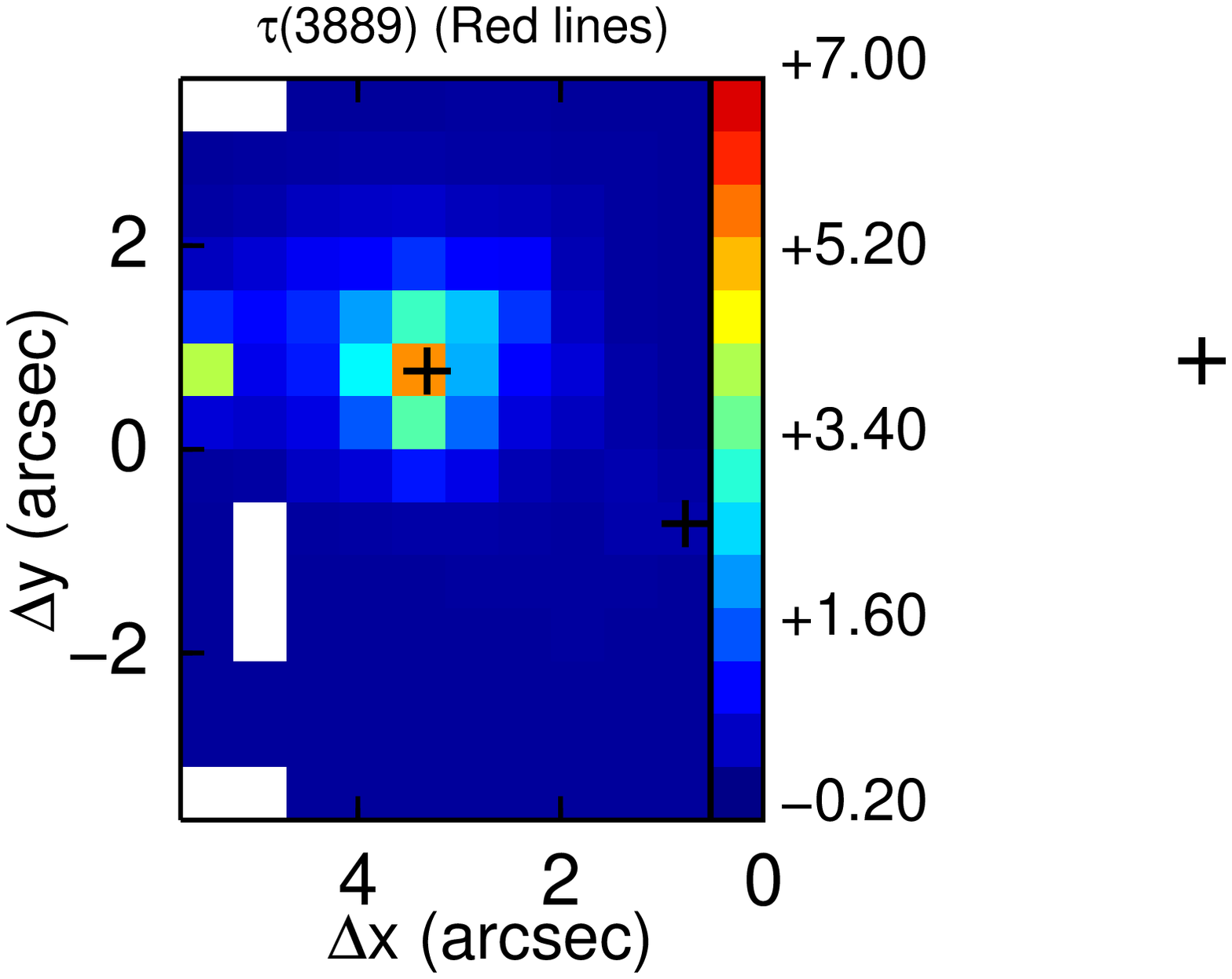}\\
   \caption[Maps for $\tau(3889)$]{Maps for $\tau(3889)$ as derived from the $\lambda$4922 and $\lambda$3889  (left) and the $\lambda$6678 and $\lambda$7065 (right) emission lines.
 \label{maptau}}
 \end{figure}

A comparison of the initial $y^+$ maps  with those obtained from a line highly sensitive to radiative transfer effects in each set ($\lambda$3889 and $\lambda$7065), allowed us to determine the respective $f_\tau(\lambda)$ map for each set, which can, in turn, be converted into $\tau(3889)$ maps. These are shown in Fig. \ref{maptau}. The optical depth has the same structure in both maps, with the peak at the main super star cluster(s). The shape of the area presenting $\tau(3889)>0$ is circular but resolved. With a FWHM=1\farcs6$\sim$30~pc, this is larger than the seeing ($\sim$0\farcs9). This is consistent with a picture where large optical depths are not restricted to the deepest core of the galaxy, associated with the supernebula, but extend over a larger region. Also, the fact that
the optical depths derived from the red set of lines are larger than those derived from the blue set is in harmony with a picture where deeper layers in the nebula, which we interpreted as denser and hotter \citep[][see also Beck et al. 2012]{mon12}, suffer from larger radiative transfer effects and suggests a link between \hei\ optical depth and dust optical depth.

\subsubsection{Final derivation of He$^+$ abundance}

\begin{figure}[th!]
   \centering
\includegraphics[angle=0,width=0.24\textwidth,clip=,bb = 45 30 405 390]{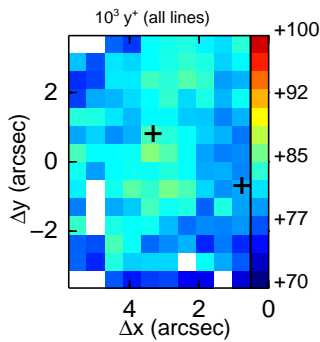}
   \caption[Final $y^+$ map]{Final map for $y^+$ derived from a weighted average of those derived for the lines in the blue and red sets.  Note we used a different scale than in Fig. \ref{abunmapsinglet}  to increase the contrast and emphasize the structure following the excitation structure. Instead, the scale is common with that in the right column of Fig. \ref{icfhe} to make easier the comparison between total and ionic helium abundances.
  \label{abundancefinal}}
  \end{figure}

Finally for both sets, the information of the  mildly sensitive line was added and abundances from each line were recalculated using their corresponding $\tau(3889)$. The final abundance maps for each set  were  made using a weighted average.  We used the mean fluxes of the \hei\ in the utilized area to determine the respective weights. These were 3:1:1 for the $\lambda$5878:$\lambda$6678:$\lambda$7065 and 25:5:1 for $\lambda$3889:$\lambda$4471:$\lambda$4922.  The mean ($\pm$ standard deviation) for the red and blue sets for $y^+$ are
79.3($\pm$2.5) and 82.0($\pm$3.8),
i.e., results from the red and the blue sets agree to within $\sim$3\%. 
Since differences between the  maps derived from the blue and red sets are consistent with tracing the same $y^+$ per spaxel, we derived a final map using the information of all the lines by averaging these two maps with a \emph{blue:red} weight of 0.8:1.0, the mean ratio between the brightest lines in the blue and red sets (i.e. $\lambda$3889 and $\lambda$5876).  This is shown in Fig. \ref{abundancefinal}. The mean ($\pm$ standard deviation) is
 80.3($\pm$2.7).
  This value compares well with other values of $y^+$ reported in the literature for this area \citep{kob97,lop07,sid10}.


To estimate the emissivities, the calculation of these abundances was made with assumed $n_e$(\sii) and $T_e$(\hei) = $T_e$(\hi) = $T_e$(\oiii). As for the mapping of the collisional effects, in order to evaluated the influence of the selection of $T_e$,  the derivation presented here was repeated assuming   $T_e$(\hei) = $T_e$(\hi) = 0.87 $T_e$(\oiii) and finding a mean  ($\pm$ standard deviation) of
79.4($\pm$3.0).
This implies that the precise selection of $T_e$ has a small effect ($\sim$1\%) in the determination of $y^+$, in comparison to other factors like e.g.  the correction for absorption or radiative transfer effects, as long as one keeps this selection within reasonable values.

Finally, it is worth mentioning that throughout all the derivation of $y^+$, we have assumed that He\,\textsc{i} singlets are formed under Case B conditions (i.e. lines are formed in the limit of infinite Lyman line  optical depth and there is no optical depth in transitions arising from excited states). This is the standard framework for the derivation of helium abundances.
However, under certain conditions, helium Ly$\alpha$ $\lambda$584 (and higher Lyman transitions) may be attenuated due to two effects: absorption of the helium Ly$\alpha$ line by dust and hydrogen absorption \citep{fer80,shi93}.
If these effects were important, the Case B assumption would no longer be applicable. The fact that we have recovered similar helium
abundances from the two sets of lines supports the Case B assumption.
 
\subsection{Determination of He abundance}

The favorite scenario to explain the extra nitrogen found within the area studied here is that this has been produced by Wolf-Rayet stars and presumedly those at the two Super Star Clusters. In this scenario, as it happens with the nitrogen, an overabundance of helium is also expected to be observed.
Specifically, typical stellar atmospheric $N/He$ ratios  would be $\sim3\times10^{-3}$ and $\sim5\times10^{-4}$ for WN and WC stars, respectively \citep{smi82}. This could be use as a reference for the expected $N/He$ ratios of the material newly incorporated into the warm ISM. Are our data supporting an enrichment in helium abundance compatible with these ratios and, therefore, supporting the Wolf-Rayet stars hypothesis?

  \begin{figure}[ht!]
   \centering
\includegraphics[angle=0,width=0.39\textwidth,clip=]{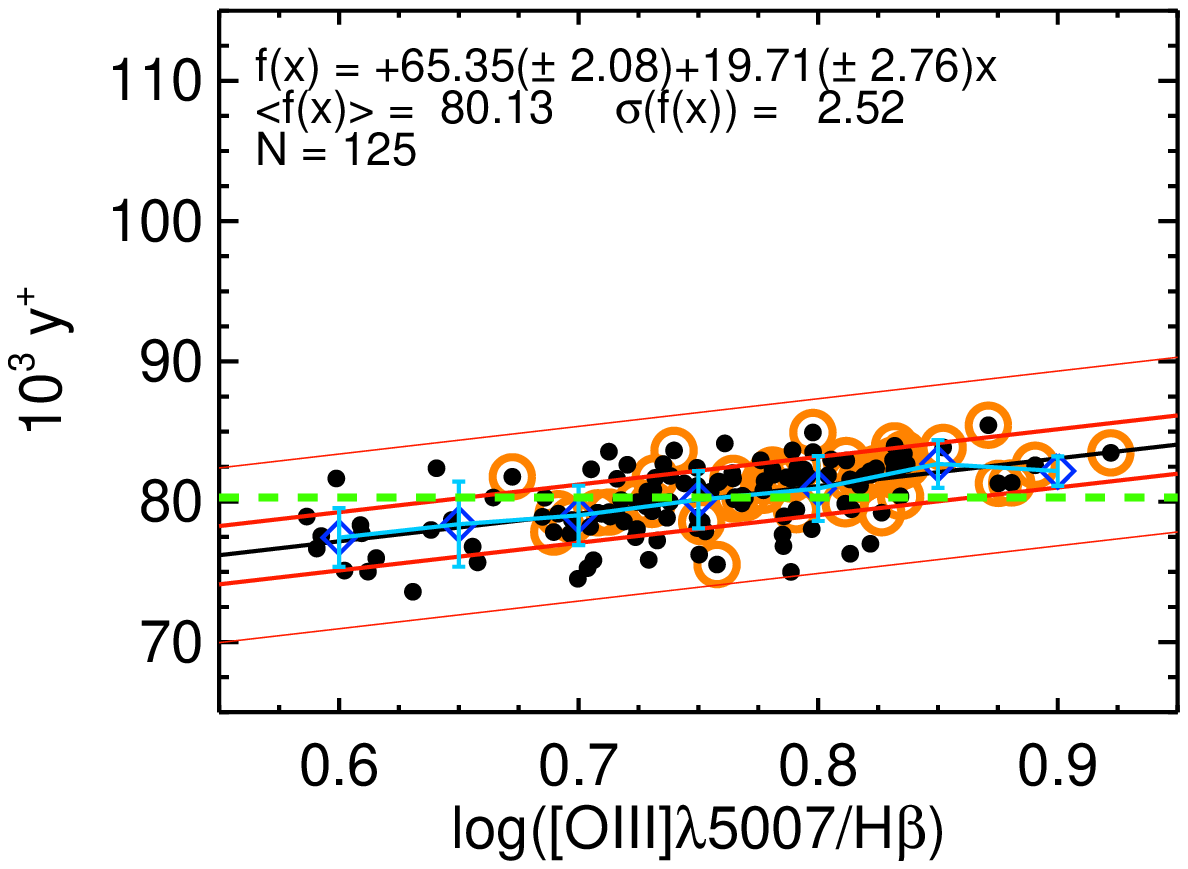}
\caption{
He$^+$ abundance vs. \ohb. The first-degree polynomial fit is shown with a black line. The 1-$\sigma$ and 3-$\sigma$ levels are marked with thick and thin red lines respectively. Mean and standard deviation of each 0.05 dex bin in $\log$(\ohb) are shown with blue diamonds and error bars respectively. The green horizontal dashed line shows the mean value. Data points corresponding to spaxels with $\log(N/O)>-1.3$ have been marked with orange circles.
}
 \label{ionidegvsymas}
 \end{figure}

As can be clearly seen in Fig. \ref{abundancefinal}, the $y^+$ map presents some structure that follows the excitation, even if the standard deviation for $y^+$ is small ($\sim$4\% of the mean value). This is better seen in Fig. \ref{ionidegvsymas} which is an updated version of Fig. 14 in \citetalias{mon10} but restricted to the area studied here. 
The figure presents $y^+$ for each individual spaxel versus  a tracer of the excitation, in this case \ohb. 
For each bin of 0.05~dex, we overplotted the mean and standard deviation of $y^+$ values with blue diamonds and error bars and fitted all our data points with a first-degree polynomial.
Considering that if we find $<y^+>$  in the bin of highest \ohb\ to be larger than $<y^+>+\sigma(y^+)$ in the bin of lowest \ohb\ indicates the presence of a gradient, then this plot is consistent with a positive gradient in $y^+$.
Moreover, spaxels with an extra amount of nitrogen have, on average higher $y^+$.
Is this increase due \emph{only} to variations in the ionization structure or are we witnessing an enrichment in the helium abundance?  
To answer this question, one needs to estimate and correct for the  unseen neutral helium, removing in this way, the dependence on the excitation.
This is not straight forward  as examplified by previous studies \citep[e.g.][]{vie00,sau02,gru02}.

  \begin{figure}[ht!]
   \centering
\includegraphics[angle=0,width=0.24\textwidth,clip=,bb = 45 30 405 390]{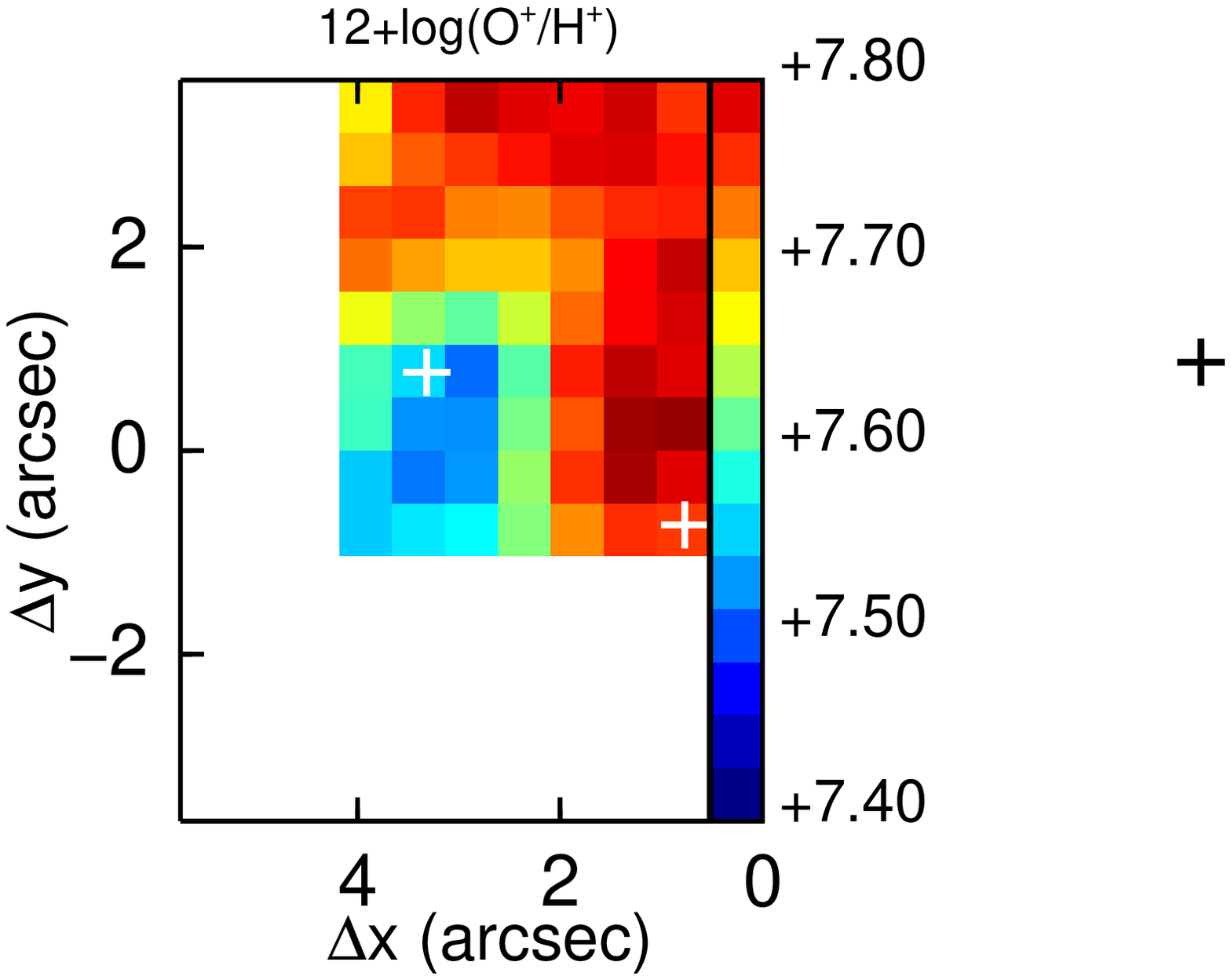}
\includegraphics[angle=0,width=0.24\textwidth,clip=,bb = 45 30 405 390]{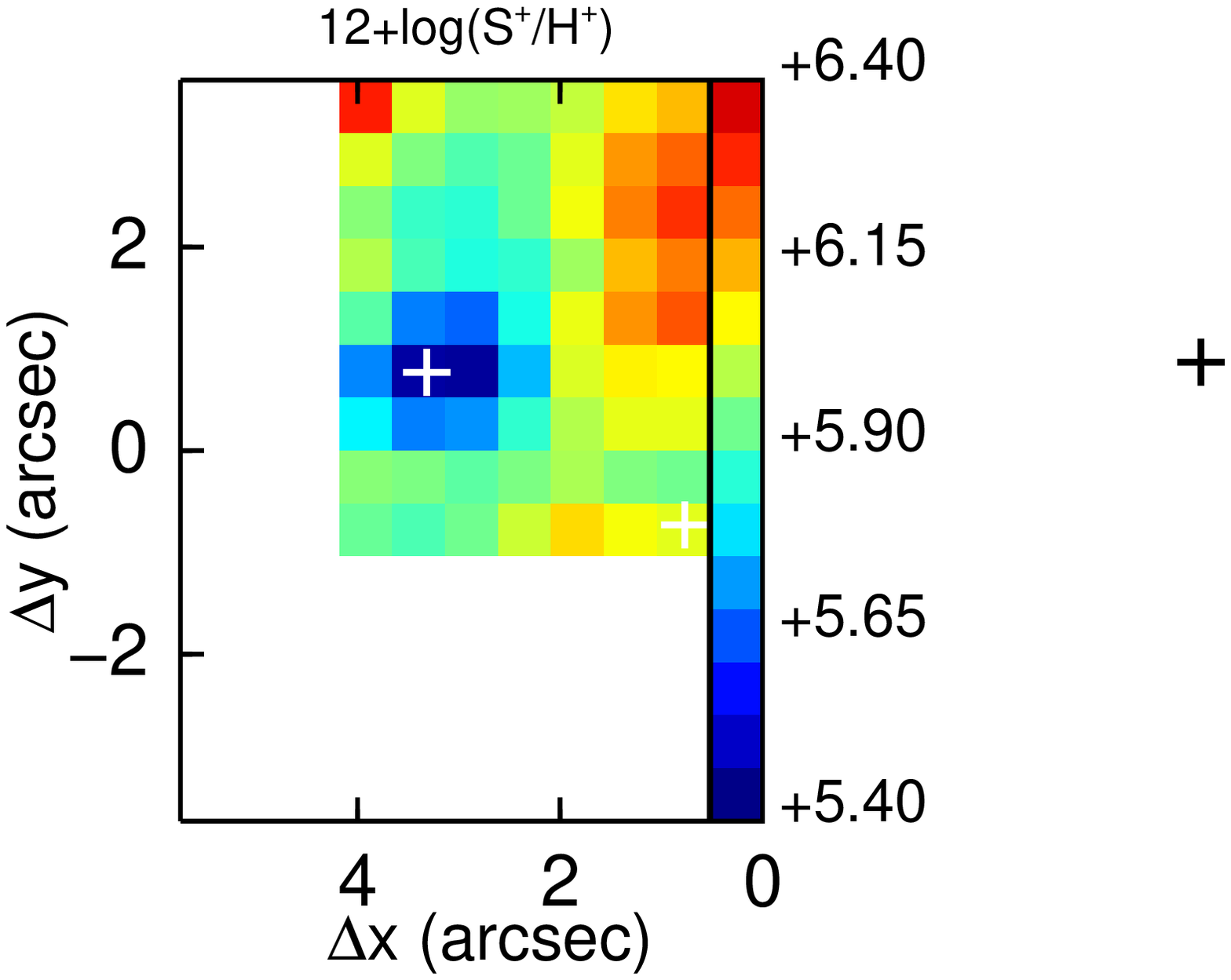}
   \caption[Ionic abundance maps]{Maps for the ionic abundances utilized in the derivation of the ionization correction factors. \emph{Left:} $O^+/H^+$. \emph{Right:} $S^+/H^+$ }
 \label{ionicabundancemap}
 \end{figure}

  \begin{figure}[th!]
   \centering
\includegraphics[angle=0,width=0.24\textwidth,clip=,bb = 45 30 405 390]{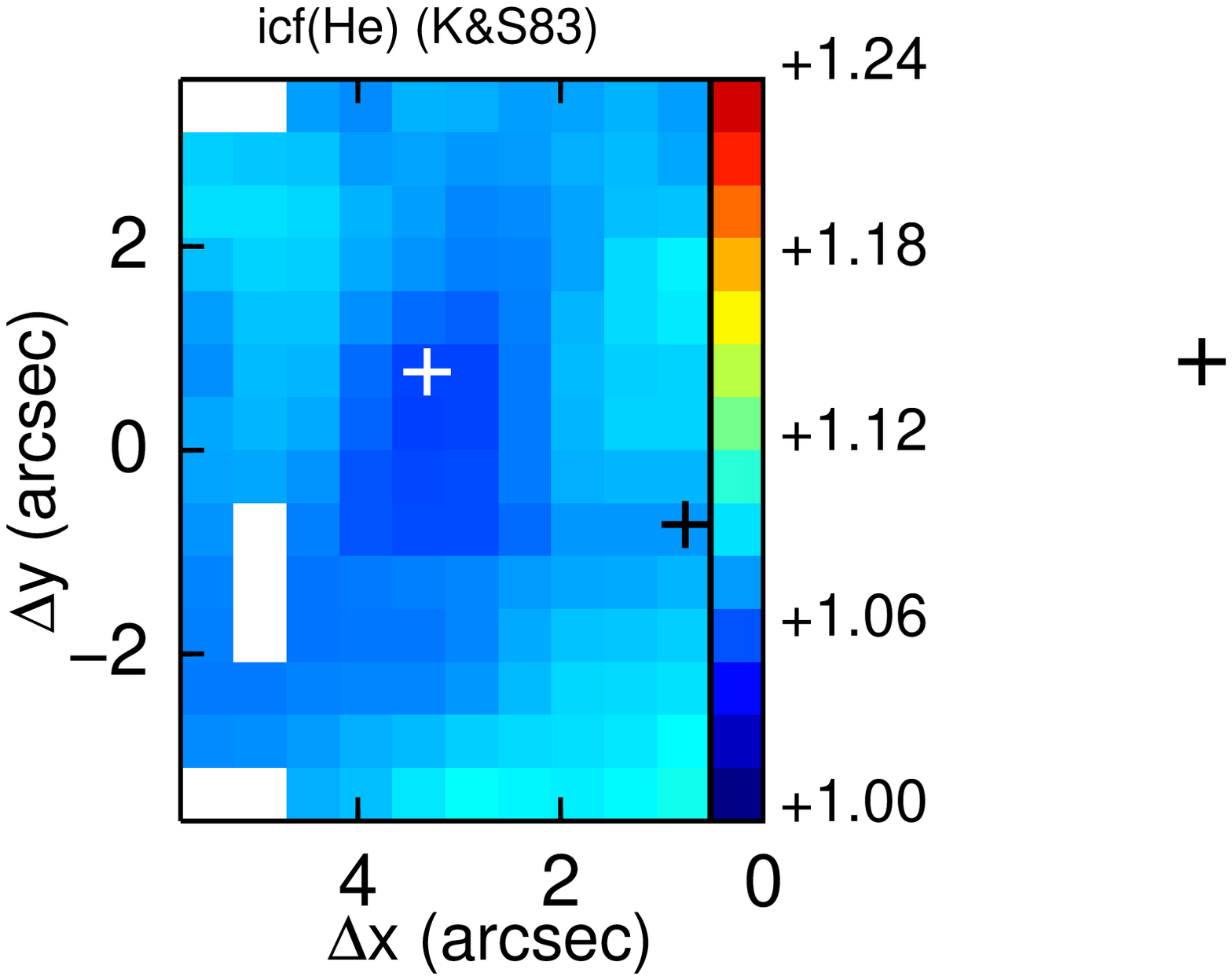}
\includegraphics[angle=0,width=0.24\textwidth,clip=,bb = 45 30 405 390]{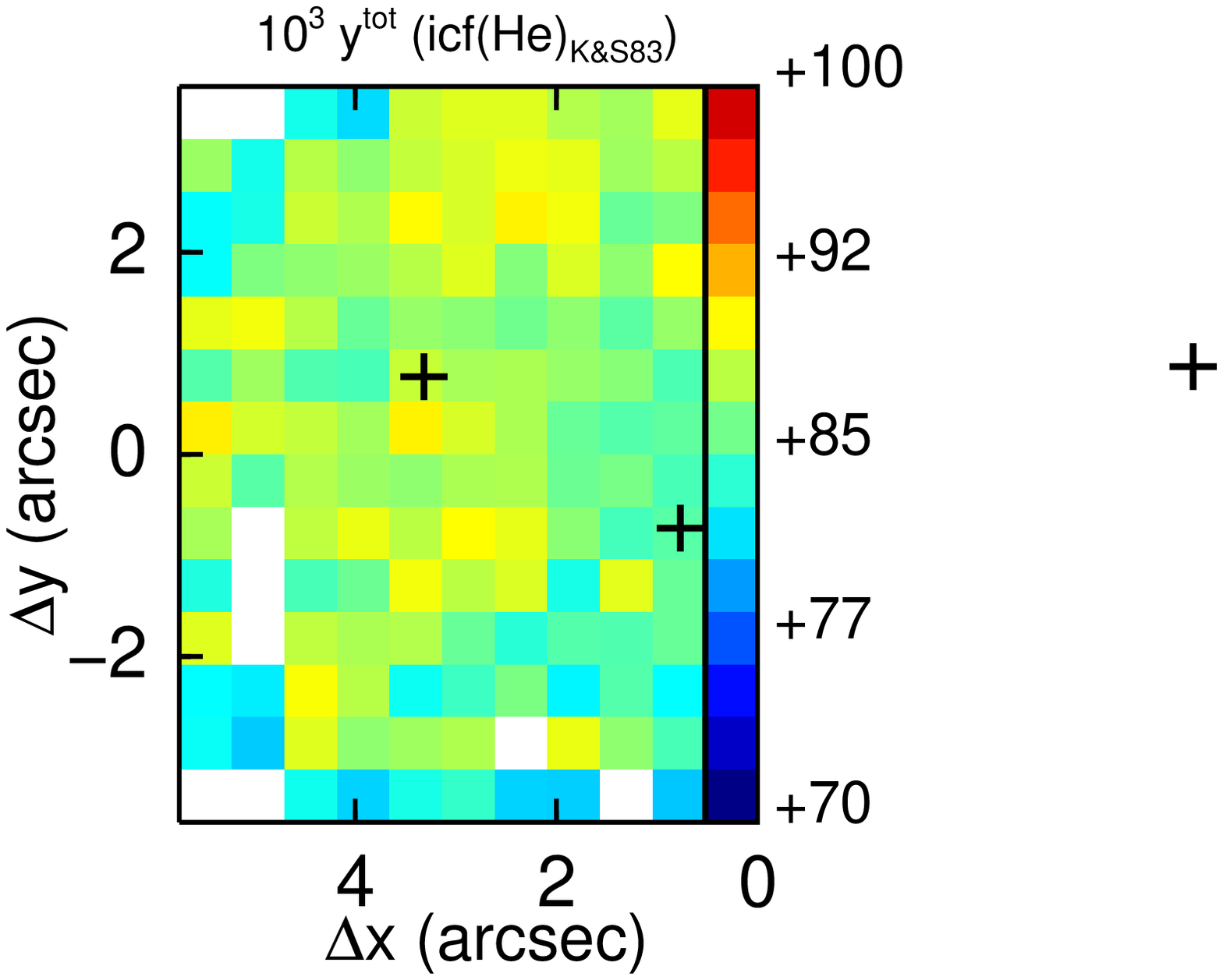}\\
\includegraphics[angle=0,width=0.24\textwidth,clip=,bb = 45 30 405 390]{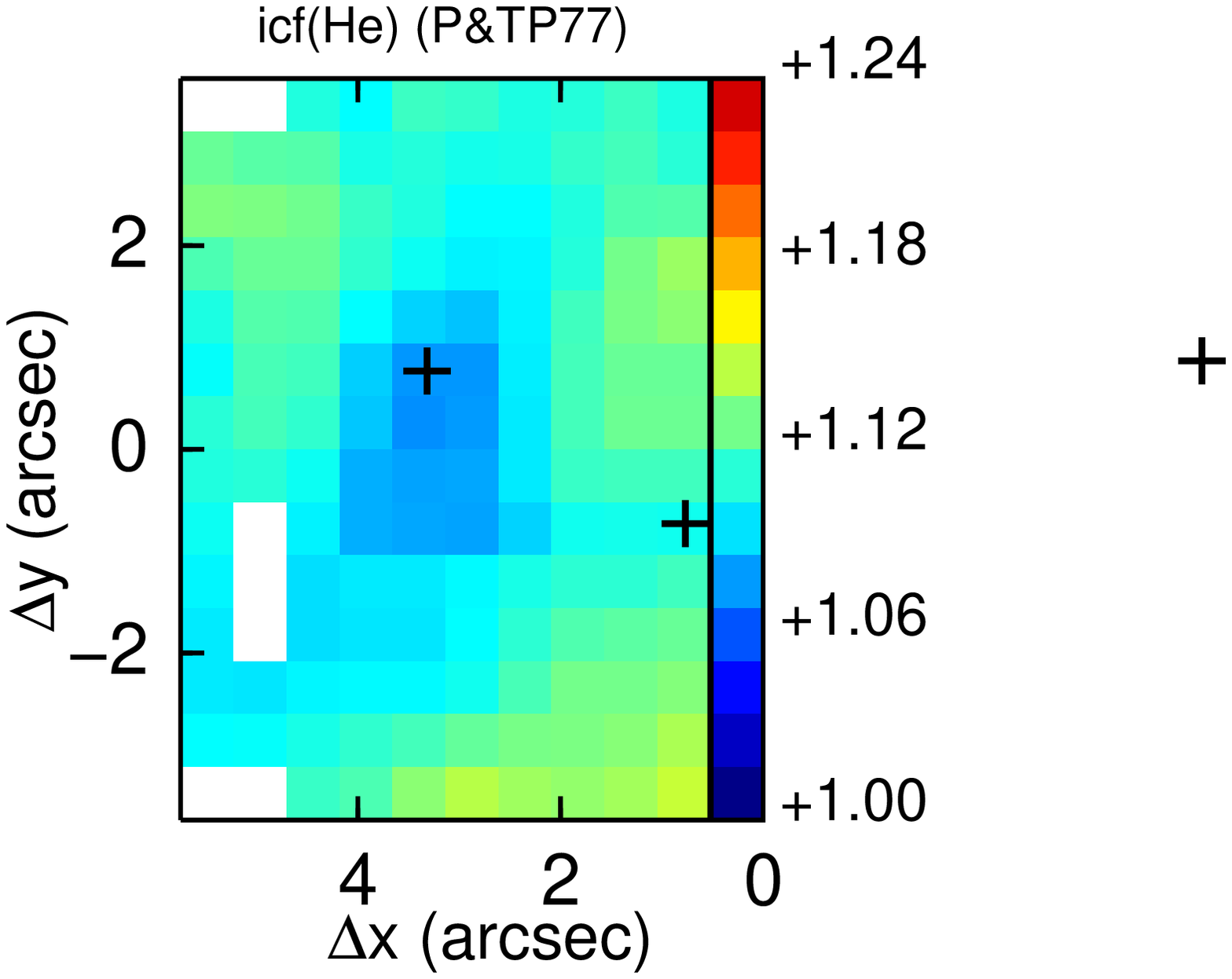}
\includegraphics[angle=0,width=0.24\textwidth,clip=,bb = 45 30 405 390]{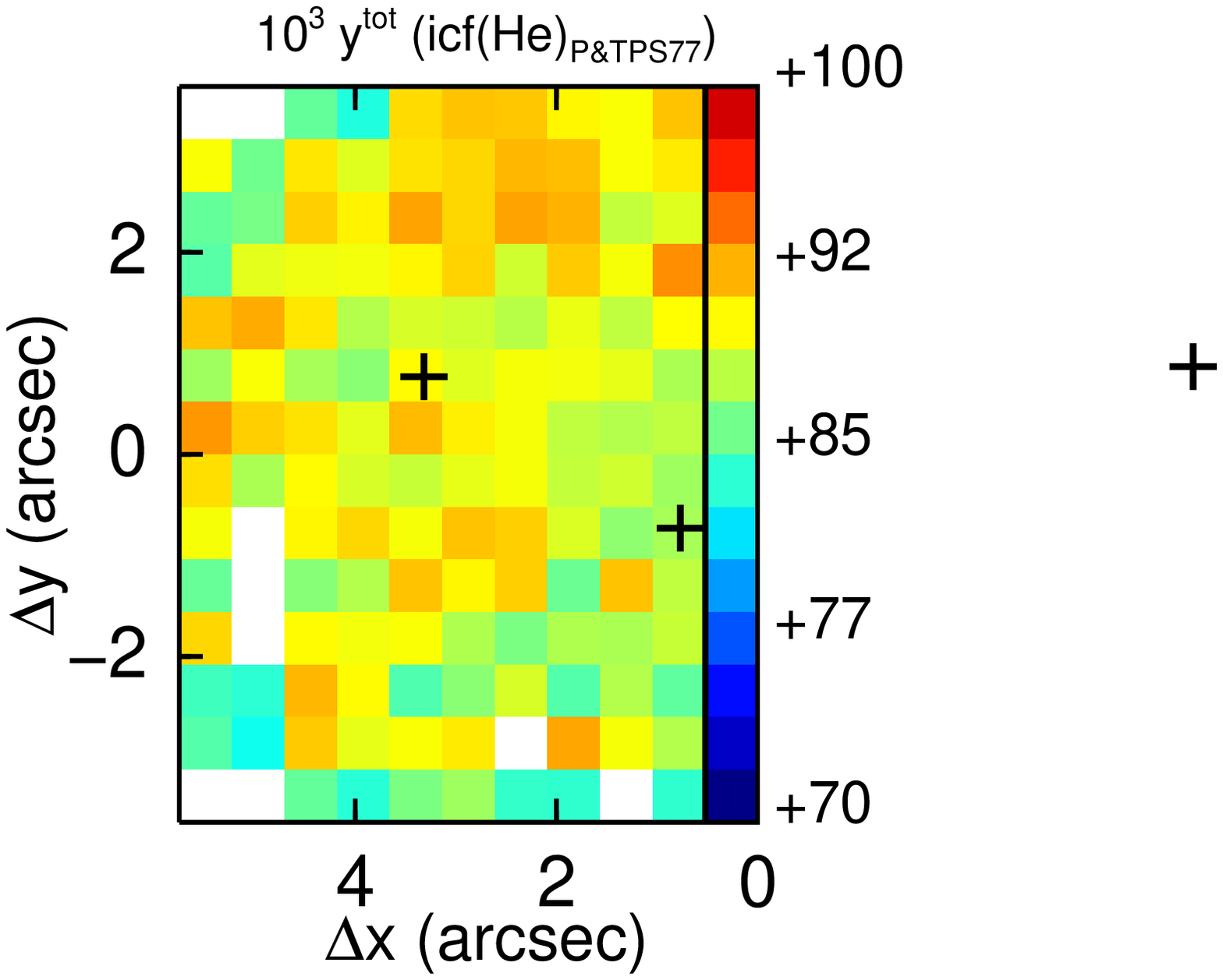}\\
\includegraphics[angle=0,width=0.24\textwidth,clip=,bb = 45 30 405 390]{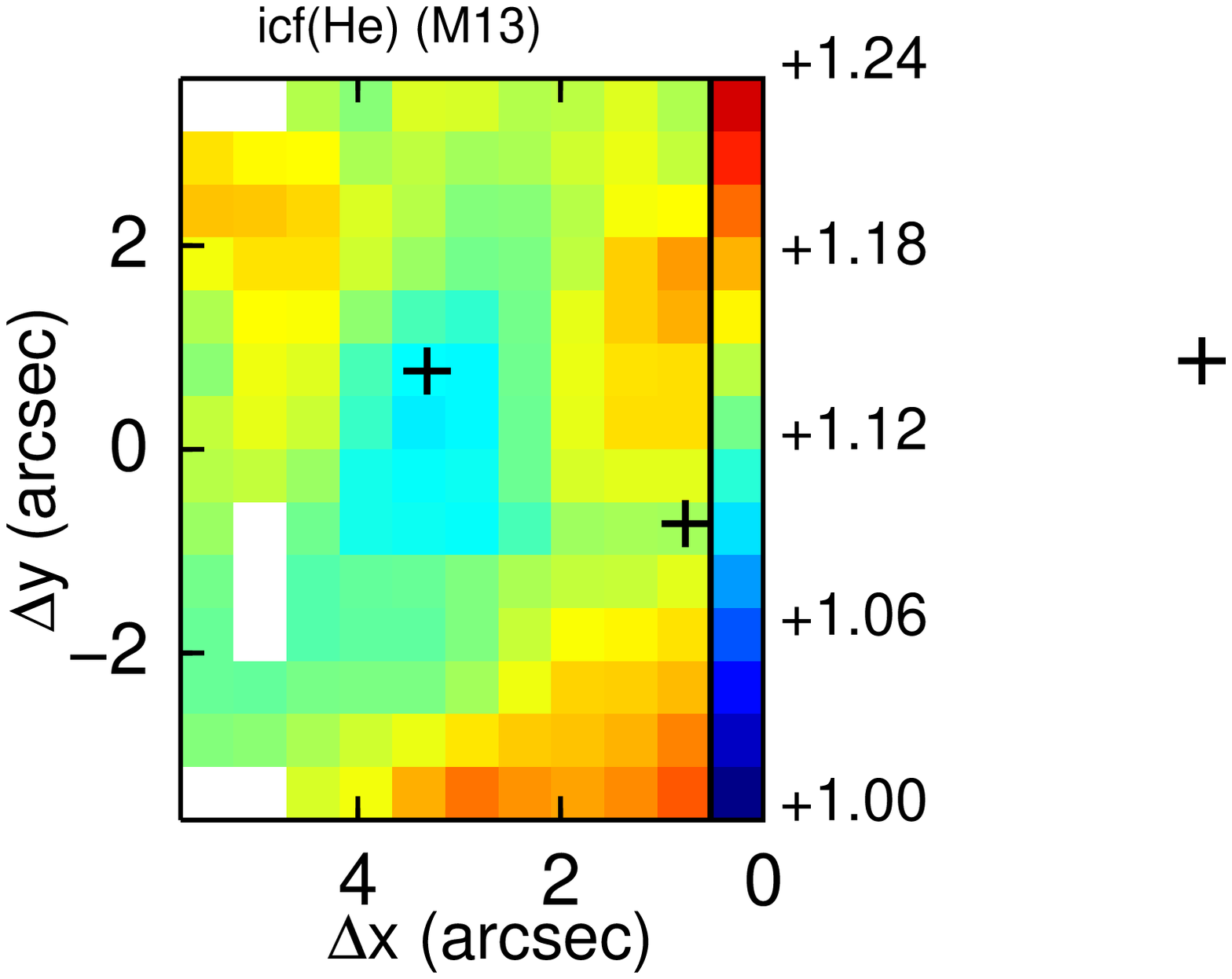}
\includegraphics[angle=0,width=0.24\textwidth,clip=,bb = 45 30 405 390]{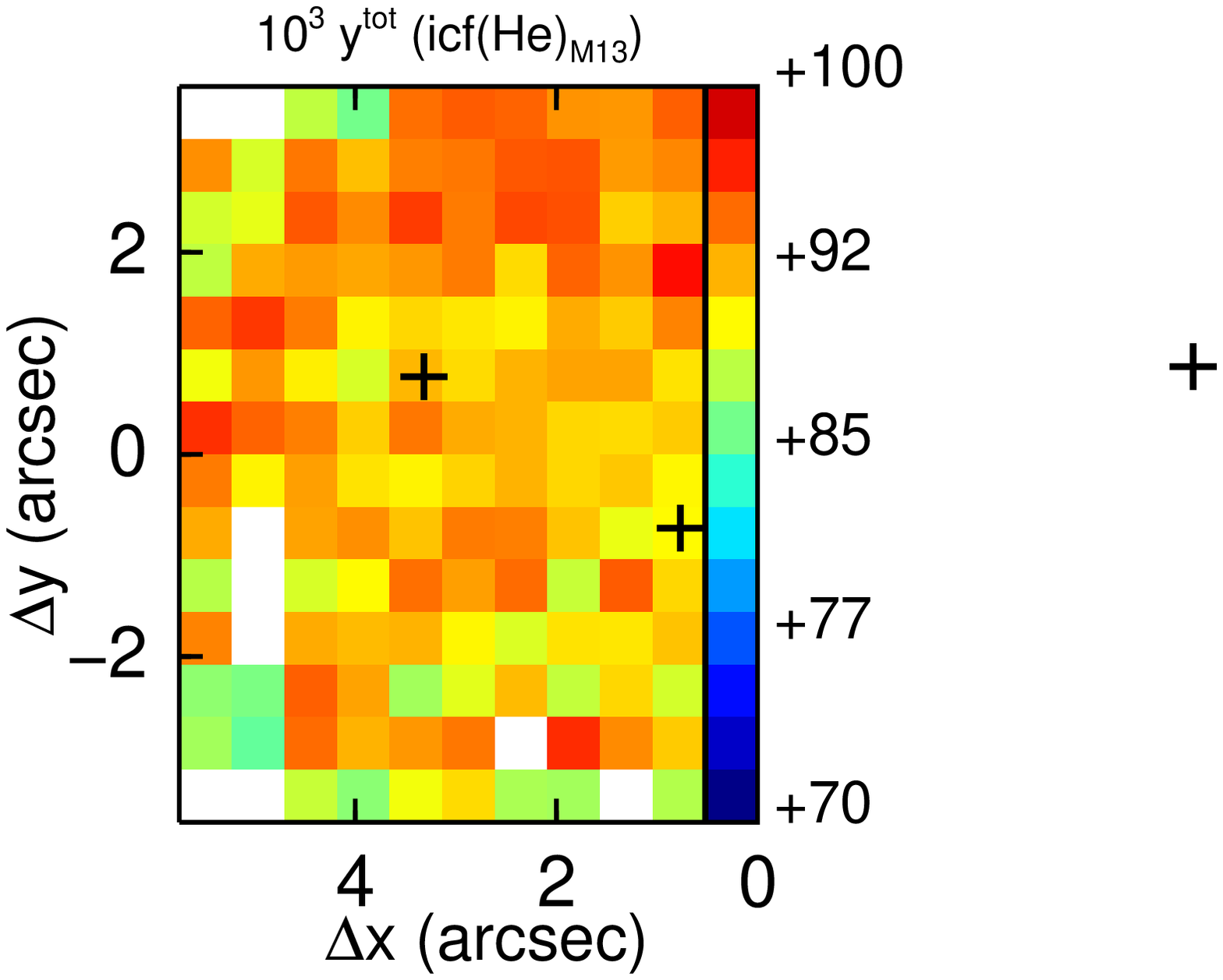}\\
   \caption[icf(He)]{\emph{Left:} Maps for the icf(He) as estimated from the expressions proposed by \citet[][\emph{top}]{kun83},  \citet[][\emph{center}]{pei77} and assuming no gradient in helium abundance \emph{bottom}. \emph{Right:} Corresponding helium abundance maps. }
 \label{icfhe}
 \end{figure}

  \begin{figure}[th!]
   \centering
\includegraphics[angle=0,width=0.39\textwidth,clip=]{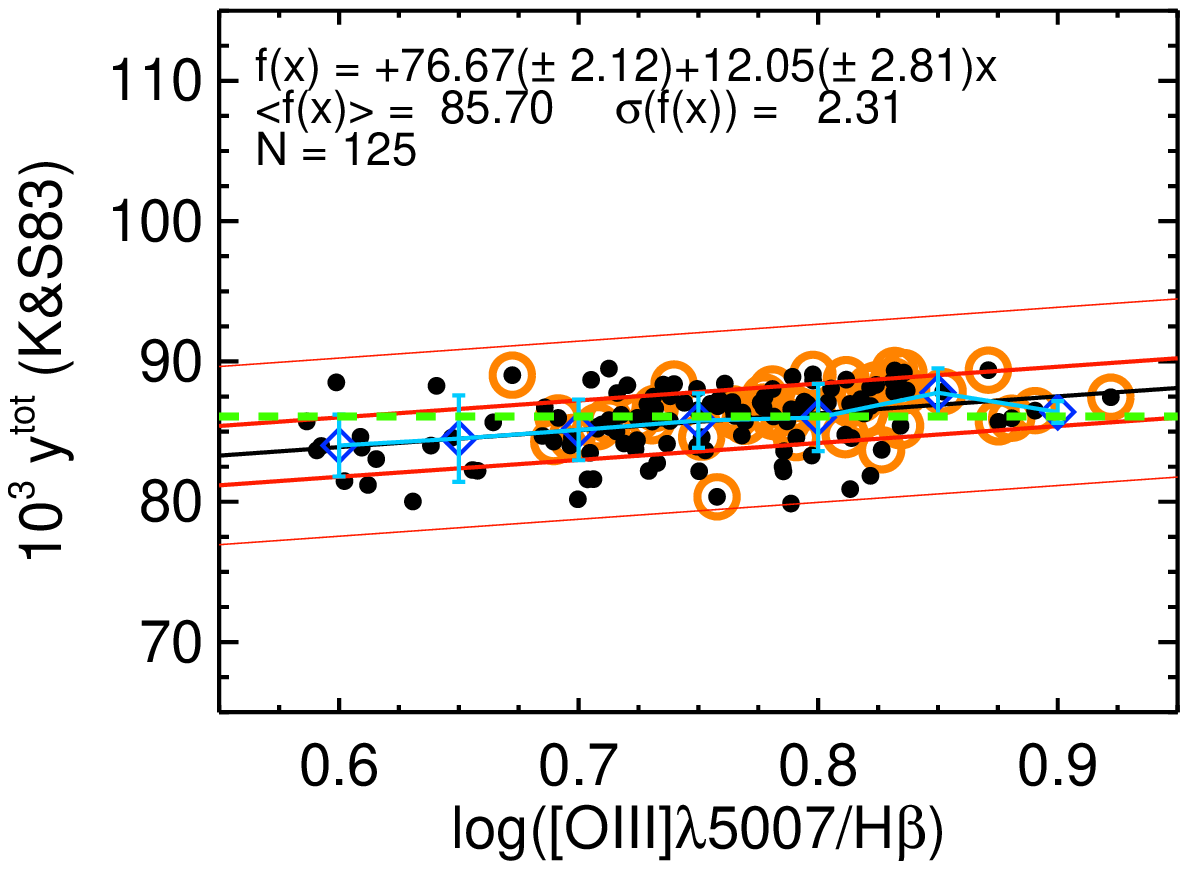}\\
\includegraphics[angle=0,width=0.39\textwidth,clip=]{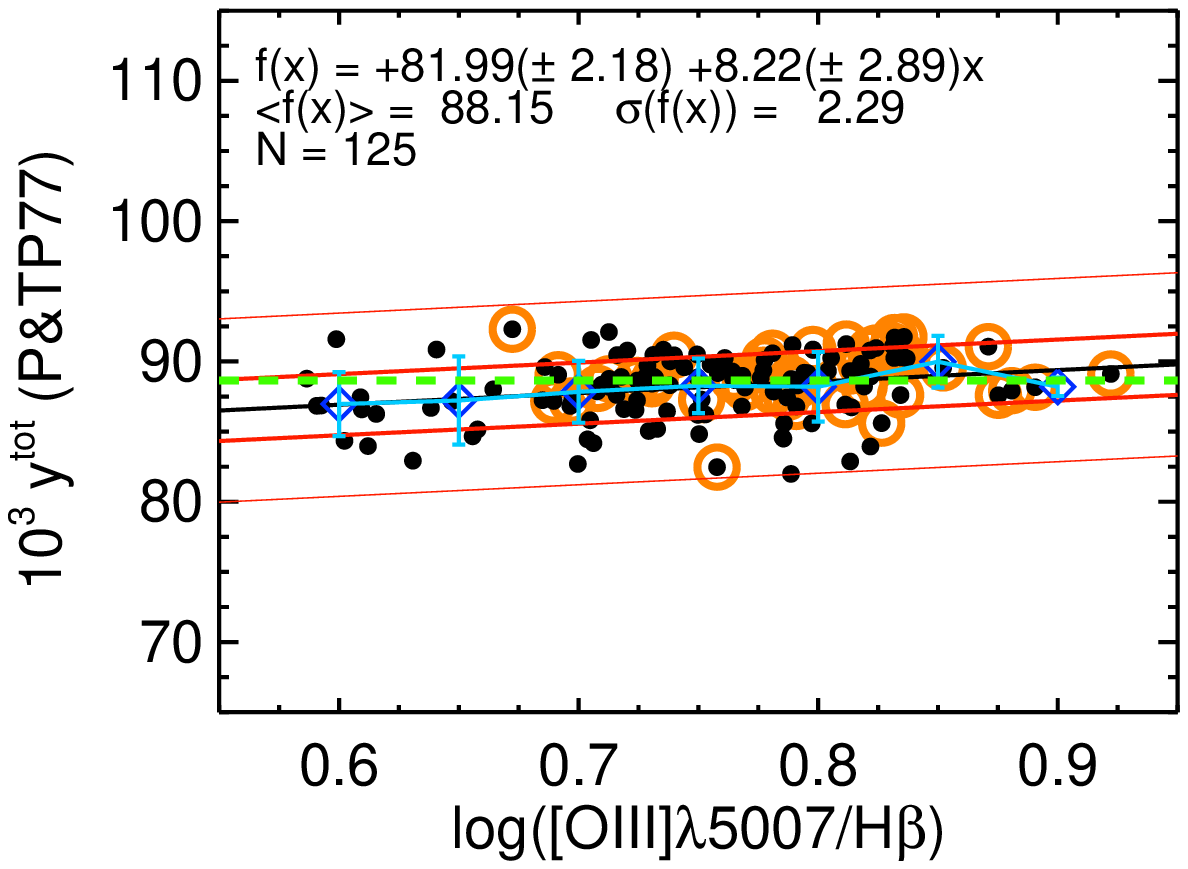}\\
\includegraphics[angle=0,width=0.39\textwidth,clip=]{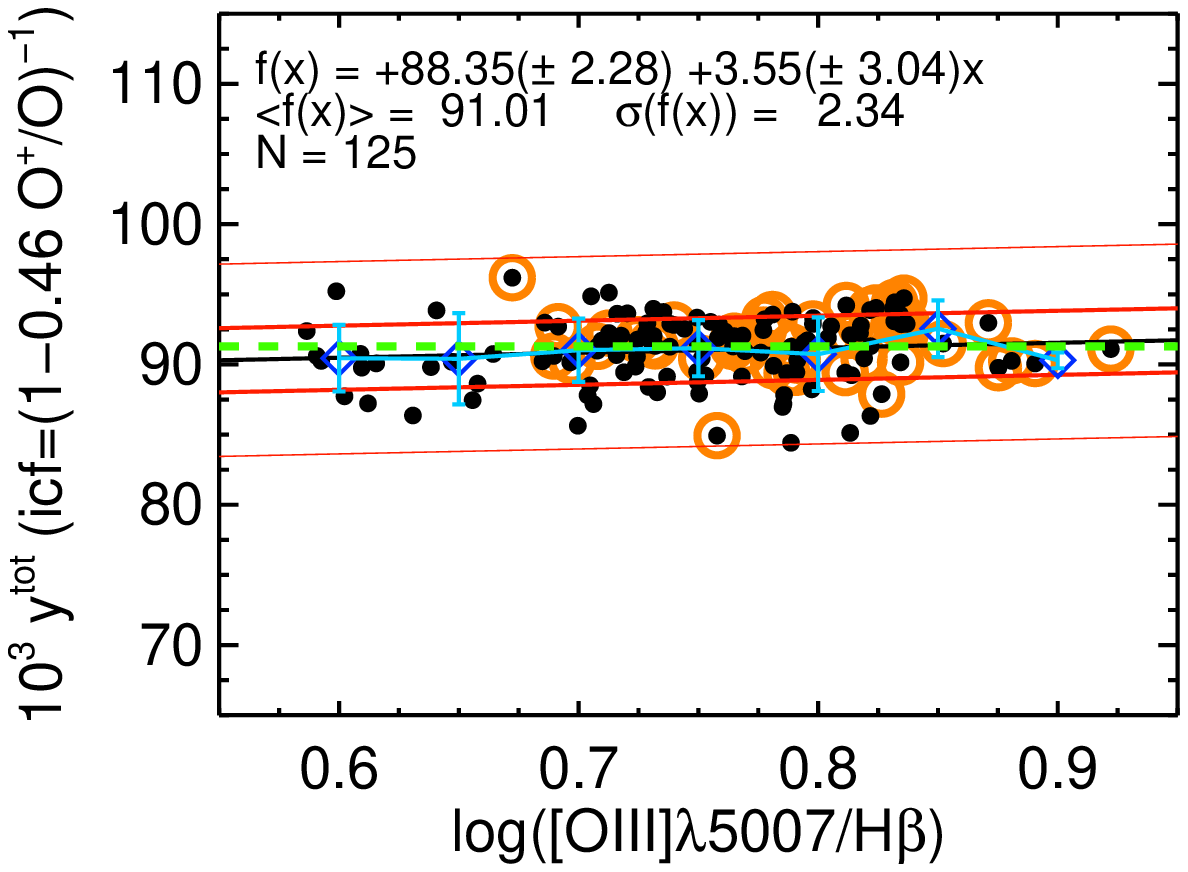}\\
\caption{
Helium abundance as derived using the icf(He) proposed by  \citet[][\emph{top}]{kun83},  \citet[][\emph{center}]{pei77} and assuming no gradient in helium abundance (\emph{bottom}) vs. \ohb. Symbol and color codes are as in Fig. \ref{ionidegvsymas}.
}
 \label{ionidegvsy}
 \end{figure}

Within the spirit of keeping the analysis simple, here we compare the results using three approaches. The first two make use the icf's proposed by  \citet{kun83} and  \citet{pei77}:

\begin{equation}
\rm icf(He)_{K\&S83} = (1 - 0. 25\,O^+ / O)^{-1}                             \label{eqks83}
\end{equation}
\begin{equation}
\rm icf(He)_{P\&TP77} = (1 - 0. 35\,O^+ / O- 0.65\,S^+ / S)^{-1}  \label{eqptp77}
\end{equation}

For the third one, we assumed \emph{a priori} a functional form similar to that of \citet{kun83} and iteratively determined the required coefficients to obtain a relation consistent with no gradient of helium abundance. This would represent the case of largest reasonable correction for unseen helium:

\begin{equation}
\rm icf(He)_{M13} = (1 - 0. 46\, O^+ / O)^{-1}                                  \label{eqm13}
\end{equation}

These icf's depend on the ionic and total abundances of oxygen and sulfur. A map for the total abundance of oxygen was presented in \citetalias{mon12}. We utilized a constant sulfur abundance determined as the mean of those presented in \citetalias{wes13}. Regarding the ionic abundances of oxygen and sulfur, they were derived as part of the work presented in \citetalias{mon12}. However, maps were not included there and are displayed in Fig. \ref{ionicabundancemap} for completeness. An extra assumption is needed to calculate the icf's in those areas where no measurement for the ionic abundance is available. For that, we utilized the information in the other spaxels to fit  a  first-degree polynomial to the relation between  the \sha\ line ratio and icf(He):

\begin{equation}
\rm icf(He)_{K\&S83} =  (0.354\pm0.012) \mathrm{[S\,\textsc{ii}]/H}\alpha + (1.033\pm0.002)
\end{equation}
\begin{equation}
\rm icf(He)_{P\&TP77} =  (0.447\pm0.026) \mathrm{[S\,\textsc{ii}]/H}\alpha + (1.054\pm0.003)
\end{equation}
\begin{equation}
\rm icf(He)_{M13} =  (0.763\pm0.025) \mathrm{[S\,\textsc{ii}]/H}\alpha + (1.057\pm0.004) 
\end{equation}

The  maps for the three icf(He) estimations are presented in the left column of Fig. \ref{icfhe}. In all three cases, the icf is smallest at the peak of emission for the ionized gas and increases outwards, following the ionization structure. 

The icf(He) values are close to one another but variable within a range $\sim1.04-1.09$,  $\sim1.06-1.12$ and $\sim1.09-1.20$ when using Eqs. \ref{eqks83}, \ref{eqptp77}, and  \ref{eqm13}, respectively.
The corresponding maps with the total helium abundance are presented in the right column of this figure, while the dependence on the excitation is presented in Fig. \ref{ionidegvsymas}.
Note that, when using the \citet{kun83} and \citet{pei77} icf's, the slope of the relation between $y$ and \ohb\ is slightly positive within the errors of the fit. However, according to the criterion presented for the $y^+$ vs. \ohb\ relation, the derived helium abundance for these icf's would be effectively consistent with no positive gradient.
The mean total helium abundance ranges between 0.086 and 0.091, depending on the assumed icf.  This range is larger by a factor $\sim$2 than  the uncertainties due to the measurements  (assumed to be traced by  the standard deviation).  This implies that the main source of uncertainty is still in the assumptions taken on the way to the derivation of the final abundance and highlight how,
even with a level of data quality as high as those utilized here (i.e. data allowing to determine locally the physical conditions of the gas, the fluxes of several helium lines and estimations of the contribution of underlying stellar populations), achieving an uncertainty $\lsim1$\% is extremely difficult.

The main conclusion of Fig. \ref{ionidegvsy} is that the relation between the total helium abundance, $y_{tot}$, and the excitation of the gas, as traced by \ohb\  is consistent with a lack of gradient in  helium abundance.
However, if the extra helium were produced by the Wolf-Rayet stars, the required amount to be detected would be tiny in comparison to the pre-existing helium. This implies that  a positive slope in Fig. \ref{ionidegvsy} consistent with this enrichment would be indistinguishable of the presented fits.
Exploiting these data to their limits, we can derive the average excess in $N/He$, by comparing the mean abundances in the N-enriched and non N-enriched areas.
A map for the $N/H$ abundance can be derived using the $O/H$ and $N/O$ maps presented in \citetalias{mon12}. Assuming a value of $\log(N/O=-1.3)$ as the limit above which the interstellar medium is enriched in nitrogen, we find a mean $N/H$ of $8.4\times10^{-6}$ and $17.4\times10^{-6}$ in the spaxels without and with enrichment. This implies a flux weighted  excess in nitrogen of $N/H_{exc}=8.9\times10^{-6}$.
Proceeding in the same manner with the different estimations of the total helium abundance, we find mean $He/H$ values ranging between $8.49\times10^{-2}$ and $9.05\times10^{-2}$ for the non-enriched spaxels and between $8.68\times10^{-2}$ and $9.14\times10^{-2}$, depending on the assumed icf(He). Differences in helium abundances between the enriched and non-enriched zones are $\sim0.9-1.9\times10^{-3}$. The lowest value was derived using our largest icf(He) (i.e. Eq. \ref{eqm13}) which, by construction, was defined to minimize any helium abundance gradient. The largest value compares well with the stellar atmospheric $N/He$ ratios for an N-type Wolf-Rayet star \citep[e.g.][]{smi82}. However, it is also at the limit of the uncertainties ($\sim2\times10^{-3}$assumed to be traced by the standard deviation for the spaxels under consideration in each separate group). 
Therefore, these data appear to be marginally in accord with the hypothesis of a putative enrichment of helium due to the Wolf-Rayet population in the main \hii\ region, but more importantly, they stress the difficulties in pushing this methodology further in order to  confirm this contamination without doubt.

\section{Discussion}

\subsection{Kinematics and radiative transfer effects}

The  standard approach in the literature to estimate radiative transfer effects assumes
a negligible influence of the movements of the gas in the \hii\ region. i.e. a static nebula \citep[$\omega=v/v_{th}=0$, in the formalism presented by ][]{rob68}. This is the approach utilized in Sec. \ref{secymas}. 
The general use of this assumption is motivated by the difficulty of obtaining data of sufficient  quality to allow, on top of measuring the flux of several helium lines with good signal to noise, tracing their profiles and clearly identify the different kinematic components in a consistent manner in all of them.
However, it is not unusual for starburst galaxies or \hii\ regions to present velocity gradients and/or relatively high velocity dispersion that can be attributed to outflows (or expanding structures).
In particular, the kinematic study presented in \citetalias{mon10} and  \citetalias{wes13} showed that movements in this \hii\ region are significant and can indeed be attributed to an outflow caused by the two embedded Super Star Clusters. Specifically, in  \citetalias{mon10}, we detected a relatively static (i.e. small velocity gradient) narrow component on top of a much broader component ($\sigma\sim20-25$ km s$^{-1}$) with a velocity gradient of $\Delta v\sim70$~km~s$^{-1}$. An additional component was also detected with the higher resolution GMOS data  \citepalias{wes13}. Therefore, these data provide a very good opportunity to explore what is the impact  of the kinematics in the derivation of the optical depths from an empirical point of view.  
%

  \begin{figure}[th!]
   \centering
\includegraphics[angle=0,width=0.24\textwidth,clip=,bb = 45 30 405 390]{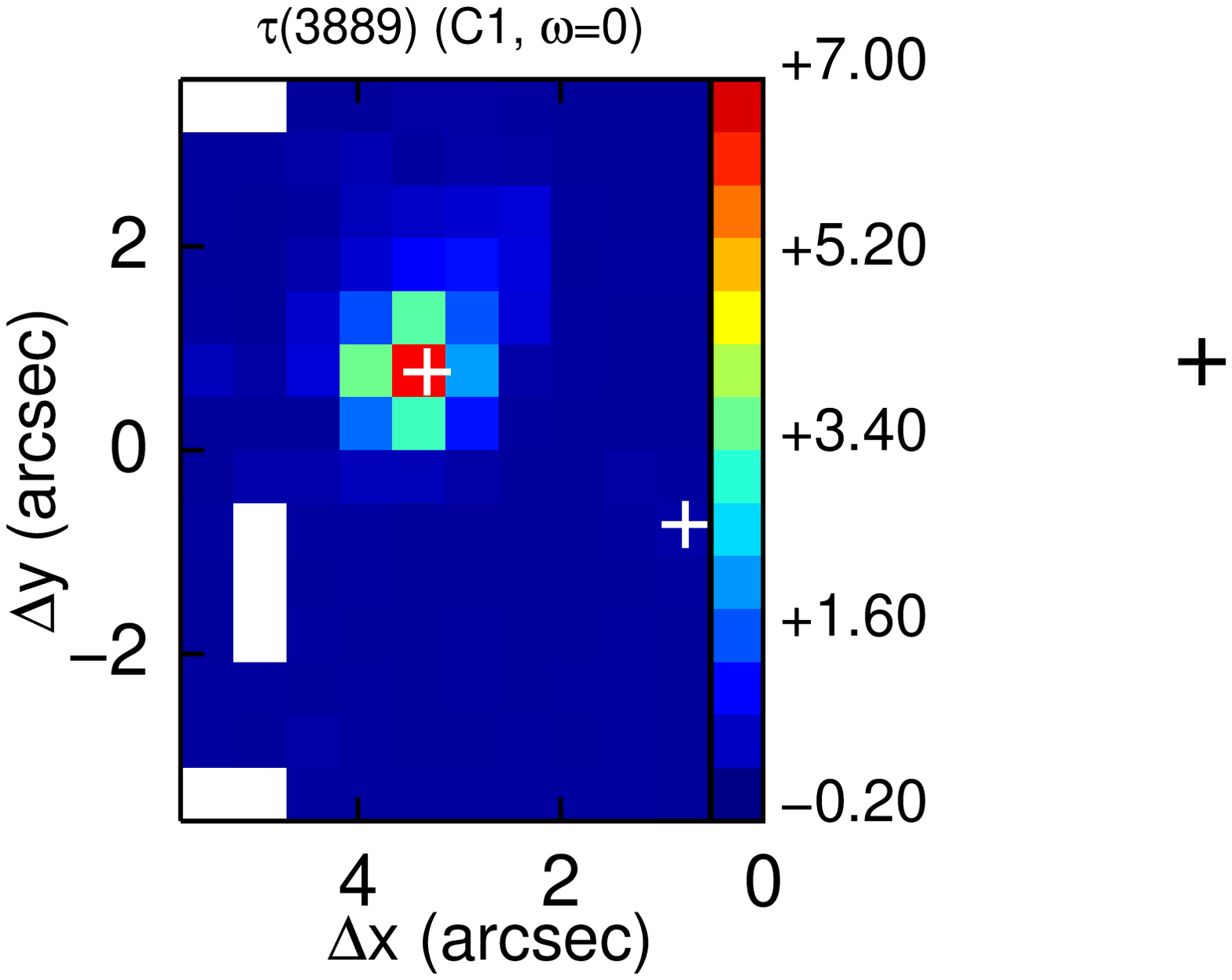}
\includegraphics[angle=0,width=0.24\textwidth,clip=,bb = 45 30 405 390]{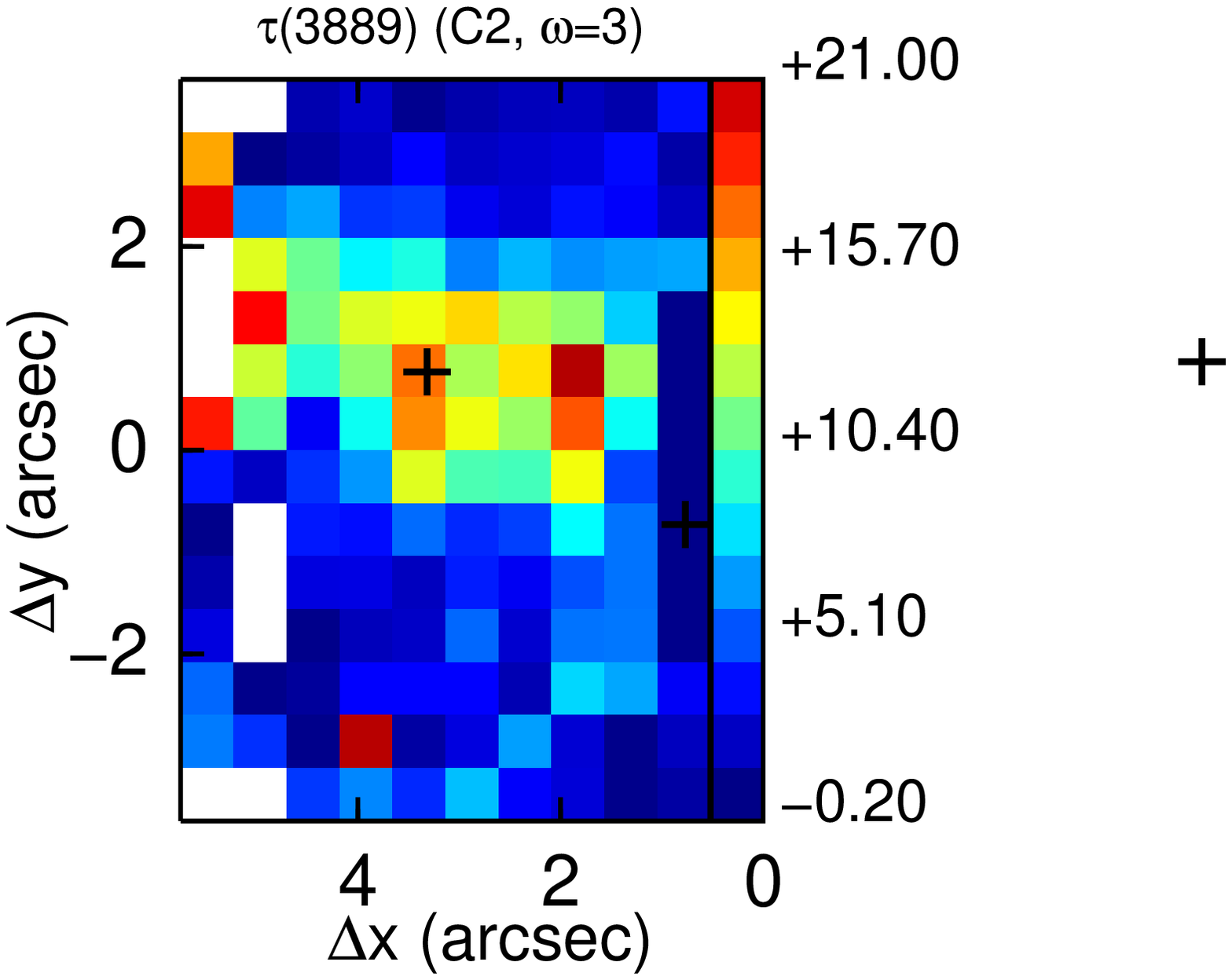}\\
   \caption[$\tau(3889)$ for the two kinematic components]{ $\tau(3889)$ for the two main kinematic components presented in Fig. 19 of \citetalias{mon10} derived from the fits to relations presented by \citet{rob68} for $T_e$=10\,000~K. \emph{Left:} Narrow kinematic component and $\omega=0$.  \emph{Right:} Broad kinematic component and $\omega=3$. The $\tau$(3889) map for this component using the relation for $\omega=0$ (not shown) displays the same structure but with values $\sim4-6$ times smaller.}
 \label{taucompos}
 \end{figure}

Since this is only an exploratory analysis and a consistent approach based on the identification and modeling of multiple kinematic components in \emph{several} emission lines would be more complex (and
not supported by the signal-to-noise of all the \hei\ data) than the one presented in Sec. \ref{secymas}, we opt for a simpler analysis based on the  \hei$\lambda$6678 and \hei$\lambda$7065 lines. These were the pair of lines utilized to estimate $\tau(3889)$ in the \emph{red set}. Also both were observed with the same instrumental set-up, which was the same as for \ha\ in the multicomponent analysis presented in  \citetalias{mon10}. Therefore, we could link on a spaxel-by-spaxel basis the multicomponent analysis utilized in \citetalias{mon10} to similar components in  \hei$\lambda$6678 and \hei$\lambda$7065.
The contribution to the total flux from each component varies from spaxel to spaxel as well as with the line in consideration. Specifically, the broad component presents a larger contribution  (i.e. $\sim35-60$\%) to \hei$\lambda$7065 than to \hei$\lambda$6678 ($\sim30-50$\%). This implies different correction factors, $f_\tau(7065)$, for the narrow and broad component that can be converted into optical depths as in Sec. \ref{secymas}.
Given the velocity gradient and the velocity dispersion measured for the broad component, values for $\omega=v/v_{th}=3$ are more appropriate for this component. Fitting the same functional form as in Eqs. \ref{eqfac7065}-\ref{eqfac3889} to the values reported in \citet{rob68} for  $\omega=v/v_{th}=3$ and $T_e=10\,000$~K, we derived the following relation:

\begin{equation}
f_\tau(7065)_{\omega=3} = 1 + 0.279\tau^{0.521}
\end{equation}

The structure of the derived optical depth for both the narrow and the broad components is shown in Fig. \ref{taucompos} and several conclusions can be drawn from this figure. Firstly, a comparison between the left panel of this figure with those presented in Fig. \ref{maptau} highlights the similarity between the optical depth maps derived for the narrow component and for the line integrated analysis. To our knowledge, this is the first empirical evidence supporting the traditional assumption that movements in extragalactic \hii\ regions with  a known outflow have a negligible effect in the estimation of the global optical depth. 

Secondly, a comparison between the two maps in Fig. \ref{taucompos}  shows how the \emph{structure} of  the two $\tau(3889)$ is different: while the narrow component $\tau(3889)$ displays a clear peak associated with the peak of emission for the \ghiir\ and decreases outwards, the broad component  remains relatively high over a large area of $\sim4^{\prime\prime}\times2^{\prime\prime}$ ($\sim$74~pc$\times$37~pc) centered on the double cluster.

Thirdly, the comparison between the specific values displayed in these two graphics shows that the broad component suffers from radiative transfer effects to a larger degree than the narrow one. This is a consequence of the adopted relation between $f_\tau(7065)$ and $\tau(3889)$, which is motivated by our kinematic results. If we had used Eq. \ref{eqfac7065} (i.e. assuming $\omega=0$ for the \emph{broad} component), we would have obtained comparable values for the $\tau(3889)$ in both components (although, still, with a different structure since this is determined by the $f_\tau(7065)$ itself, which in turns depends on the relative flux between \hei$\lambda$6678 and \hei$\lambda$7065).

There is certainly room for improvement in this analysis. For example, although the multicomponent analysis treats in a consistent way the line profiles of \hei$\lambda$6678 and \hei$\lambda$7065, the same physical conditions (i.e. $T_e$, $n_e$, extinction) had to be assumed for both components. Moreover, we used here only two helium lines, in contrast with the more canonical methodology for determining helium abundances based in a larger set of helium lines. These improvements suggest a new approach to the study of the helium content in galaxies with IFS. Specifically,  the 2D analysis presented in the previous sections together with the separate analysis of different kinematic components along the line of sight (i.e. in a given spaxel) constitute a 3D view of the radiative transfer effects in the nebula.

\subsection{Relation between collisional and radiative transfer effects and other properties of the ionized gas} 

We have evaluated  locally both collisional and radiative transfer effects in the helium lines, as well as several important physical conditions ($n_e$, $T_e$, excitation) and chemical properties (i.e. relative abundance $N/O$) of the ionized gas. Therefore, the central part of \object{NGC~5253} constitutes a good case study to explore whether there is a (strong) spatial relation of collisional and/or radiative transfer effects with other gas properties.

\begin{figure}[th!]
   \centering
\includegraphics[angle=0,width=0.24\textwidth,clip=]{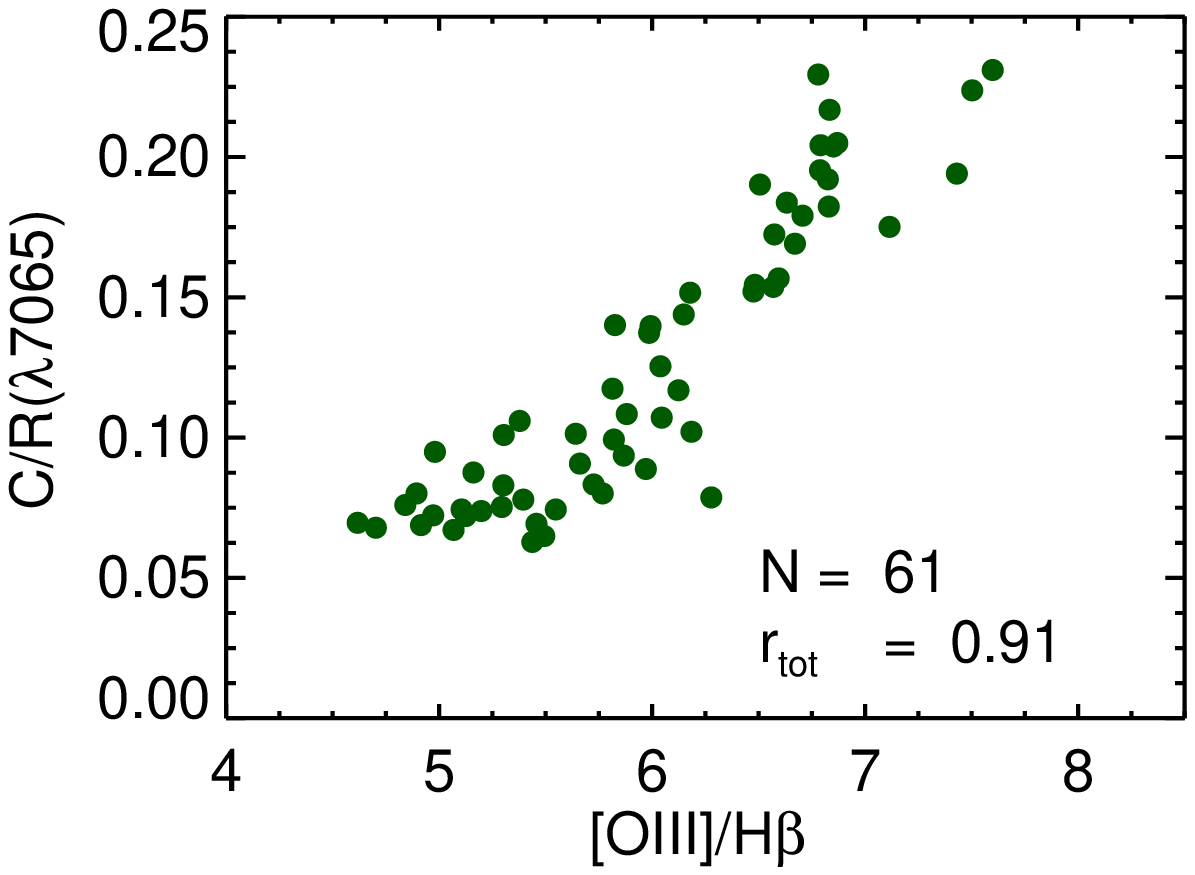}
\includegraphics[angle=0,width=0.24\textwidth,clip=]{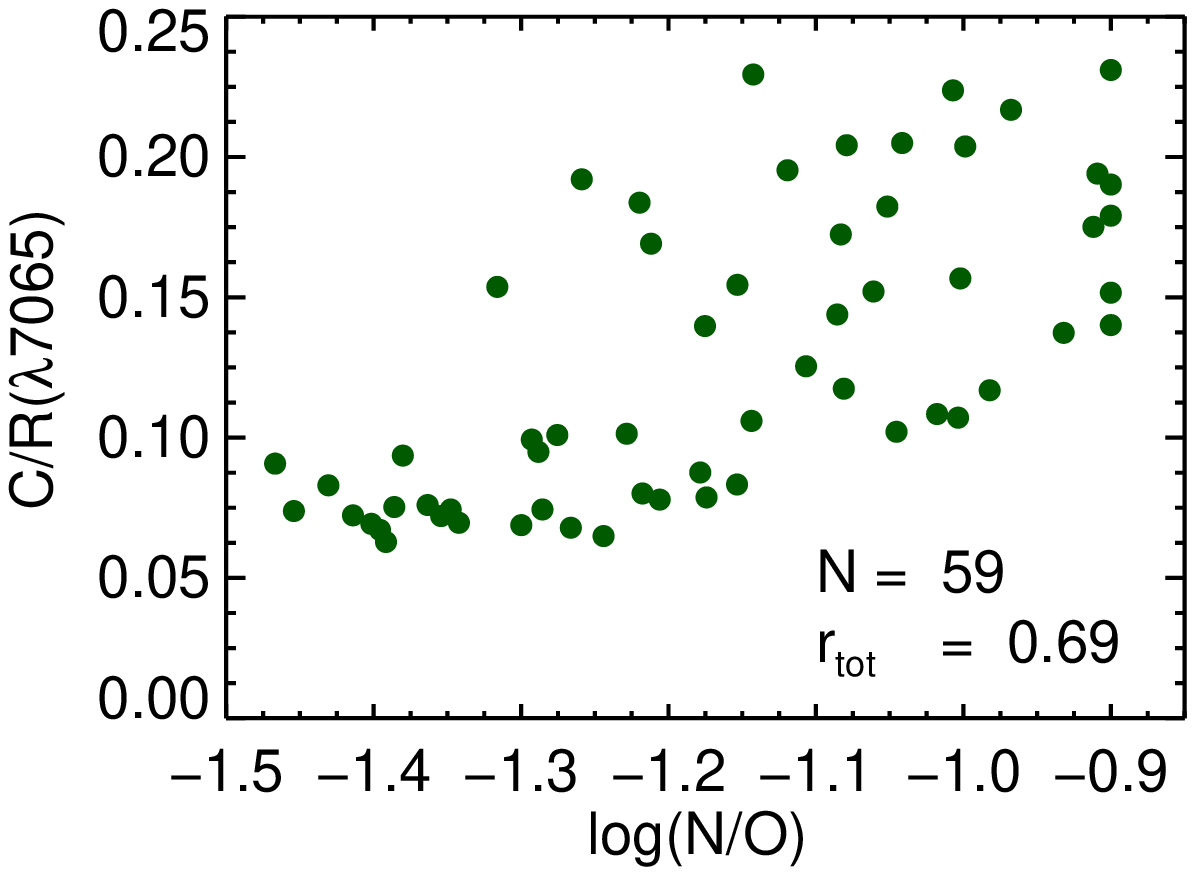}\\
   \caption[C/R ratios]{Relation between collisional effects as traced by the C/R ratio for the $\lambda$7065 line and the excitation (\emph{left}) and relative abundance of nitrogen, N/O (\emph{right}). The Pearson's correlation coefficients are indicated in the corners of the individual plots.
 \label{c2rvscosas}}
 \end{figure}

As discussed in Sec. \ref{seccoli}, theory predicts a strong dependence of the collisional effects on the electron density and, to a lesser extent, on the electron temperature. This implies a strong spatial correlation with the these properties. Indeed, when all the spaxels in the area under study are considered, the relations between $C/R(\lambda7065)$ vs.  $n_e$ and $T_e$ have Pearson's correlation coefficients of 0.99 and 0.71, respectively. Regarding the other two properties,  Fig. \ref{c2rvscosas} presents the relation between the collisional effects (as traced by $C/R$(7065)) and the excitation (as traced by \ohb) and the relative abundance of nitrogen: $N/O$.  The excitation presents a degree of correlation similar to  $n_e$. The correlation for $N/O$ is not as strong as that for $n_e$ but still comparable to that for $T_e$.

  \begin{figure}[th!]
   \centering
\includegraphics[angle=0,width=0.24\textwidth,clip=]{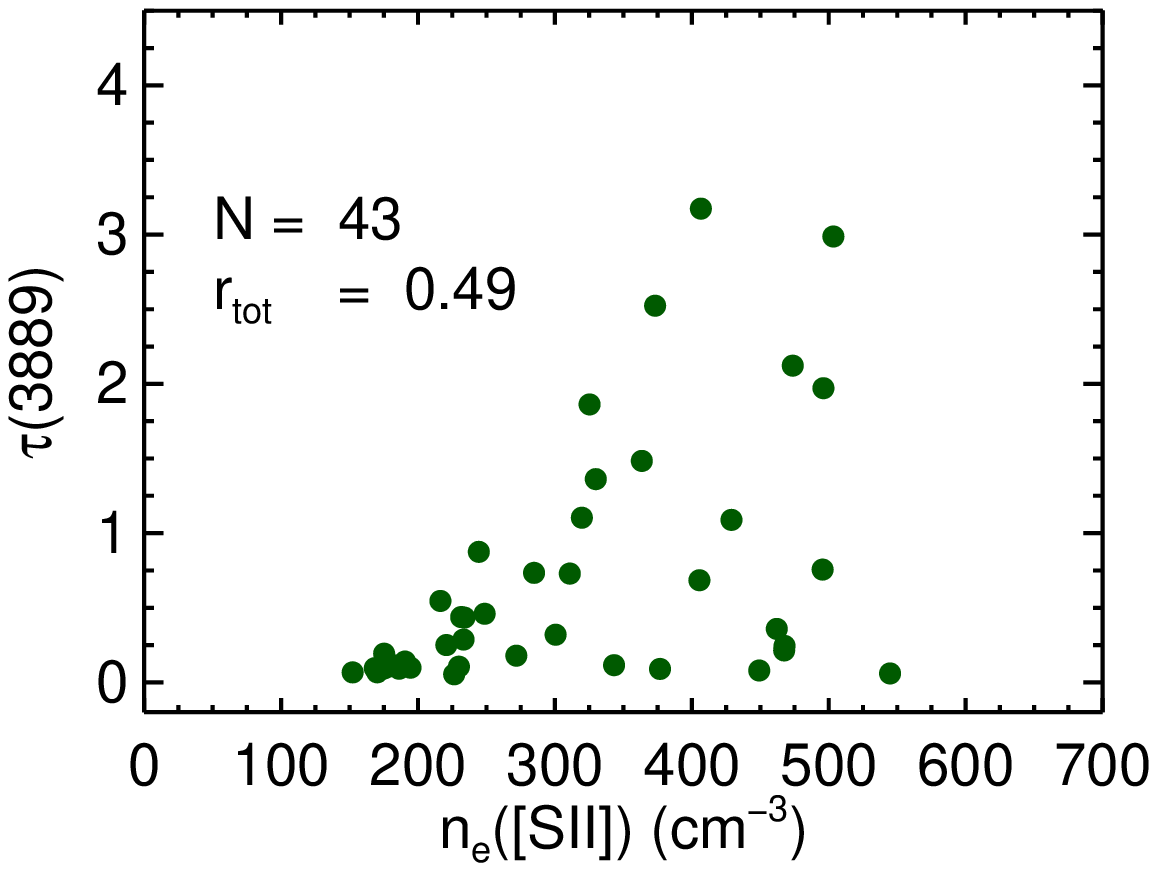}
\includegraphics[angle=0,width=0.24\textwidth,clip=]{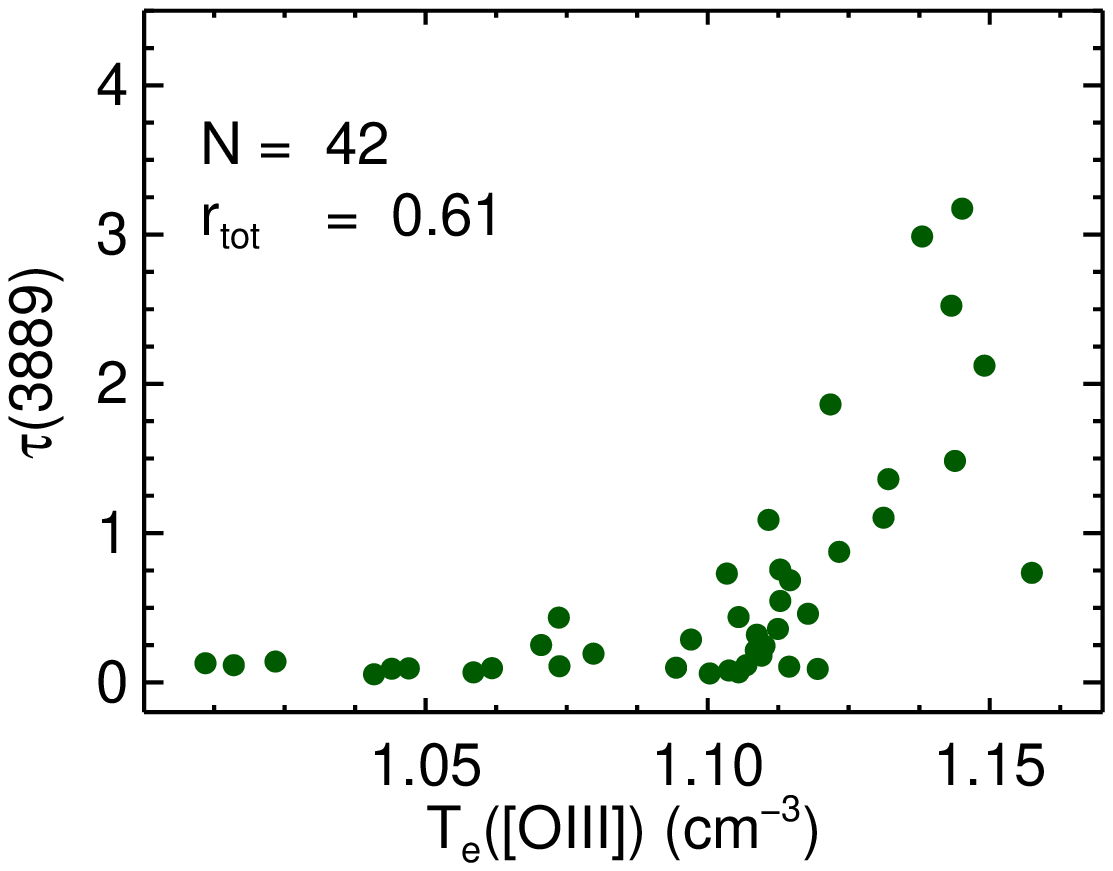}\\
\includegraphics[angle=0,width=0.24\textwidth,clip=]{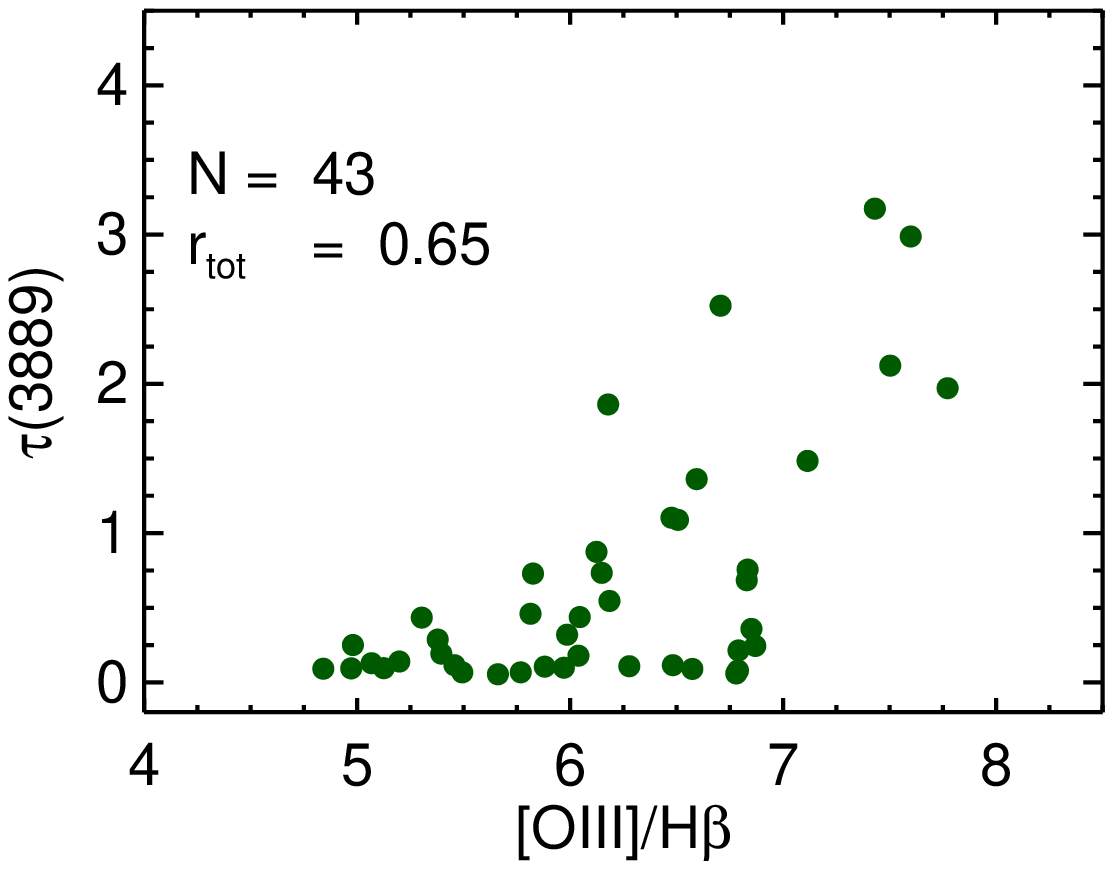}
\includegraphics[angle=0,width=0.24\textwidth,clip=]{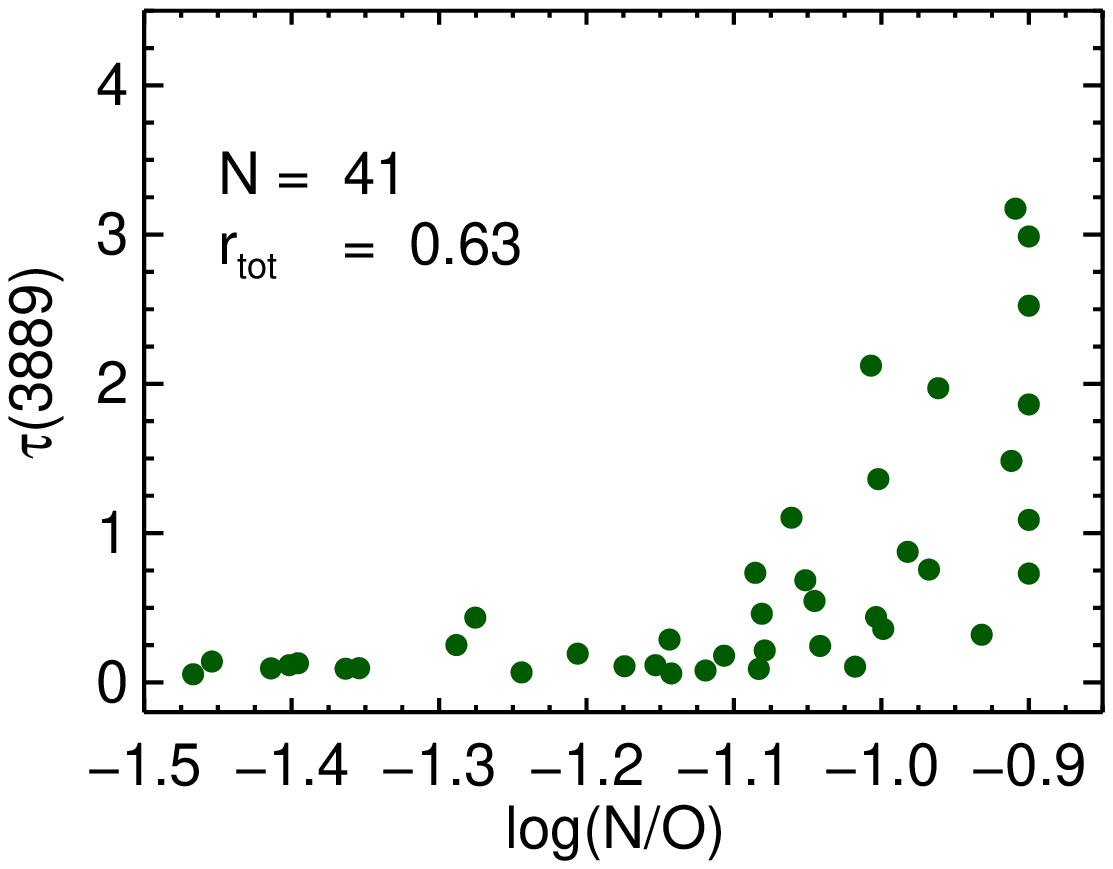}\\
   \caption[$\tau(3889)$ vs. physical and chemical properties]{Relation between the self-absorption effects as traced by $\tau(3889)$  and different physical and chemical properties in \object{NGC~5253}.  From left to right and from top to bottom, these are electron density and temperature, excitation and relative abundance of nitrogen, N/O.  The correlation coefficients  are indicated in the corners of the individual plots.
 \label{tauvscosas}}
 \end{figure}

Likewise, Fig. \ref{tauvscosas} shows the relation between radiative effects, as traced by $\tau$(3889) and different properties of the gas. As expected, there is a good correlation with electron density. For electron temperatures lower than $\sim$10\,700~K, \hei\ emission lines do not suffer from radiative transfer effects in a significant way. However, above this threshold there is a clear correlation between the temperature and the relevance of the radiative transfer effects. 

More interestingly, as happened with the collisional effects, there is a good correlation between the excitation and the relevance of radiative transfer effects (lower left panel in Fig. \ref{tauvscosas}).  It is not clear why the excitation should correlate with the contribution of  collisional effects (mainly) or the radiative transfer effects (to a lesser extent). 

Something similar occurs with the relative abundance, $N/O$. At the typical $\log(N/O)$ values for this galaxy ($\lsim-1.3$), \hei\ radiative transfer effects appear to be negligible. However, those locations displaying extra nitrogen appear to show significant radiative transfer effects. Rather than a discernible correlation between individual spaxels, there appears a correlation between the upper limit  to the contribution of the radiative transfer effects and the amount of extra nitrogen. To our knowledge, there is no reason to link these two properties from the theoretical point of view. The fact that both the relations with $T_e$ and $N/O$ present bimodal behavior (i.e. with "turning points"), and that there is a similar structure for the electron temperature and relative abundance \citepalias{mon12}, implying a good correlation (i.e. $r=0.91$) between $T_e$ and $N/O$, points towards the local electron temperature as a common cause.

\section{Conclusions}

This is the fourth in a series of articles that make use of IFS-based data to study in detail the 2D physical and chemical properties of the gas in the main \ghiir\ of the nearby BCD \object{NGC~5253}. The main goal of this article was to estimate the contribution of the collisional and radiative transfer effects on the helium emission lines and to map, in turn, the helium abundance.
%
%
The major conclusions can be summarized as follows:

   \begin{enumerate}
      \item The collisional effects on the different helium transitions have been mapped for the first time in an extragalactic object. As expected, they reproduce the electron density structure. They are negligible (i.e. $\sim$0.1-0.6\%) for transitions in the singlet cascade while relatively important  for those transitions in the triplet cascade. In particular, they can contribute up to 20\% of the flux in the \hei$\lambda$7065 line.
      
      \item The contribution of the collisional effects is sensitive to the assumed $T_e$ for helium. Specifically, we found  differences $\sim25-35$\% for the $\lambda$3889 and $\lambda$7065 lines two reasonable assumptions for the $T_e$ sensitivity. Relative differences for the other lines were larger. However, this does not have important consequences  since the contribution of the collisional effect to the observed spectrum for these lines is negligible.

      \item We present a map for the optical depth at $\lambda$3889 in the main \ghiir\ of \object{NGC~5253}. $\tau(3889)$ is elevated over an extended and circular area of $\sim$30~pc in diameter, centered at the Super Star Cluster(s), where it reaches its maximum. 

      \item The singly ionized helium abundance, $y^+$, has been mapped using extinction corrected fluxes of six \hei\ lines, realistic assumptions for $T_e$, $n_e$, and the stellar absorption equivalent width as well as the most recent emissivities. We found a mean($\pm$ standard deviation of $10^3 y^+ \sim80.3(\pm2.7)$ over the mapped area.
      
      \item We derived total helium abundance maps using three possible icf(He)'s. 
The relation between the excitation and the total helium abundances is consistent with no abundance gradient.
Differences between the derived total abundances according to the three methods are \emph{larger} than statistical errors associated with the data themselves, emphasizing how uncertainties in the derivation of helium abundances are dominated by the adopted assumptions.
      
      \item We illustrated the difficulty of detecting a putative helium enrichment due to the presence of Wolf-Rayet stars in the main \ghiir. This is due to the comparatively large amount of preexisting helium. The data are marginally consistent with an excess in the $N/He$ ratio in the nitrogen enriched area of the order of the atmospheric $N/He$ ratios in W-R stars. However, this excess is also of the same order of the uncertainty estimated for the $N/He$ ratios in the nitrogen enriched and non-enriched areas.
      
      \item We explored the influence of the kinematics in the evaluation of the \hei\ radiative transfer effects. Our data empirically support  the use of the traditional assumption that motions in an extragalactic \hii\ region have a negligible effect in the estimation of the global optical depths. However, individually, the broad kinematic component (associated with an outflow) is affected by radiative transfer effects in a much more significant way than the narrow one.
      
      \item The local relationships between the contribution of collisional and radiative transfer effects to the helium lines and different physical and chemical properties of the gas have been explored. Interestingly, we found a relation between the amount of extra nitrogen and the upper limit of the contribution from radiative transfer effects that requires further investigation. We suggest the electron temperature as perhaps a common agent causing this relation.
      
 \end{enumerate}

\begin{acknowledgements}

We are very grateful to the referee for the careful and diligent reading of the manuscript as  well as for the useful comments that helped us to clarify and improve the first submitted version of this paper.
Also, we thank R. L. Porter for advice on use of his tabulated \hei\ emissivities and for so promptly informing us of the Corrigendum.

Based on observations carried out at the European Southern
Observatory, Paranal (Chile), programmes 078.B-0043(A) and
383.B-0043(A). This paper uses 
the plotting package \texttt{jmaplot}, developed by Jes\'us
Ma\'{\i}z-Apell\'aniz,
\texttt{http://dae45.iaa.csic.es:8080/$\sim$jmaiz/software}. This 
research made use of the NASA/IPAC Extragalactic 
Database (NED), which is operated by the Jet Propulsion Laboratory, California
Institute of Technology, under contract with the National Aeronautics and Space
Administration.

A.~M.-I. is supported by the Spanish Research Council within the program JAE-Doc, Junta para la Ampliaci\'on de Estudios, co-funded by the FSE.
A.M-I is also grateful to ESO - Garching, where part of this work was carried out, for their hospitality and funding via their visitor program. 
This work has been partially funded by the Spanish PNAYA, project AYA2010-21887 of the Spanish MINECO. 
The research leading to these results has received funding from the European Community's Seventh Framework Programme (/FP7/2007-2013/) under grant agreement No 229517.
\end{acknowledgements}

\bibliography{mybib_aa}{}

\begin{thebibliography}{49}
\expandafter\ifx\csname natexlab\endcsname\relax\def\natexlab#1{#1}\fi

\bibitem[{{Allington-Smith} {et~al.}(2002){Allington-Smith}, {Murray},
  {Content}, {Dodsworth}, {Davies}, {Miller}, {Jorgensen}, {Hook}, {Crampton},
  \& {Murowinski}}]{all02}
{Allington-Smith}, J., {Murray}, G., {Content}, R., {et~al.} 2002, \pasp, 114,
  892

\bibitem[{{Alonso-Herrero} {et~al.}(2010){Alonso-Herrero},
  {Garc{\'{\i}}a-Mar{\'{\i}}n}, {Rodr{\'{\i}}guez Zaur{\'{\i}}n},
  {Monreal-Ibero}, {Colina}, \& {Arribas}}]{alo10}
{Alonso-Herrero}, A., {Garc{\'{\i}}a-Mar{\'{\i}}n}, M., {Rodr{\'{\i}}guez
  Zaur{\'{\i}}n}, J., {et~al.} 2010, \aap, 522, A7

\bibitem[{{Alonso-Herrero} {et~al.}(2004){Alonso-Herrero}, {Takagi}, {Baker},
  {Rieke}, {Rieke}, {Imanishi}, \& {Scoville}}]{alo04}
{Alonso-Herrero}, A., {Takagi}, T., {Baker}, A.~J., {et~al.} 2004, \apj, 612,
  222

\bibitem[{{Aver} {et~al.}(2010){Aver}, {Olive}, \& {Skillman}}]{ave10}
{Aver}, E., {Olive}, K.~A., \& {Skillman}, E.~D. 2010, \jcap, 5, 3

\bibitem[{{Beck} {et~al.}(2012){Beck}, {Lacy}, {Turner}, {Kruger}, {Richter},
  \& {Crosthwaite}}]{bec12}
{Beck}, S.~C., {Lacy}, J.~H., {Turner}, J.~L., {et~al.} 2012, \apj, 755, 59

\bibitem[{{Benjamin} {et~al.}(1999){Benjamin}, {Skillman}, \& {Smits}}]{ben99}
{Benjamin}, R.~A., {Skillman}, E.~D., \& {Smits}, D.~P. 1999, \apj, 514, 307

\bibitem[{{Cid Fernandes} {et~al.}(2005){Cid Fernandes}, {Mateus}, {Sodr{\'e}},
  {Stasi{\'n}ska}, \& {Gomes}}]{cid05}
{Cid Fernandes}, R., {Mateus}, A., {Sodr{\'e}}, L., {Stasi{\'n}ska}, G., \&
  {Gomes}, J.~M. 2005, \mnras, 358, 363

\bibitem[{{Cid Fernandes} {et~al.}(2009){Cid Fernandes}, {Schoenell}, {Gomes},
  {Asari}, {Schlickmann}, {Mateus}, {Stasinska}, {Sodr{\'e}}, \&
  {Torres-Papaqui}}]{cid09}
{Cid Fernandes}, R., {Schoenell}, W., {Gomes}, J.~M., {et~al.} 2009, in Revista
  Mexicana de Astronomia y Astrofisica, vol. 27, Vol.~35, Revista Mexicana de
  Astronomia y Astrofisica Conference Series, 127--132

\bibitem[{{Ferland}(1980)}]{fer80}
{Ferland}, G.~J. 1980, \mnras, 191, 243

\bibitem[{{Fluks} {et~al.}(1994){Fluks}, {Plez}, {The}, {de Winter},
  {Westerlund}, \& {Steenman}}]{flu94}
{Fluks}, M.~A., {Plez}, B., {The}, P.~S., {et~al.} 1994, \aaps, 105, 311

\bibitem[{{Fukugita} \& {Kawasaki}(2006)}]{fuk06}
{Fukugita}, M. \& {Kawasaki}, M. 2006, \apj, 646, 691

\bibitem[{{Gonz{\'a}lez Delgado} {et~al.}(2005){Gonz{\'a}lez Delgado},
  {Cervi{\~n}o}, {Martins}, {Leitherer}, \& {Hauschildt}}]{gon05}
{Gonz{\'a}lez Delgado}, R.~M., {Cervi{\~n}o}, M., {Martins}, L.~P.,
  {Leitherer}, C., \& {Hauschildt}, P.~H. 2005, \mnras, 357, 945

\bibitem[{{Gruenwald} {et~al.}(2002){Gruenwald}, {Steigman}, \&
  {Viegas}}]{gru02}
{Gruenwald}, R., {Steigman}, G., \& {Viegas}, S.~M. 2002, \apj, 567, 931

\bibitem[{{Harbeck} {et~al.}(2012){Harbeck}, {Gallagher}, \&
  {Crnojevi{\'c}}}]{har12}
{Harbeck}, D., {Gallagher}, J., \& {Crnojevi{\'c}}, D. 2012, \mnras, 422, 629

\bibitem[{{Harris} {et~al.}(2004){Harris}, {Calzetti}, {Gallagher}, {Smith}, \&
  {Conselice}}]{har04}
{Harris}, J., {Calzetti}, D., {Gallagher}, III, J.~S., {Smith}, D.~A., \&
  {Conselice}, C.~J. 2004, \apj, 603, 503

\bibitem[{{Izotov} \& {Thuan}(2010)}]{izo10}
{Izotov}, Y.~I. \& {Thuan}, T.~X. 2010, \apjl, 710, L67

\bibitem[{{Izotov} {et~al.}(2007){Izotov}, {Thuan}, \& {Stasi{\'n}ska}}]{izo07}
{Izotov}, Y.~I., {Thuan}, T.~X., \& {Stasi{\'n}ska}, G. 2007, \apj, 662, 15

\bibitem[{{Karachentsev} {et~al.}(2007){Karachentsev}, {Tully}, {Dolphin},
  {Sharina}, {Makarova}, {Makarov}, {Sakai}, {Shaya}, {Kashibadze},
  {Karachentseva}, \& {Rizzi}}]{kar07}
{Karachentsev}, I.~D., {Tully}, R.~B., {Dolphin}, A., {et~al.} 2007, \aj, 133,
  504

\bibitem[{{Kobulnicky} {et~al.}(1997){Kobulnicky}, {Skillman}, {Roy}, {Walsh},
  \& {Rosa}}]{kob97}
{Kobulnicky}, H.~A., {Skillman}, E.~D., {Roy}, J.-R., {Walsh}, J.~R., \&
  {Rosa}, M.~R. 1997, \apj, 477, 679

\bibitem[{{Kunth} \& {Sargent}(1983)}]{kun83}
{Kunth}, D. \& {Sargent}, W.~L.~W. 1983, \apj, 273, 81

\bibitem[{{L{\'o}pez-S{\'a}nchez} {et~al.}(2007){L{\'o}pez-S{\'a}nchez},
  {Esteban}, {Garc{\'{\i}}a-Rojas}, {Peimbert}, \& {Rodr{\'{\i}}guez}}]{lop07}
{L{\'o}pez-S{\'a}nchez}, {\'A}.~R., {Esteban}, C., {Garc{\'{\i}}a-Rojas}, J.,
  {Peimbert}, M., \& {Rodr{\'{\i}}guez}, M. 2007, \apj, 656, 168

\bibitem[{{Luridiana}(2009)}]{lur09}
{Luridiana}, V. 2009, \apss, 324, 361

\bibitem[{{Markwardt}(2009)}]{mar09}
{Markwardt}, C.~B. 2009, in Astronomical Society of the Pacific Conference
  Series, Vol. 411, Astronomical Society of the Pacific Conference Series, ed.
  {D.~A.~Bohlender, D.~Durand, \& P.~Dowler}, 251--+

\bibitem[{{Martins} {et~al.}(2005){Martins}, {Gonz{\'a}lez Delgado},
  {Leitherer}, {Cervi{\~n}o}, \& {Hauschildt}}]{mar05b}
{Martins}, L.~P., {Gonz{\'a}lez Delgado}, R.~M., {Leitherer}, C.,
  {Cervi{\~n}o}, M., \& {Hauschildt}, P. 2005, \mnras, 358, 49

\bibitem[{{Monreal-Ibero} {et~al.}(2010){Monreal-Ibero}, {V{\'{\i}}lchez},
  {Walsh}, \& {Mu{\~n}oz-Tu{\~n}{\'o}n}}]{mon10}
{Monreal-Ibero}, A., {V{\'{\i}}lchez}, J.~M., {Walsh}, J.~R., \&
  {Mu{\~n}oz-Tu{\~n}{\'o}n}, C. 2010, \aap, 517, A27+, (Paper I)

\bibitem[{{Monreal-Ibero} {et~al.}(2012){Monreal-Ibero}, {Walsh}, \&
  {V{\'{\i}}lchez}}]{mon12}
{Monreal-Ibero}, A., {Walsh}, J.~R., \& {V{\'{\i}}lchez}, J.~M. 2012, \aap,
  544, A60, (Paper II)

\bibitem[{{Olive} \& {Skillman}(2001)}]{oli01}
{Olive}, K.~A. \& {Skillman}, E.~D. 2001, \na, 6, 119

\bibitem[{{Osterbrock} \& {Ferland}(2006)}]{ost06}
{Osterbrock}, D.~E. \& {Ferland}, G.~J. 2006, {Astrophysics of gaseous nebulae
  and active galactic nuclei}, ed. D.~E. {Osterbrock} \& G.~J. {Ferland}

\bibitem[{{Pagel} {et~al.}(1986){Pagel}, {Terlevich}, \& {Melnick}}]{pag86}
{Pagel}, B.~E.~J., {Terlevich}, R.~J., \& {Melnick}, J. 1986, \pasp, 98, 1005

\bibitem[{{Pasquini} {et~al.}(2002){Pasquini}, {\'Avila}, {Blecha}, {Cacciari},
  {Cayatte}, {Colless}, {Damiani}, {de Propris}, {Dekker}, {di Marcantonio},
  {Farrell}, {Gillingham}, {Guinouard}, {Hammer}, {Kaufer}, {Hill}, {Marteaud},
  {Modigliani}, {Mulas}, {North}, {Popovic}, {Rossetti}, {Royer}, {Santin},
  {Schmutzer}, {Simond}, {Vola}, {Waller}, \& {Zoccali}}]{pas02}
{Pasquini}, L., {\'Avila}, G., {Blecha}, A., {et~al.} 2002, The Messenger, 110,
  1

\bibitem[{{Peimbert} {et~al.}(2002){Peimbert}, {Peimbert}, \&
  {Luridiana}}]{pei02}
{Peimbert}, A., {Peimbert}, M., \& {Luridiana}, V. 2002, \apj, 565, 668

\bibitem[{{Peimbert} {et~al.}(2007){Peimbert}, {Luridiana}, \&
  {Peimbert}}]{pei07}
{Peimbert}, M., {Luridiana}, V., \& {Peimbert}, A. 2007, \apj, 666, 636

\bibitem[{{Peimbert} \& {Torres-Peimbert}(1976)}]{pei76}
{Peimbert}, M. \& {Torres-Peimbert}, S. 1976, \apj, 203, 581

\bibitem[{{Peimbert} \& {Torres-Peimbert}(1977)}]{pei77}
{Peimbert}, M. \& {Torres-Peimbert}, S. 1977, \mnras, 179, 217

\bibitem[{{Porter} {et~al.}(2007){Porter}, {Ferland}, \& {MacAdam}}]{por07}
{Porter}, R.~L., {Ferland}, G.~J., \& {MacAdam}, K.~B. 2007, \apj, 657, 327

\bibitem[{{Porter} {et~al.}(2012){Porter}, {Ferland}, {Storey}, \&
  {Detisch}}]{por12}
{Porter}, R.~L., {Ferland}, G.~J., {Storey}, P.~J., \& {Detisch}, M.~J. 2012,
  \mnras, L487

\bibitem[{{Porter} {et~al.}(2013){Porter}, {Ferland}, {Storey}, \&
  {Detisch}}]{por13}
{Porter}, R.~L., {Ferland}, G.~J., {Storey}, P.~J., \& {Detisch}, M.~J. 2013,
  ArXiv e-prints

\bibitem[{{Robbins}(1968)}]{rob68}
{Robbins}, R.~R. 1968, \apj, 151, 511

\bibitem[{{Sakai} {et~al.}(2004){Sakai}, {Ferrarese}, {Kennicutt}, \&
  {Saha}}]{sak04}
{Sakai}, S., {Ferrarese}, L., {Kennicutt}, Jr., R.~C., \& {Saha}, A. 2004,
  \apj, 608, 42

\bibitem[{{Sauer} \& {Jedamzik}(2002)}]{sau02}
{Sauer}, D. \& {Jedamzik}, K. 2002, \aap, 381, 361

\bibitem[{{Shields}(1993)}]{shi93}
{Shields}, J.~C. 1993, \apj, 419, 181

\bibitem[{Sidoli(2010)}]{sid10}
Sidoli, F. 2010, PhD thesis, University of London, UK

\bibitem[{{Smith} \& {Willis}(1982)}]{smi82}
{Smith}, L.~J. \& {Willis}, A.~J. 1982, \mnras, 201, 451

\bibitem[{{Storey} \& {Hummer}(1995)}]{sto95}
{Storey}, P.~J. \& {Hummer}, D.~G. 1995, \mnras, 272, 41

\bibitem[{{Turner} {et~al.}(2000){Turner}, {Beck}, \& {Ho}}]{tur00}
{Turner}, J.~L., {Beck}, S.~C., \& {Ho}, P.~T.~P. 2000, \apjl, 532, L109

\bibitem[{{Viegas} {et~al.}(2000){Viegas}, {Gruenwald}, \& {Steigman}}]{vie00}
{Viegas}, S.~M., {Gruenwald}, R., \& {Steigman}, G. 2000, \apj, 531, 813

\bibitem[{{Walsh} \& {Roy}(1989)}]{wal89}
{Walsh}, J.~R. \& {Roy}, J.-R. 1989, \mnras, 239, 297

\bibitem[{{Westmoquette} {et~al.}(2013){Westmoquette}, {James},
  {Monreal-Ibero}, \& {Walsh}}]{wes13}
{Westmoquette}, M.~S., {James}, B., {Monreal-Ibero}, A., \& {Walsh}, J.~R.
  2013, \aap, 550, A88, (Paper III)

\bibitem[{{Zhang} {et~al.}(2005){Zhang}, {Liu}, {Liu}, \& {Rubin}}]{zha05}
{Zhang}, Y., {Liu}, X.-W., {Liu}, Y., \& {Rubin}, R.~H. 2005, \mnras, 358, 457

\end{thebibliography}
\bibliographystyle{./aa}

\end{document}